\documentclass[floatfix,aps,reprint,superscriptaddress,amsmath,amssymb,footnotebib,onecolumn]{revtex4-2}
\usepackage{amsmath}

\usepackage{graphicx}
\graphicspath{{./}{figs/.}}

\usepackage{dcolumn}
\usepackage{bm}
\usepackage{float}
\usepackage[english]{babel} 
\usepackage[T1]{fontenc}
\usepackage[usenames,dvipsnames]{xcolor}
\usepackage[colorlinks=true,citecolor=Cerulean,linkcolor=RubineRed,urlcolor=Cerulean]{hyperref}
\usepackage{booktabs}
\usepackage{mathtools}
\usepackage{physics}
\usepackage{txfonts}
\usepackage{framed}
\usepackage{dsfont}
\usepackage{soul}

\usepackage{tabularray}
\usepackage{array}
\usepackage{calc}

\usepackage{pifont}

\newcommand{\steve}[1]{#1}
\newcommand{\callum}[1]{\textcolor{blue}{[#1]}}

\graphicspath{{./figs/}, {../visual_elements/figs/generated_figs/} }

\begin{document}

\title{Taming Quantum Systems: A Tutorial for Using Shortcuts-To-Adiabaticity, \\ 
Quantum Optimal Control, and Reinforcement Learning 
}

\author{Callum W. Duncan}
\affiliation{Aegiq Ltd., Cooper Buildings, Arundel Street, Sheffield, S1 2NS, United Kingdom}
\affiliation{Department of Physics, SUPA and University of Strathclyde, Glasgow G4 0NG, United Kingdom}
\author{Pablo M. Poggi}
\affiliation{Department of Physics, SUPA and University of Strathclyde, Glasgow G4 0NG, United Kingdom}
\affiliation{Center for Quantum Information and Control, Department of Physics and Astronomy, University of New Mexico, Albuquerque, New Mexico 87131, USA}
\author{Marin Bukov}
\affiliation{Max Planck Institute for the Physics of Complex Systems, Nöthnitzer Str. 38, 01187 Dresden, Germany}
\author{Nikolaj Thomas Zinner}
\affiliation{Department of Physics and Astronomy, Aarhus University, DK-8000 Aarhus C, Denmark}
\affiliation{Kvantify Aps, DK-2100 Copenhagen, Denmark}
\author{Steve Campbell}
\affiliation{School of Physics, University College Dublin, Belfield, Dublin 4, Ireland}
\affiliation{Centre for Quantum Engineering, Science, and Technology, University College Dublin, Belfield, Dublin 4, Ireland}
\affiliation{Dahlem Center for Complex Quantum Systems, Freie Universit\"at Berlin, Arnimallee 14, 14195 Berlin, Germany}

\begin{abstract}
\vskip1cm
Precise manipulation of quantum effects at the atomic and nanoscale has become an essential task in ongoing scientific and technological endeavours. Quantum control methods are thus routinely exploited for research in areas such as quantum materials, quantum chemistry, and atomic and molecular physics, as well as in the development of quantum technologies like computing, simulation, and sensing.  Here, we present a pedagogical introduction to the basics of quantum control methods in tutorial form, with the aim of providing newcomers to the field with the core concepts and practical tools to use these methods in their research. We focus on three areas: shortcuts to adiabaticity, quantum optimal control, and machine-learning-based control. We lay out the basic theoretical elements of each area in a pedagogical way and describe their application to a series of example cases. For these, we include detailed analytical derivations as well as extensive numerical results.
As an outlook, we discuss quantum control methods in the broader context of quantum technologies development and complex quantum systems research, outlining potential connections and synergies between them.
\end{abstract}
\date{\today}
\maketitle

\vskip1cm


 \begin{figure}[h!]
 \centering
 \includegraphics[width=1.5\columnwidth]{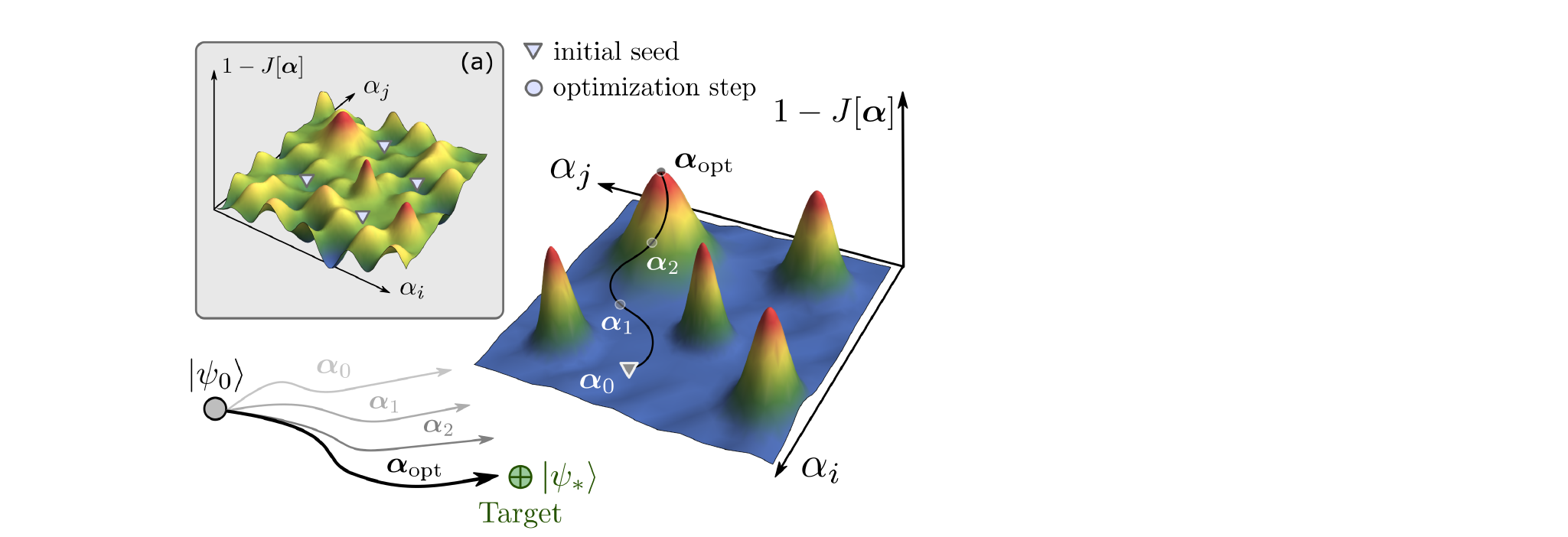}
 \end{figure}

\clearpage

\tableofcontents
\clearpage


\section{Introduction}\label{sec:Introduction}
Ever since experiments gained access to the quantum regime, there has been a continuing need to develop and perfect efficient techniques to manipulate quantum systems. These manipulations occur in real time, and the corresponding time-dependent control parameter can be described by a function of time, called the \textit{control protocol}. The recent development of quantum-enhanced devices has led to a significant surge in activity which has witnessed some remarkable advances, both theoretical and experimental, in the use of these control techniques; for example quantum control has found fertile application in assessing the thermodynamics of quantum systems, providing a means to boost the performance of nanoscale quantum heat engines, to develop ultraprecise quantum sensors, and to robustly implement quantum gates~\cite{Koch2022, STAreview}. 

Unlike classical mechanics, where the most common way of achieving control is to apply mechanical forces, in quantum control electromagnetic fields prevail, both in the DC and the AC regime. This can be traced back to the necessity to couple to Angstrom-sized atoms, and even smaller objects, where quantum effects become prevalent. Moving beyond the control of a single system to few and multi-particle settings introduces other control knobs besides external potentials, in particular, varying the interaction strength. For instance, in ultracold atoms this can be achieved using so-called Feshbach resonances~\cite{bloch2008manybody}, while in systems of optical tweezer arrays, one can directly control the positions of the trapped atoms which effectively fine-tunes the strength of the position-dependent interaction potential~\cite{Kaufman2021}. The ability to manipulate an arbitrary quantum system then becomes a multifaceted problem, one which must also be cognisant of other physically relevant constraints, for example: a finite control pulse energy and/or bandwidth, smoothness of the control protocol, or the intrinsic decoherence time which will bound the protocol time duration that can vary greatly between different physical platforms~\cite{Koch2022, STAreview, PRXQtutorial, baldwin2021optimal}. These constraints can feed directly into characterising the efficacy of a control protocol with the most common figure of merit being the fidelity (quantum overlap), between the protocol-evolved final state, $\ket{\psi(t)}$ and the desired target state, $\ket{\psi_*}$, 
\begin{equation}
\label{Eq1Fidelity}
\mathcal{F} = \vert \bra{\psi_*} \psi(t)\rangle \vert^2
\end{equation}
where we have assumed the states in Eq.~\eqref{Eq1Fidelity} are pure (as will be the case for all examples in this tutorial); however, we remark that these states could also be eigenstates or mixed ensembles. 
While we will focus on the use of fidelity as our key figure of merit throughout this tutorial, it is important to note that when it comes to many-body states this should be done with care since it often decreases exponentially with the number of particles. Indeed, if we consider a product state with single particle fidelity of $0.99$, the many-body fidelity of a system with $100$ particles is $0.99^{100}\approx 0.36\ll 0.99$. In general there are many possible cost functions that can be used when optimising the dynamics of a quantum system, for example, the energy (density) is typically minimized in the process of preparing the ground state of a quantum system. More generally, one might be interested in maximizing and/or minimizing the expectation values of a specific observable, e.g., the magnetization, the momentum distribution, the strength of a correlation function or less frequently accessible quantities, such as quantum entanglement~\cite{kraus2001optimal,goerz2011,watts2015optimizing,tashev2024reinforcement}.  Regardless of the specific choice, the techniques presented below can be readily adapted for a given situation.

In large, spatially extended systems, it is often useful to consider two kinds of control protocols: global and local. Global protocols usually act (almost) homogeneously on the entire system, while local protocols can either be applied to a smaller part of the system or to individual constituents, e.g., the atoms of a chain using single-site addressability. Whatever the means of controlling a quantum system, it is important to keep in mind that the control protocols that can be emulated in experiments typically correspond to local (or sums of local) operators in the Hamiltonian description of the system; this is particularly relevant for many-body control. Thus, locality imposes a formidable, yet very physical, constraint on the accessibility of control fields~\cite{bukov2019geometric}. The various control techniques that are known to be effective for single, isolated quantum systems must be carefully analyzed in view of this constraint when the system size is scaled up. Providing an introduction to the tools and techniques necessary to understand the interplay between implementability and scalability of control protocols for quantum systems is the focus of this tutorial. 


\subsection{Scope of this tutorial}

We aim to provide a succinct entry point to some of the state-of-the-art approaches to controlling quantum systems. There are many excellent reviews available on these and related topics, see for example Refs~\cite{Deffner2017, Glaser2015, Koch2022, STAreview, Stefanatos2020, PRXQtutorial, GiannelliPLA, Hatomura2024, ansel2024_arxiv, deffner2020thermodynamic}, which the interested reader can delve into. Here, we aim to achieve two goals: Firstly, to provide a highly pedagogical introduction to the most prevalent techniques used to control many-body systems; thus, in addition to the technical details of the control protocols themselves, we also provide extensive details on other important tools for solving paradigmatic many-body systems. Secondly, by introducing a suite of techniques arising from the various subcommunities that work on quantum control, our aim is to highlight the benefits that can arise from employing hybridized control techniques, i.e. those that take advantage of aspects of several control strategies.

\begin{figure}[t]
\includegraphics[width=0.9\columnwidth]{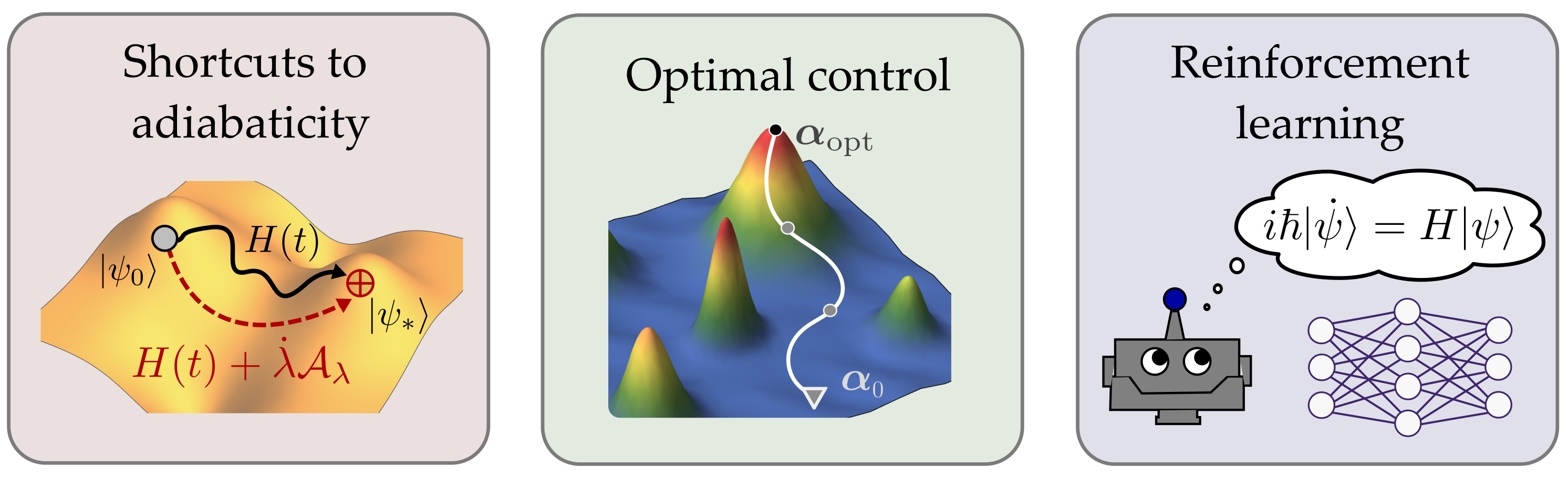}
\label{fig:scope}
\caption{Schematic of the main three approaches to quantum control that are discussed in this tutorial article: shortcuts to adiabaticity (Sec. \ref{sec:STA}, optimal control (Sec. \ref{sec:QOC}), and reinforcement learning (Sec. \ref{sec:RL_theory}). }
\end{figure}

With these aims in mind, in what follows we present the basic ingredients for three of the leading approaches to achieving non-adiabatic control (Fig.~\ref{fig:scope}): shortcuts-to-adiabaticity, quantum optimal control, and reinforcement learning. They are presented in isolation (and in no particular order) to ensure that the core theoretical underpinnings of each approach are presented in a suitably pedagogical manner. As such, the interested reader can dive straight into a given technique without the need to work through any preceding sections. For a given task, the choice of which approach to employ depends on a range of competing factors. The experimental architecture clearly imposes certain constraints, e.g. the available set of controls and/or measurements. Beyond such hardware limitations, however, other requirements necessarily requested from the dynamics will also impact the choice of control technique to use. For instance, if the system must follow a specific dynamical path in Hilbert space, e.g. remaining in the ground state, then the suite of tools from shortcuts-to-adiabaticity are likely to be the most suitable choice. However, in many situations the intermediate dynamics is not of any particular relevance and we are only interested in e.g. state transfer or state preparation. In these cases, often the goal is simply to find a protocol that works under the given constraints and, therefore, the task becomes principally about pulse shape optimisation where optimal control provides several efficient approaches. If, on the other hand, one is interested in automating the procedure of finding optimal protocols, then adopting a deep learning approach is advisable, as it allows trained RL agents to generalize and produce control protocols even for parameter values not encountered during optimization; moreover, RL algorithms are particularly useful for on-line feedback control based on (partial) measurements. In Sec.~\ref{sec:comparison} we provide further discussion along these lines, aiming to highlight some of the strengths and weaknesses of the techniques that we discuss in this tutorial and highlight the benefit that comes from combining techniques to tackle a problem.

Another motivation for this tutorial, that we believe sets this work apart from other reviews and pedagogical resources on the topic, is to provide an ``introductory one-stop-shop'' for the various control subcommunities. To facilitate this, we explicitly demonstrate the application of all protocols for the manipulation of a single two-level quantum system, in addition to several other relevant examples. \steve{These additional examples focus on the some of most common areas to which a given technique is applied. Therefore in Sec.~\ref{sec:STA} we consider the exact and approximate control of critical many-body systems, Sec.~\ref{sec:QOC} provides examples relevant to quantum computation focussing on the implementation of gates and the generation of multipartite entanglement, while Sec.~\ref{sec:RL_theory} explores universal single-qubit state preparation and active qubit feedback control using an ancilla.} In the hope of bringing some consistency to the nomenclature, we also provide tables summarising the notation employed when introducing a given technique in Table~\ref{table:AGPsSymbols} for shortcuts-to-adiabaticity, Table~\ref{table:QOCSymbols} for optimal control, and Table~\ref{table:RLSymbols} for reinforcement learning. Finally, the reader can refer to Table~\ref{table:Abbreviations} for a list of abbreviations commonly used throughout this tutorial.
We provide Jupyter notebooks for interested readers to explore further the control techniques introduced on GitHub~\cite{github_code}.

\subsection{Methodologies of quantum control}

We can broadly define three classes of quantum systems of increasing control complexity: Few (including single) particle systems are often used to demonstrate the basics of control methods and can also benchmark numerical quantum control algorithms against exactly solvable problems. However, they also bear relevance to real systems such as qubits (two-level systems) and qutrits or $\Lambda$-systems (three-level systems). In these settings common control problems include state population transfer or finding optimal single- or two-qubit gates. The relative simplicity of the models typically can allow for the inclusion of dissipation effects~\cite{Koch2016OpenSys}. The next class of systems that we will be primarily interested in are fermionic band models and weakly interacting bosons described by quadratic Hamiltonians. Examples in this category include topological insulators, weakly-interacting bosons, the transverse-field Ising model, and the Kitaev honeycomb model. The translation invariant counterparts of such models are exactly solvable in momentum space, which offers the possibility to explore some control techniques analytically. In particular, two-band models can be viewed as a collection of independent two-level systems labelled by (quasi)momentum. As we will demonstrate, control over the individual momentum modes allows for optimal control of the entire system, however, the momentum-dependence ultimately compromises the locality of the applied control fields in real space. Nevertheless, useful insight into the most relevant operators for achieving control can be learned which can prove valuable in moving to the final class of systems. The most complex control class comprises many-body systems with no known exact solutions, e.g., nonintegrable models with Hubbard- and Heisenberg-type Hamiltonians. In general we can expect here optimal control techniques to work whenever the state of interest is gapped in energy; for gapless models much less is known regarding their controllability, and there are many open questions at present.

Broadly speaking, we distinguish between adiabatic and non-adiabatic techniques when classifying quantum control protocols. Adiabatic protocols (which we will rigorously define for quantum systems in the proceeding sections) make use of the discrete structure of quantized spectra and high-fidelity control timescales are inversely proportional to the energy gaps of the controlled level to its neighboring levels. A famous example is the Landau-Zener model, where an exponentially small fraction is left in the excited state after the completion of the adiabatic process~\cite{LZreview}. Stimulated Rapid Adiabatic Passage (STIRAP)~\cite{STIRAPreview} is another technique which exploits spectral characteristics to achieve robust state transfer. In this tutorial, we will focus explicitly on non-adiabatic techniques. This class encompasses counter-diabatic (CD) protocols that generate a transitionless time evolution with respect to the instantaneous basis of the Hamiltonian describing the system of interest, which is typically far from the adiabatic regime. It also captures the broader set of Hamiltonians that are tailored to bring the controlled system into a target state within a fixed protocol duration (but allows for the creation of excitations during the evolution so long as they are removed before the protocol comes to an end) as is typical in many quantum optimal control approaches.\\

\section{Shortcuts to Adiabaticity}
\label{sec:STA}

Many practical tasks involve quantum adiabatic processes and, therefore, a significant body of work has amassed in developing techniques that either exploit the inherent adiabatic dynamics characteristic of slowly driven systems or, as will be the focus of this section, achieve an effective adiabatic dynamics in a finite time. We will first carefully define the concept of adiabaticity before focusing on counterdiabatic driving, which is arguably the conceptually simplest control method for achieving the desired evolution and works by directly suppressing any non-adiabatic transitions. We will review some modern reformulations and extensions of this idea, in particular discussing how to construct the control fields through the adiabatic gauge potential and how to determine local approximations to achieving the control using variational techniques. We will explicitly consider the control of a simple two level system described by the Landau-Zener model in Sec.~\ref{sec:LandauZener}, which will serve as a paradigmatic example that will also be considered when discussing the other control techniques that are a focus of this tutorial, namely optimal control in Sec.~\ref{subsec:QOC_example_state} and reinforcement learning in Sec.~\ref{sec:RL_1q} and~\ref{sec:RL_2q}. We will also consider the application of these techniques to simple many-body settings embodied by three variations of the Ising model: (i) nearest neighbor, (ii) infinite range, and (iii) non-integrable.
\vskip1cm

\subsection{The Adiabatic Theorem}

It is important from the outset to give a precise definition of an adiabatic process to avoid some common confusion: we recall that, in thermodynamics, adiabatic processes are those that do not transfer heat between the system and its environment, in the sense that they occur on timescales faster than the scales required for heat exchange; they may be slow or fast. By contrast, in quantum mechanics, as we shall see below, adiabatic evolution is defined by the lack of population transfer between an eigenstate of interest undergoing unitary evolution and the rest of the states in the instantaneous spectrum. Note that the energy of the state of interest can change in the process of adiabatic evolution, as the drive does work on the system, but this does not give rise to any heat; heat in quantum mechanics is defined microscopically as arising from any transitions between the instantaneous eigenstates of the Hamiltonian, caused by a nonadiabatic evolution. Adiabaticity in quantum mechanics is thus different from the notion of adiabaticity in classical thermodynamics; in fact, quantum adiabaticity is closer to what is called a quasistatic process in thermodynamics where the system remains in equilibrium throughout the evolution~\cite{dalessio2016quantum}. In this tutorial, the concept of adiabaticity refers to quantum mechanics.

To understand control protocols that achieve an effectively adiabatic evolution, let us now formalise the concept of adiabaticity for quantum dynamics. The suite of techniques termed ``shortcuts-to-adiabaticity'' arise precisely from exploring approaches that mimic an adiabatic dynamics on, in principle, arbitrary timescales. Thus, we begin with a recapitulation of the adiabatic theorem, which already provides a wealth of knowledge regarding the types of generators that allow for the control of a quantum system. There are several rigorous derivations of the adiabatic theorem, starting from the earliest incarnations arising from Erhenfest's analysis and the proof by Born and Fock~\cite{Born1928} which have later been refined and extended by Kato to the case of degenerate eigenstates~\cite{Kato1950}. Herein, we follow a similar treatment found in many standard texts and reviews, e.g.~\cite{BudichReview}.

Consider a quantum system described by the time-dependent non-degenerate Hamiltonian $H(t)$ which evolves according to the time-dependent Schr\"odinger equation
\begin{equation}
\label{TDSE}
i\hbar \partial_t \ket{\psi(t)} = H(t) \ket{\psi(t)}.
\end{equation}
Furthermore, it is clear that at every instant in time the system also satisfies the time-independent Schr\"odinger equation
\begin{equation}
\label{TISE}
H(t) \ket{n(t)} = E_n(t) \ket{n(t)}
\end{equation}
where $\{ \ket{n(t)} \}$ and $\{ E_n(t) \}$ are the {\it instantaneous} eigenstates and eigenenergies of the system, i.e. this is simply the eigenvalue equation for the matrix $H(t)$ treating $t$ as a parameter. Since the states $\{ \ket{n(t)} \}$ form a valid basis for the Hilbert space of the system at every instant of time, we can use them to express the solution to Eq.~\eqref{TDSE}, i.e. we can assume an ansatz solution of the form
\begin{equation}
\label{ansatz}
\ket{\psi(t)} = \sum_n c_n(t) \ket{n(t)}
\end{equation}
with $c_n(t)$ some complex coefficients. Putting Eq.~\eqref{ansatz} into Eq.~\eqref{TDSE} and using the spectral decomposition of $H(t)$ we have 
\begin{eqnarray}
i \hbar \left[ \partial_t \left( \sum_n c_n(t) \ket{n(t)} \right) \right] & = & H(t) \left[  \sum_n c_n(t) \ket{n(t)} \right] \nonumber \\
& = & \sum_k E_k(t) \ketbra{k(t)}{k(t)} \left[  \sum_n c_n(t) \ket{n(t)} \right].
\end{eqnarray}
Dropping the explicit time-dependence for brevity (i.e. $H(t)\!\equiv\! H$, $\ket{n(t)}\!\equiv\! \ket{n}$, and $E_n(t)\!\equiv \! E_n$) and denoting the time derivative of any quantity $x(t)$ by $\partial_t x(t) \equiv \dot{x}$
\begin{eqnarray}
i\hbar \left[ \sum_n \left( \dot{c}_n \ket{n} + c_n \ket{\dot{n}} \right) \right] & = & \sum_{n,k} c_n E_k \ketbra{k}{k} n\rangle \nonumber \\
&=& \sum_n c_n E_n \ket{n}. 
\end{eqnarray}
To get a set of differential equations for the coefficients $c_n$ we project onto the $m^\text{th}$ eigenstate, $\bra{m}$, of $H$, 
\begin{equation}
i\hbar \left[ \sum_n \Big( \dot{c}_n \braket{m}{n} + c_n \braket{m}{\dot{n}} \Big) \right] = \sum_n c_n E_n \braket{m}{n},
\end{equation}
which again due to orthonormality allows us to simplify most of the sums. Separating terms with $m=n$ from $m\neq n$ we find
\begin{equation}
i\hbar \left(\dot{c}_m + c_m \braket{m}{\dot{m}} + \sum_{n\neq m} c_n \braket{m}{\dot{n}} \right) = c_m E_m
\end{equation}
\begin{eqnarray}
\implies i\hbar \dot{c}_m &=& c_m E_m - i \hbar c_m \braket{m}{\dot{m}} - i\hbar \sum_{m\neq n} c_n \braket{m}{\dot{n}} \nonumber \\
&=& \Big(E_m -i\hbar \braket{m}{\dot{m}} \Big)c_m - i\hbar \sum_{n\neq m}c_n \braket{m}{\dot{n}}. \label{eq:Adiabaticity1}
\end{eqnarray}
The key to the adiabatic theorem lies in the last term of Eq.~\eqref{eq:Adiabaticity1}. If this term is negligible, the evolution of the coefficients, $c_m$, obeys a simple first-order differential equation
\begin{eqnarray}
\frac{\dot{c}_m}{c_m} &=& - \frac{i}{\hbar} \Big( E_m - i\hbar \braket{m}{\dot{m}} \Big) \nonumber \\
 \implies c_m &=& \exp\left[ - \frac{i}{\hbar} \int_0^t E_m - i\hbar \braket{m}{\dot{m}}~\mathrm ds  \right].
\end{eqnarray} 
This can be conveniently cast as 
\begin{equation}
c_m(t)=\exp[-i (\phi_d(t) + \phi_g(t))],
\end{equation}
with the dynamic phase $\phi_d(t)=\frac{1}{\hbar} \int_0^t E_m(s)\mathrm d s$ and the geometric phase $\phi_g(t) = -\int_0^t i \braket{m(s)}{\partial_s{m(s)}}\mathrm ds$, where we have briefly reintroduced the explicit time-dependence.
Clearly, if the system begins in the $n^\text{th}$ eigenstate of $H(0)$, then the only non-trivial equation for the coefficients will correspond to the initial condition $c_n(0)=1$, and it will remain in this state only accumulating the dynamical and geometric phases~\cite{Berry1984}.

Thus, to understand the adiabatic theorem we must examine the last term in Eq.~\eqref{eq:Adiabaticity1} more carefully. Notice that this term effectively captures the instantaneous coupling between eigenstates due to the rate of change of the Hamiltonian. In order for this term to be negligible, we require $\braket{m}{\dot{n}}\to0$. Let us analyse this term more closely by taking the time derivative of Eq.~\eqref{TISE}
\begin{eqnarray}
\partial_t \Big( H \ket{n} \Big) & = & \partial_t \Big( E_n \ket{n} \Big) \nonumber \\
\dot{H} \ket{n} + H \ket{\dot{n}} &=& \dot{E}_n \ket{n} + E_n \ket{\dot{n}}.
\end{eqnarray}
Projecting again onto the $m^\text{th}$ state and recalling that the sum in Eq.~\eqref{eq:Adiabaticity1} involves only $m\neq n$
\begin{eqnarray}
&&\bra{m} \dot{H} \ket{n} + \bra{m} H \ket{\dot{n}} =  \dot{E}_n \braket{m}{n} + E_n \braket{m}{\dot{n}} \nonumber \\
&&\bra{m} \dot{H} \ket{n} + E_m \braket{m}{\dot{n}}   = E_n \braket{m}{\dot{n}} \nonumber \\
&\implies& \braket{m}{\dot{n}} = \frac{\bra{m} \dot{H} \ket{n}}{E_n - E_m}. \label{eq:AdiabaticCond}
\end{eqnarray}
This equation captures the adiabatic condition by precisely expressing the error accumulated due to a non-adiabatic drive. While we shall focus on the insight into the controllability of quantum systems, it is worth stressing that the adiabatic theorem provides a versatile starting point for a range of physical settings, including models of quantum computation~\cite{AdiabaticComp}, topological states of matter~\cite{BudichReview}, and for understanding the dynamics of critical systems according to the Kibble-Zurek mechanism~\cite{polkovnikov2005universal, Uwe1, Uwe2, DziarmagaPRL, Damski2005, DelCampo2014, Puebla2020}. \steve{While the preceding derivation applies to a very broad range of settings, we note that it requires modification when e.g. the system exhibits degeneracies in its energy spectrum~\cite{Kato1950} or for periodically driven systems~\cite{schindler2024counterdiabatic, eckardt2008avoided, weinberg2017adiabatic}.}

\subsection{Exact counterdiabatic term}
\label{sec:CD}
While Eq.~\eqref{eq:AdiabaticCond} has been well known for almost a century, only recently has been shown that this term can be used to determine precisely the correction that needs to be added to the generator of the dynamics in order to suppress any non-adiabatic transitions for an arbitrary evolution governed by some time-dependent process, $\lambda(t)$. Demirplak and Rice~\cite{Demirplak2003, Demirplak2008}, and independently Berry~\cite{Berry2009}, showed that augmenting the system Hamiltonian with the so-called counterdiabatic (CD) term ensures that the dynamics precisely tracks an instantaneous adiabatic evolution. This CD term follows directly from Eq.~\eqref{eq:AdiabaticCond} and is given by
\begin{equation}
\label{HCD}
\mathcal{A} \equiv \mathcal{A}_\lambda \equiv \mathcal{A}(t)= i\hbar {\sum_{n} \sum_m} \frac{\ketbra{m}{m} \dot{H} \ketbra{n}{n}}{E_n - E_m},\qquad {n\neq m}.
\end{equation} 
An important note: in order to bring some consistency to notation, here and throughout the symbol $\mathcal{A}$ (including any parameter dependent variations) will be reserved for the {\it exact} counterdiabatic term, while (variations) of the symbol $A$ will be used to denote approximations. We refer to Tables~\ref{table:AGPsSymbols}, \ref{table:QOCSymbols}, and~\ref{table:RLSymbols}, which should serve as a reference point to clarify the conventions used throughout this tutorial. 
\begin{table}[!h]
\centering
\begin{tabular}{|p{12.5cm}|c|}
\hline
Object & Proposed Symbol \\
\hline\hline
Original system Hamiltonian & $H(t)$ \\
\hline
Berry correction (no diagonal elements, also known as Kato potential)& $\mathcal{A}$, $\mathcal{A}_\lambda$, $\mathcal{A}(t)$ \\
\hline
Gauge potential (table point iv) [U's] & $\dot{\lambda}(i \partial_\lambda U)U^{\dagger}$ \\
\hline
Total generator (original system hamiltonian + controls) for Kato/Berry & $H_\text{CD}(t)=H(t) + \mathcal{A}_\lambda$ \\
\hline
Total generator (original system hamiltonian + controls) for AGP & $H_\text{CD}(t)=H(t) + \dot{\lambda}(i \partial_\lambda U)U^{\dagger}$ \\
\hline
Approximate AGP & $A_\lambda^{(x)}$ or $A_\lambda^{[x]}$ \\
\hline
\end{tabular}
\caption{Summary of notation for the most common quantities in shortcuts-to-adiabaticity. Note that the symbol $\mathcal{A}$ (and any of parameter dependent variations) always refers to the same object, i.e. the exact counterdiabatic term.}
\label{table:AGPsSymbols}
\end{table}

We assume that there are no degeneracies in the spectrum, so the denominator in Eq.~\eqref{HCD} is finite for all times, $t$. If we consider evolving a specific initial eigenstate of $H(0)$, the unitary dynamics governed by the Hamiltonian
\begin{equation}
H_\text{CD} = H + \mathcal{A},
\end{equation}
ensures that at every instant of time, the state of the system is guaranteed to be found (up to an overall phase) in the corresponding instantaneous eigenstate of the original Hamiltonian, $H$. Since this holds for any choice of eigenstate, it follows that this Hamiltonian will also realise an adiabatic dynamics for an arbitrary state of the system. 

While we have arrived at Eq.~\eqref{HCD} by carefully deriving the adiabatic condition, Eq.~\eqref{eq:AdiabaticCond}, this term can be directly determined starting from a more ``traditional" approach of inverse engineering the dynamics such that a given target state is exactly achieved~\cite{Berry2009}. Assuming the system starts in the eigenstate $\ket{n(0)}$ (where for clarity we have explicitly included the time-dependence) an adiabatic evolution implies that the state at some final time, $T$, will have accumulated the dynamical and geometric phases only, i.e., it will be of the form
\begin{equation}
\ket{\psi_n(T)} =  \text{exp}\left[-i(\phi_d(T) + \phi_g(T) \right] \ket{n(T)} = \text{exp}\left[ - \frac{i}{\hbar} \int_0^T E_n(t) - i\hbar \braket{n(t)}{\partial_t n(t)}~\mathrm dt  \right] \ket{n(T)}.
\end{equation}
The goal of inverse engineering is to determine the Hamiltonian that will evolve the initial state to precisely this adiabatically connected target state, i.e.,
\begin{equation}
\ket{\psi(T)} =  U(T) \ket{n(0)} \equiv \ket{\psi_n(T)}.
\end{equation}
To determine what Hamiltonian generates such dynamics, we use the fact that any unitary, $U(t)$, satisfies
\begin{equation}
\label{Uidentity}
i\hbar \partial_t U(t) = H' U(t) \implies H' = i\hbar \left( \partial_t U(t) \right) U^{\dagger}(t).
\end{equation}
where we have omitted the explicit time dependence of $H'$ for clarity. Given that our target is $\ket{\psi_n(T)}$ our required unitary is therefore
\begin{equation}
\label{eq:evo-op}
U(T) = \sum_n U_n(T) = \sum_n  \text{exp}\left[ - \frac{i}{\hbar} \int_0^T E_n(t) - i\hbar \braket{n(t)}{\partial_t n(t)}~\mathrm dt  \right] \ketbra{n(T)}{n(0)}.
\end{equation}
We can now use Eq.~\eqref{Uidentity} to determine the corresponding Hamiltonian. Recalling that $\partial_t e^{f(t)} = f'(t) e^{f(t)}$ and for brevity defining $f(T)=\text{exp}\left[ - \frac{i}{\hbar} \int_0^T E_n(t) - i\hbar \braket{n(t)}{\partial_t n(t)}~\mathrm dt \right]$, we can determine the ingredients necessary for evaluating Eq.~\eqref{Uidentity} by focusing on a single value of $n$,
\begin{eqnarray}
\partial_t U_n(t) &=& \left( -\frac{i}{\hbar} E_n - \braket{n(t)}{\partial_t n(t)} e^{f(t)} \right) \ketbra{n(t)}{n(0)} + e^{f(t)}\ketbra{\partial_t n(t)}{n(0)} \nonumber \\
U^{\dagger}_n (t) &= & e^{f(t)^*}\ketbra{n(0)}{n(t)} \nonumber \\
\implies i\hbar(\partial_t U_n)U_n^\dagger &=& i\hbar \Big(  \left( -\frac{i}{\hbar} E_n(t) - \braket{n(t)}{\partial_t n(t)} e^{f(t)} \right) \ketbra{n(t)}{n(0)} + e^{f(t)}\ketbra{\partial_t n(t)}{n(0)}   \Big) \times e^{f(t)^*}\ketbra{n(0)}{n(t)} \nonumber \\
&=& E_n \ketbra{n(t)}{n(t)} - i\hbar \braket{n(t)}{\partial_t n(t)}e^{f(t)}e^{f(t)^*} \ketbra{n(t)}{n(t)} + i\hbar e^{f(t)}e^{f(t)^*} \ketbra{\partial n(t)}{n(t)}.
\end{eqnarray}
While not immediately obvious, the term $\braket{n(t)}{\partial_t n(t)}$ is purely imaginary~\cite{BudichReview}, and therefore the function defined by $f(t)$ is imaginary; to see this, one can take the derivative of the normalization condition $\bra{n(t)}\ket{n(t)}=1$ and rearrange terms. We can simplify the above expression since $e^{f(t)}e^{f(t)^*}=1$, and therefore the Hamiltonian that gives rise to a perfectly adiabatic dynamics is given by
\begin{eqnarray}
\label{eq:IE}
H' = H_\text{CD} &= & \sum_n E_n(t) \ketbra{n(t)}{n(t)} + i\hbar \sum_n \Big( \ketbra{\partial_t n(t)}{n(t)} - \braket{n(t)}{\partial_t n(t)} \ketbra{n(t)}{n(t)} \Big) \nonumber\\
& =& H+ \mathcal{A}.
\end{eqnarray}
where the second term on the RHS is identical to Eq.~\eqref{HCD}~\cite{Berry2009}. 

This formulation is the starting point for a wealth of studies regarding the control of quantum systems~\cite{STAreview}. As we shall see, when directly applied to many-body systems, the CD term generally gives rise to a highly non-local interaction term thus limiting its direct experimental applicability to single or few-body systems~\cite{delCampo2012Assisted, CampbellPRL}. Nevertheless, the determination of Eq.~\eqref{HCD} still provides useful insight into the types of interactions that render a complex system controllable and this insight can then be used to restrict the control Hamiltonian to experimentally relevant interactions. As we will now discuss, the adiabatic gauge potential provides an alternative approach to deriving the CD term that allows us to achieve this aim.

\subsection{Adiabatic gauge potential}

An alternative method to think about the non-adiabatic terms in Eq.~\eqref{eq:AdiabaticCond} is through the adiabatic gauge potential (AGP)~\cite{sels2017minimizing,kolodrubetz2017geometry}. We begin by considering a state $\ket{\psi(\lambda(t))}$ evolving under the time-dependent Schr\"odinger equation of Eq.~\eqref{TDSE} where we change slightly notation such that we describe the time dependence by the parameter $\lambda(t)$. From here, we will drop the explicit time-dependence and retain only the $\lambda$-dependence. In principle, we can always construct the unitary transformation $U(\lambda)$ to a ``co-moving" reference frame which diagonalises the Hamiltonian for all times. Given that a diagonal Hamiltonian cannot induce any diabatic transitions, states evolving under it simply acquire a phase with frequencies corresponding to the eigenenergies. However, we know that physics must be consistent between reference frames. Therefore, performing the unitary transformation on the equation of motion in this case must produce additional non-diagonal terms and these are given by the adiabatic gauge potential. This is indeed the case, as the transformation of a Hamiltonian by a general unitary for which $\partial_t U \neq 0$ gives
\begin{equation}
    H(t) \rightarrow U^\dagger(t) H(t) U(t) - i \hbar U^\dagger(t)\partial_t U(t).
\end{equation}
Converting this to the specific problem at hand, using the unitary which diagonalises the Hamiltonian we get
\begin{equation}
    H(\lambda) \rightarrow U^\dagger(\lambda) H(\lambda) U(\lambda) - i \hbar \dot{\lambda} U^\dagger(\lambda)\partial_\lambda U(\lambda) = \tilde{H}(\lambda) - \dot{\lambda} \tilde{\mathcal{A}}_\lambda,
\end{equation}
where we have used the fact that $\partial_t = \dot{\lambda} \partial_\lambda$, with tildes denoting operators that are in the co-moving frame and we have defined the AGP in the co-moving frame as
\begin{equation}
    \tilde{\mathcal{A}}_\lambda = i \hbar U^\dagger(\lambda)\partial_\lambda U(\lambda).
\end{equation}

We can find the adiabatic gauge potential in the original reference frame by inspecting its matrix elements with respect to a fixed parameter independent basis, $\{\ket{n}_0\}$, with $\ket{m (\lambda)} = \sum_n U_{nm}(\lambda)\ket{n}_0$, giving~\cite{kolodrubetz2017geometry}
\begin{equation}
    \bra{m}_0 \tilde{\mathcal{A}}_\lambda \ket{n}_0 = \steve{i \hbar \bra{m}_0 U^\dagger(\lambda) \partial_\lambda U(\lambda) \ket{n}_0 }= i \hbar  \bra{m (\lambda)} \partial_\lambda \ket{n (\lambda)} =  \bra{m (\lambda)} \mathcal{A}_\lambda \ket{n (\lambda)},
\end{equation}
meaning that the gauge potential in the lab frame can be thought of as proportional to the derivative along the parameter path,
\begin{equation}
\label{eq:Alambda}
    \mathcal{A}_\lambda = i \hbar \partial_\lambda =  i \hbar [\partial_\lambda U(\lambda) ]U^\dagger(\lambda).
\end{equation}


If we modify the original Hamiltonian by adding the AGP then we obtain 
\begin{equation}
\label{eq:H_tilde}
    H(\lambda) + \dot{\lambda} \mathcal{A}_\lambda \rightarrow \tilde{H}(\lambda) + \dot{\lambda} \tilde{\mathcal{A}}_\lambda - \dot{\lambda} \tilde{\mathcal{A}}_\lambda = \tilde{H}(\lambda),
\end{equation}
and since $\tilde{H}(\lambda)$ is diagonal, it cannot cause diabatic transitions between the eigenstates. Thus, addition of the AGP is equivalent to CD driving up to a dynamical global phase (see Table~\ref{table:AGPs}). This can also be seen by considering the differentiation with respect to $\lambda$ of the condition
\begin{equation}
    \bra{m (\lambda)} H(\lambda) \ket{n(\lambda)} = 0 \text{ for } m \neq n,
\end{equation}
where $\ket{n(\lambda)}$ and $\ket{m(\lambda)}$ are two eigenstates, i.e., the above is nothing more than the orthogonality condition. The differentiation of this orthogonality condition gives us
\begin{eqnarray}
    \partial_\lambda \bra{m (\lambda)} H(\lambda) \ket{n(\lambda)} & = & \left( \partial_\lambda \bra{m (\lambda)} \right) H(\lambda) \ket{n(\lambda)} + \bra{m (\lambda)} \left( \partial_\lambda H(\lambda) \right) \ket{n(\lambda)} + \bra{m (\lambda)} H(\lambda) \left( \partial_\lambda \ket{n(\lambda)} \right) \nonumber \\ & = & \left(E_m(\lambda) - E_n(\lambda)\right) \braket{m (\lambda)}{\partial_\lambda n(\lambda)} + \bra{m (\lambda)} \left( \partial_\lambda H(\lambda) \right) \ket{n(\lambda)} = 0,
\end{eqnarray}
where we have used the fact that $\braket{\partial_\lambda m (\lambda)}{ n(\lambda)} = - \braket{m (\lambda)}{\partial_\lambda n(\lambda)}$ which follows from differentiating the condition $\bra{n(\lambda)}\ket{m(\lambda)}=0$. By rearranging above we can find the matrix elements for the adiabatic gauge potential as 
\begin{equation}
    \bra{m (\lambda)}\mathcal{A}_\lambda\ket{n(\lambda)} = i \hbar \braket{m (\lambda)}{\partial_\lambda n(\lambda)} = i \hbar \frac{\bra{m (\lambda)} \left( \partial_\lambda H(\lambda) \right) \ket{n(\lambda)}}{E_m(\lambda)-E_n(\lambda)},
\end{equation}
which clearly is closely related to Eq.~\eqref{eq:AdiabaticCond}. It therefore follows that we can write the off-diagonal part of the AGP in the instantaneous eigenbasis of the Hamiltonian as exactly the counterdiabaitic driving term, i.e., 
\begin{equation}\label{eq:agpcd}
    \mathcal{A}(t) = \dot{\lambda} \sum_{m\neq n} \ketbra{m (\lambda)}\mathcal{A}_\lambda\ketbra{n(\lambda)} = i \hbar \dot{\lambda} \sum_{m\neq n} \frac{\ketbra{m (\lambda)} \left( \partial_\lambda H(\lambda) \right) \ketbra{n(\lambda)}}{E_m(\lambda)-E_n(\lambda)}. 
\end{equation}
While Eq.~\eqref{eq:agpcd} itself does not provide any additional information regarding the controllability of complex quantum systems, it provides the starting point for taking a variational approach to constructing the control Hamiltonian. This provides a means to enforce the locality constraints in an insightful manner as well as allowing to assess the relevance of non-local interaction terms~\cite{COLD_PRXQ}.

The diagonal elements of the gauge potential, $\bra{n(\lambda)}\mathcal{A}_\lambda\ket{n(\lambda)}=i\bra{n(\lambda)}\ket{\partial_\lambda n(\lambda)}$, are known as the Berry connection. Importantly, they are not physical, i.e., they cannot be observed in experiments. To see this, consider the gauge transformation $\ket{n(\lambda)}\to \exp(i\chi^{(n)}(\lambda))\ket{n(\lambda)}$ which depends on the parameter $\lambda$ via the phase $\chi^{(n)}(\lambda)$: since it attaches an overall global phase $\chi^{(n)}(\lambda)$, it cannot change the values of physical observables; on the other hand, it leads to a shift in the Berry connection: 
\begin{equation}
\label{eq:gauge_transf}
    i\bra{n(\lambda)}\ket{\partial_\lambda n(\lambda)} \to i\bra{n(\lambda)}\ket{\partial_\lambda n(\lambda)} - \partial_\lambda \chi^{(n)}(\lambda).
\end{equation}
A physically observable quantity is the Berry phase: for an adiabatic trajectory along a \textit{closed} curve $\mathcal{C}$, the Berry phase can be written as $\gamma = \phi_g(T) = \oint_\mathcal{C} i\bra{n(\lambda)}\ket{\partial_\lambda n(\lambda)} \mathrm d\lambda$, which is gauge-invariant.

\subsection{Counterdiabatic driving and eigenstate phase accumulation}
\label{sec:phases}

Counterdiabatic time evolution generated by $H_\text{CD}$ [Eq.~\eqref{eq:IE}] or $\mathcal{A}_\lambda$ [Eq.~\eqref{eq:Alambda}] and starting from an eigenstate of the Hamiltonian leaves the population in the adiabatically connected instantaneous eigenstate. However, as we mentioned above, such evolution often results in a global phase; we then find the system in the instantaneous eigenstate at time $t$, \textit{up to a phase}. In specific cases, we can decompose this global phase into a dynamical phase $\phi_d$ and a geometric phase $\phi_g$.
Table~\ref{table:AGPs} below summarizes this behavior and helps us distinguish between the zoo of different Hamiltonians $\mathcal{H}$ we introduced above, which all solve the Schr\"odinger initial value problem
\begin{equation}
    i\partial_t\ket{\psi(t)} = \mathcal{H}(t)\ket{\psi(t)},\qquad
    \ket{\psi(0)}=\ket{n(0)}.
\end{equation}

In the following, it will be useful to define the co-moving reference frame via the unitary
\begin{equation}
    V(t_2,t_1) = \mathcal{W}(t_2, t_1)S_{\lambda(t_1)},\qquad \mathcal{W}(t_2, t_1)=\mathcal{T}_\lambda\exp\left(-i \int^{\lambda(t_2)}_{\lambda(t_1)}\mathrm d \lambda\;\mathcal{A}_{\lambda} \right)
\end{equation}
where $\mathcal{T}_\lambda$ is the path-ordered exponential, and $S_\lambda$ diagonalizes the Hamiltonian $\mathcal{H}(\lambda)$, i.e.,
$S^\dagger_{\lambda} \mathcal{H}(\lambda) S_\lambda = \sum_n \mathcal{E}_n(\lambda) \ket{n_0}\bra{n_0} = \tilde{\mathcal{D}}(\lambda)$. The operator $\mathcal{W}(t_2, t_1)$ is known as the Wilson line operator, and it evolves the eigenstates at $\lambda(t_1)$ to the eigenstates at $\lambda(t_2)$. If we define $V^\dagger(t,0)$ to transform from the lab frame to the co-moving frame, we have
\begin{eqnarray}
    \tilde{\mathcal{H}}(t) &=& V^\dagger(t,0)\mathcal{H}(t)V(t,0) - i V^\dagger(t,0)\partial_t V(t,0)  = \tilde{\mathcal{D}}(\lambda(t)) - \dot\lambda \tilde{\mathcal{A}}_\lambda,
\end{eqnarray}
where in the first term we used that the combination $\mathcal{W}(t,0)S_{\lambda(0)}$ diagonalizes the Hamiltonian $\mathcal{H}(\lambda(t))$ [$S_{\lambda(0)}$ diagonalizes the Hamiltonian at $t=0$, and $\mathcal{W}(t,0)$ propagates the eigenstates to time $t$]. In the second (Gallilean) term, note that the derivative $\partial_t$ leaves $S_{\lambda(0)}$ intact, and the expression follows using the unitarity of $S$ and the definition of $\mathcal{W}$ above.
Finally, recall that the evolution operator $U(t_2,t_1)$ in the lab frame can be obtained from the evolution operator in the co-moving frame $\tilde U(t_2,t_1)$, using the relation
\begin{equation}
\label{eq:lab-rot}
    U(t_2,t_1) = V(t_2,t_1) \tilde U(t_2,t_1) V^\dagger(t_1,0) = \mathcal{W}(t_2,t_1) S_{\lambda(t_1)} \tilde U(t_2,t_1) S^\dagger_{\lambda(t_1)} \mathcal{W}^\dagger(t_1,0)
\end{equation}

Let us now consider six different cases for $\mathcal{H}(t)$, and discuss the phases accumulated by the eigenstates:
\begin{enumerate}
    \item[(i)] As we have shown in Sec.~\ref{sec:CD} above, when $\mathcal{H}(t)=H(t)$ \textit{and} we consider evolution in the perfect adiabatic limit ($T\to\infty, \dot\lambda\to 0, \dot\lambda T \to const.$, and we are slower compared to the inverse energy gap to adjacent eigenstates), at time $t$ we find the system in the state $\ket{\psi(t)}=\exp[-i(\phi_d(t)+\phi_g(t))]\ket{n(t)}$. This is nothing else but the content of the adiabatic theorem we proved above;

    \item[(ii)] If we evolve using the counterdiabatic Hamiltonian $\mathcal{H}(t)=H(t)+\dot\lambda\mathcal{A}_\lambda=H_\text{CD}(t)$, the evolved state accumulates both a dynamical and a geometric phase. The difference to (i) is that we have relaxed the condition of working in the adiabatic limit. This statement is what counterdiabatic driving is designed for. 
    This is easiest to see in the co-moving frame, where 
    \begin{equation}
        \tilde{\mathcal H}(t)=\tilde H(t) +\dot\lambda \tilde{\mathcal{A}}_\lambda(t) - V^\dagger(t,0)i\partial_\lambda V(t,0) = \tilde H(t) +\dot\lambda \tilde{\mathcal{A}}_\lambda(t) - \dot\lambda \tilde{\mathcal{A}}_\lambda(t) = \sum_n E_n(t)\ket{n_0}\bra{n_0},
    \end{equation}
    see Eq.~\eqref{eq:H_tilde}, which is diagonal since $\tilde{\mathcal{A}}_\lambda$ is canceled by the Galilean term $V^\dagger(t,0)i\partial_\lambda V(t,0)$.
    Time evolution in the co-moving frame then amounts to integrating the diagonal elements, and recovers precisely the dynamical phase $\phi_{d,n}(t)=\hbar^{-1}\int^t_0 E_n(s)\mathrm d s$:
    \begin{equation}
        \tilde U(t,0) = \exp\left( -i \sum_n \phi_{d,n}(t) \ket{n_0}\bra{n_0} \right).
    \end{equation}
    Going back to the lab frame, cf.~Eq.~\eqref{eq:lab-rot}, we have
    \begin{equation}
        U(t,0) = \mathcal{W}(t,0)S_{\lambda(0)}\exp\left( -i \sum_n \phi_{d,n}(t) \ket{n_0}\bra{n_0} \right) S^\dagger_{\lambda(0)} \mathcal{W}(0,0).
    \end{equation}
    We can simplify this expression by noticing that: 
    (a) $\mathcal{W}(0,0)=1$, and
    (b) recalling that $S_{\lambda(0)}\ket{n_0} = \ket{n[\lambda(0)]}$ by definition.
    This gives
    \begin{equation}
    \label{eq:evo_frames_phases}
        U(t,0) = \mathcal{W}(t,0)\sum_n e^{-i  \phi_{d,n}(t,0)} \ket{n[\lambda(0)]}\bra{n[\lambda(0)]}.
    \end{equation}
    If we now consider a closed loop, $\lambda(0)=\lambda(T)$, the Wilson line operator contains the Berry phases~\cite{Kato1950,bradlyn2022lecture}:
    \begin{equation}
        \mathcal{W}(T,0) = \mathcal{T}_\lambda \exp\left(-i \oint \mathrm d\lambda \;\mathcal{A}_{\lambda} \right) = \sum_n e^{-i  \phi_{g,n}} \ket{n[\lambda(0)]}\bra{n[\lambda(0)]},
    \end{equation}
    where the geometric phase is $\phi_{g,n}=-i\hbar\int^T_0 \bra{n(t)}\ket{\partial_t n(t)}\dot\lambda(t)\mathrm d t$. Equation~\eqref{eq:evo_frames_phases} then coincides with Eq.~\eqref{eq:evo-op} we derived before, since for a closed loop $\ket{n(0)} = \ket{n(T)}$. 
    Thus, we find that the total accumulated phase by each eigenstate is $\phi_d(t)+\phi_g(t)$ -- precisely as in the adiabatic limit, but without the restriction of working in the adiabatic regime. This is counterdiabatic driving. 
    
    \item[(iii)] If instead we evolve using another counterdiabatic Hamiltonian $\mathcal{H}(t)=H(t)+\dot\lambda (i\partial_\lambda U(\lambda))U^\dagger(\lambda)=H_\text{CD}(t)$, the evolved state accumulates only a dynamical phase, i.e., $\ket{\psi(t)}=\exp(-i\phi_d(t))\ket{n(t)}$, because the geometric phase cancels out. To see this, consider the co-moving frame, where 
    \begin{equation}
        \tilde{\mathcal H}= \tilde H(t) +  \dot\lambda U^\dagger(\lambda)(i\partial_\lambda U(\lambda)) - 
    \dot\lambda \tilde{\mathcal{A}}_\lambda =\sum_n \left(E_n(t)+i\bra{n(s)}\ket{\partial_s n(s)}\dot\lambda(s) \right) \ket{n_0}\bra{n_0},
    \end{equation}
    which is again diagonal since the off-diagonal parts in $U^\dagger(\lambda)(i\partial_\lambda U(\lambda))$ and $\tilde{\mathcal{A}}_\lambda$ cancel leaving only the diagonal piece $i\bra{n(s)}\ket{\partial_s n(s)}\dot\lambda(s)$. 
    Time evolution in the co-moving frame then amounts to integrating the eigenenergies of $\tilde{\mathcal H}$, which is precisely the difference between the dynamical and the geometric phases, $\phi_d(t)-\phi_{g}(t)$.
    Going back to the lab frame and considering a closed loop as in example (ii) above, the Wilson line operator adds an extra geometric phase $+\phi_{g}$. Summing all contributions, we find that the geometric phases cancel out, and we are left with the dynamical phase only. 

    \item[(iv)] If, on the other hand, we use only the gauge potential $\mathcal{H}(t)=\dot\lambda\mathcal{A}(t)$ to evolve the eigenstates (also known as the Kato gauge potential~\cite{Kato1950}), the system does not accumulate a dynamical phase, but it does accumulate a geometric phase, so the state at time $t$ is $\ket{\psi(t)}=\exp(-i\phi_g(t))\ket{n(t)}$. To see the absence of a dynamical phase, note that in the co-moving frame, the Hamiltonian $\tilde{\mathcal H}(t) = \dot\lambda\tilde{\mathcal{A}}_\lambda - \dot\lambda\tilde{\mathcal{A}}_\lambda \equiv 0$ vanishes identically.
    The gauge choice that corresponds to the Kato potential is the so-called parallel-transport gauge. It is a proper gauge along the trajectory $\lambda(t)$. 

    \item[(v)] Whenever we evolve with the gauge potential $\mathcal{H}(t)=\dot\lambda(i\partial_\lambda U(\lambda))U^\dagger(\lambda)$ only, we do not accumulate any global phases, and hence $\ket{\psi(t)}=\ket{n(t)}$. The best way to think about this is that the effective dynamical phase is equal in magnitude and opposite in sign to the geometric phase generated by $\dot\lambda(i\partial_\lambda U(\lambda))U^\dagger(\lambda)$, so their sum vanishes. This is seen most easily in the co-moving frame from above, where the Hamiltonian $\tilde{\mathcal{H}}(t)=\tilde{\mathcal H}(t)= - \sum_{n} i\bra{n(t)}\ket{\partial_t n(t)}\ket{n_0}\bra{n_0}$ produces a dynamical phase which is the negative of the geometric phase form the Wilson line operator in the lab frame. Note that, in this case, the gauge choice $\chi(\lambda)$ is fixed such that the Berry connection is given by $i\bra{n(\lambda)}\ket{\partial_\lambda n(\lambda)}$. 

    \item[(vi)] And finally, if we consider evolution with the adiabatic gauge potential in a generic gauge $\chi^{(n)}(\lambda)$, we will find the state $\ket{\psi(t)}=\exp(-i\chi(t))\ket{n(t)}$, where $\chi(t)$ is neither the dynamical nor the geometric phase. 
\end{enumerate}

\begin{table}[t!]
\begin{center}
\begin{tabular}{ |p{0.09\textwidth-2\tabcolsep}||p{0.12\textwidth-2\tabcolsep}|p{0.16\textwidth-2\tabcolsep}|p{0.19\textwidth-2\tabcolsep}|p{0.12\textwidth-2\tabcolsep}|p{0.18\textwidth-2\tabcolsep}|p{0.12\textwidth-2\tabcolsep}|  }
 \hline
  & $\mathcal{H}{=}H(t)$ in the adiabatic limit:
 $T{\to}\infty$, $\dot\lambda{\to} 0$, $\dot\lambda T {\to} const.$ & 
 $\mathcal{H}{=}H(t) {+} \dot\lambda\mathcal{A}_\lambda {=} H_\text{CD}$ (Kato gauge potential) parallel-transport gauge & 
 $\mathcal{H}{=}H(t) {+} (i\partial_t U)U^\dagger {=} H_\text{CD}$, $\chi(\lambda){=}0$ gauge &
 $\mathcal{H}{=}\mathcal{A}_\lambda$ (Kato gauge potential), parallel-transport gauge &
 $\mathcal{H}{=}(i\partial_t U)U^\dagger{=}{-}i U\partial_t U^\dagger$, $\chi(\lambda){=}0$ gauge &
 gauge potential in generic gauge $\chi(\lambda)$
 \\
 \hline
 dynamical phase $\phi_d$  & $\phi_d(t)$     & $\phi_d(t)$ &   $\phi_d(t)-\phi_g(t)$ & $0$    & $-\phi_g(t)$ & $-\phi_g(t)+\chi(\lambda(t))$ \\
 \hline
 geometric phase $\phi_g$ &   $\phi_g(t)$  & $\phi_g(t)$   & $\phi_g(t)$ & $\phi_g(t)$    & $\phi_g(t)$ & $\phi_g(t)$ \\
 \hline
 total phase $\phi_d+\phi_g$ & $\phi_d(t)+\phi_g(t)$ & $\phi_d(t)+\phi_g(t)$ & $\phi_d(t)$ & $\phi_g(t)$ & $0$ & $\chi(\lambda(t))$ \\
 \hline
\end{tabular}
\caption{The table summarizes the acquired dynamical and geometric phases following different types of adiabatic evolution $i\partial_t |\psi(t)\rangle = \mathcal{H}(t)|\psi(t)\rangle $ with $\ket{\psi(t)}=\ket{n(0)}$, and explains the relation between the generators of counterdiabatic evolution introduced in the text. The geometric phase is a property of the eigenstate manifold and remains the same irrespective of the Hamiltonian $\mathcal{H}$ used to generate the evolution; however, the choice of $\mathcal{H}$ modifies the dynamics and hence the total phase accumulated by the wavefunction.  
The dynamical and geometric phases are defined by 
$\phi_d^{(n)}(t)=\int_0^t E_n(\lambda(s)) \mathrm ds$ and 
$\phi_g^{(n)}(t)=-\int_0^t \langle n(\lambda(s))|i\partial_s n(\lambda(s)\rangle \dot\lambda(s) \mathrm ds$, respectively; the gauge 
$\chi(\lambda(t))=\int_0^t \partial_s \chi(\lambda(s)) \dot\lambda(s) \mathrm ds$ is defined in Eq.~\eqref{eq:gauge_transf}.
}
\end{center}
\label{table:AGPs}
\end{table}

\subsection{Variational adiabatic gauge potential}
\label{sec:varcd}
We will now outline a variational approach which allows us to calculate approximate adiabatic gauge potentials for complex many-body problems, even when the eigentstates are not known. This approach starts from noting that the adiabatic gauge potential defined by Eq.~\eqref{eq:agpcd} can be alternatively given by the condition \cite{jarzynski2013generating,sels2017minimizing,kolodrubetz2017geometry}
\begin{equation}
    \left[ \mathcal{A}_\lambda, H_0 \right] = i \hbar \left( \partial_\lambda H_0(\lambda) + F_\text{ad} \right)
\end{equation}
with the generalised adiabatic force operator
\begin{equation}
    F_\text{ad} = - \sum_n \partial_\lambda E_n (\lambda) \ketbra{n(\lambda)}.
\end{equation}
Note, that $F_\text{ad}$ is equivalent to $\partial_\lambda H(\lambda)$ in the energy eigenbasis. We can therefore define an operator
\begin{equation}\label{eq:glambda}
    G_\lambda (A_\lambda) = \partial_\lambda H_0(\lambda) + \frac{i}{\hbar} \left[ A_\lambda, H_0 \right],
\end{equation}
where now $A_\lambda$ is an {\it approximation} of the adiabatic gauge potential; clearly if $A_\lambda = \mathcal{A}_\lambda$ then $G_\lambda (\mathcal{A}_\lambda) = -F_\text{ad}$. This allows us to take a variational approach to finding the adiabatic gauge potential by minimising the distance between $G_\lambda(A_\lambda)$ and $-F_\text{ad}$. Choosing the Frobenius norm this distance is given by
\begin{equation}
    D^2(A_\lambda) = \Tr \left[ \left(G_\lambda(A_\lambda) + F_\text{ad} \right)^2 \right] = \Tr \left[G_\lambda^2 (A_\lambda)\right] + \Tr \left[F_\text{ad}^2\right] + 2\Tr \left[G_\lambda (A_\lambda) F_\text{ad}\right].
\end{equation}
We can use the cyclic and linear mapping properties of the trace in the eigenbasis to simplify the last term
\begin{equation}
    \Tr\left[G_\lambda (A_\lambda) F_\text{ad}\right] = \Tr\left[ F_\text{ad} \partial_\lambda H_0(\lambda) \right] + \frac{i}{\hbar} \Tr \left[ F_\text{ad} \left[ A_\lambda, H_0 \right] \right] = - \Tr \left[ F_\text{ad}^2 \right] - \frac{i}{\hbar} \Tr \left[ \left[ F_\text{ad}, A_\lambda\right] H_0 \right] =  - \Tr \left[ F_\text{ad}^2 \right],
\end{equation}
where we have used the fact that $F_\text{ad}$ will commute with the Hamiltonian. Therefore, in order to construct an approximation to the AGP we must minimise the distance given by 
\begin{equation}
    D^2(A_\lambda) = \Tr \left[G_\lambda^2 (A_\lambda)\right] - \Tr \left[F_\text{ad}^2\right].
\end{equation}
Since the adiabatic force does not explicitly depend on $A_\lambda$, the minimisation procedure is equivalent to simply minimising the norm of $G_\lambda (A_\lambda)$. We can simplify the procedure by minimising the associated action, i.e. define
\begin{equation}
\label{eq:AGPaction}
    S  (A_\lambda) = \Tr \left[G_\lambda^2 (A_\lambda) \right] \qquad \text{and~minimize}\qquad   \frac{\partial S (A_\lambda)}{\partial A_\lambda} = 0.
\end{equation}
This will give an approximate adiabatic gauge potential, or equivalently, an approximate counterdiabatic term. The power of this approach stems from the freedom in choosing what operators appear in $A_\lambda$. By restricting to, for example, only local operators we can construct a control term that will perform as well as possible for a many-body system under such a constraint and we can readily expand the operator set to more complex terms, introducing higher order interactions. As a corollary, this approach therefore also provides valuable insight into the relevance of the non-local interaction terms that the exact CD term typically requires~\cite{COLD_PRXQ, CarolanPRA2022}.

Using the approach of variational CD for an arbitrary many-body quantum system, we may have simply moved the complexity from knowing the instantaneous eigenstates to finding a suitable ansatz for $A_\lambda$. Note, in general, there will be exponentially many operators on which the adiabatic gauge potential could potentially have support, and this could still be unwieldy even when truncating the number of bodies involved or the range of the operators. However, a physically motivated option does exist~\cite{Claeys2019Floquet}, first, we can express the adiabatic gauge potential as 
\begin{equation}
    \mathcal{A}_\lambda = \lim_{\epsilon\rightarrow 0^+} \int_0^\infty \mathrm{d}t \; \mathrm{e}^{-\epsilon t} (\mathrm{e}^{-i H(\lambda)t} \partial_\lambda H(\lambda) \mathrm{e}^{i H(\lambda)t} - \mathcal{M}_\lambda ),
\end{equation}
with $\mathcal{M}_\lambda$ representing the diagonal terms, which are not relevant. The first term of the integral can be expanded via the Baker-Campbell-Hausdorff formula, taking a representive exponent $X$ and operator $Y$,
\begin{equation}
    e^X Y e^{-X} = Y + [X,Y] + \frac{1}{2} [X,[X,Y]] + \dots + \frac{1}{n!} [X, \underbrace{[ X, \dots [X,}_{n} Y]\dots] + \dots ,
\end{equation}
with the commutator $[X,\cdot]$ repeated $n$ times in the general term. Using this we obtain 
\begin{equation}
    \mathcal{A}_\lambda = i\sum_{n=1}^\infty \alpha_n \underbrace{\Big[H(\lambda),\big[H(\lambda),\ldots[H(\lambda)}_{n},\partial_\lambda H(\lambda)]\big]\Big], 
\end{equation}
where the $\alpha_n$ coefficients still need to be solved for using the variational approach. \steve{While not immediately obvious, we find} that $\mathcal{A}_\lambda$ is off-diagonal and that only odd terms in $n$ actually contribute to the off-diagonal~\cite{Claeys2019Floquet}. Therefore, we can trivially set all $\alpha_n=0$ for all $n = 2m$ with $m \in \mathbb{Z}$ and write the adiabatic gauge potential as
\begin{equation}\label{eq:commansatz}
    \mathcal{A}_\lambda = i\sum_{l=1}^\infty \alpha_l \underbrace{\Big[H(\lambda),\big[H(\lambda),\ldots[H(\lambda)}_{2l-1},\partial_\lambda H(\lambda)]\big]\Big].
\end{equation}
Now, to take an informed ansatz we can consider a truncation of the commutation relation. Note, that while this commutator ansatz will restrict to those operators that $\mathcal{A}_\lambda$ can physically have support on, \steve{there still may exist} exponentially many terms for an arbitrary many-body Hamiltonian~\cite{lawrence2024numerical}. It should also be noted that a finite truncation of $l=1,\dots,p$ does not mean that you have all operators for up to $p$-body interactions or that it does not contain operators of more than $p$ bodies. In addition, each order of the expansion of the commutator can contain repeating terms, and they are not orthogonal to each other, though often in practice one orthogonalises each term to avoid having an overly expressive ansatz \cite{lawrence2024numerical,hatomura2021controlling,morawetz2024efficient}. The use of the commutator ansatz at finite truncation and its connection to Krylov space is an active area of study which we will briefly discuss in Sec.~\ref{sec:Outlook}.

\subsection{Examples}

With the necessary groundwork established we now move to the main purpose of this tutorial: applying these techniques to several systems including those with direct relevance for many-body systems. We take a deliberately pedagogical approach providing explicit calculations of the various control methods. In doing so we therefore also detail some well known techniques for solving paradigmatic many-body models, in particular, Jordan-Wigner, Holstein-Primakoff, and Bogoliubov transformations. 

\subsubsection{Example: The Landau-Zener model}
\label{sec:LandauZener}
We begin by considering the two-level Landau-Zener (LZ) model with Hamiltonian given by
\begin{equation}
\label{eq:LZham}
H\equiv H(t)=\Delta \sigma^x + \nu(t) \sigma^z,
\end{equation}
where $\sigma^a$, $a \in {x,y,z} $ are the Pauli matrices
\begin{equation}
\sigma^x = \begin{pmatrix}
0 & 1 \\ 1 & 0
\end{pmatrix} \qquad \sigma^y = \begin{pmatrix}
0 & -i \\ i & 0
\end{pmatrix} \qquad \sigma^z = \begin{pmatrix}
1 & 0 \\ 0 & -1
\end{pmatrix},
\end{equation}
and we assume units such that $\hbar\!=\!1$. Its simplicity notwithstanding, the LZ model captures a remarkably diverse range of physical phenomena~\cite{LZreview}. Relevant to many-body physics, it captures the central aspects of critical systems and in particular two-band models~\cite{Damski2005}. The corresponding eigenstates and eigenenergies can be written
\begin{eqnarray}
\ket{\phi_g} &= \cos \left(\theta/2\right) \ket{0} + \sin \left(\theta/2\right) \ket{1}, \qquad E_g &= -\sqrt{\Delta^2 + \nu(t)^2}  \nonumber \\
\ket{\phi_e} &= \sin \left(\theta/2\right) \ket{0} - \cos \left(\theta/2\right) \ket{1}, \qquad E_e &= \sqrt{\Delta^2 + \nu(t)^2}
\end{eqnarray}
with $\tan \theta = \tfrac{\Delta}{\nu(t)}$. Employing Eq.~\eqref{HCD} we see that the counterdiabatic Hamiltonian is given by
\begin{eqnarray}
\mathcal{A} &=& i \left( \frac{\ketbra{\phi_g}{\phi_g} \partial_t \nu(t) \sigma^z \ketbra{\phi_e}{\phi_e}}{E_g - E_e} + \frac{\ketbra{\phi_e}{\phi_e} \partial_t \nu(t) \sigma^z \ketbra{\phi_g}{\phi_g}}{E_e - E_g}  \right)  \nonumber \\
&=& i \frac{\partial_t \nu(t)}{2\sqrt{\Delta^2+\nu(t)^2}}\Big(\ketbra{\phi_e}{\phi_e}\sigma^z\ketbra{\phi_g}{\phi_g} - \ketbra{\phi_g}{\phi_g}\sigma^z\ketbra{\phi_e}{\phi_e} \Big) \nonumber  \\
&=& \steve{- \frac{\partial_t \nu(t) \sin \theta}{2\sqrt{\Delta^2+\nu(t)^2}} \sigma^y} = - \frac{\Delta~\partial_t \nu(t)}{2(\Delta^2+\nu(t)^2)} \sigma^y. 
\end{eqnarray}
We can arrive at the same expression for $\mathcal{A}$ using Eq.~\eqref{eq:IE}
\begin{eqnarray}
\mathcal{A} & = & i \Big( \ketbra{\dot{\phi}_g}{\phi_g} - \braket{\phi_g}{\dot{\phi}_g}\ketbra{\phi_g}{\phi_g} + \ketbra{\dot{\phi}_e}{\phi_e} - \braket{\phi_e}{\dot{\phi}_e}\ketbra{\phi_e}{\phi_e}\Big) \nonumber \\
& = & i(i \dot{\theta} \sigma^y) =- \frac{\Delta~\partial_t \nu(t)}{2(\Delta^2+\nu(t)^2)} \sigma^y.
\end{eqnarray}

\begin{figure}[t]
\includegraphics[width=0.5\columnwidth]{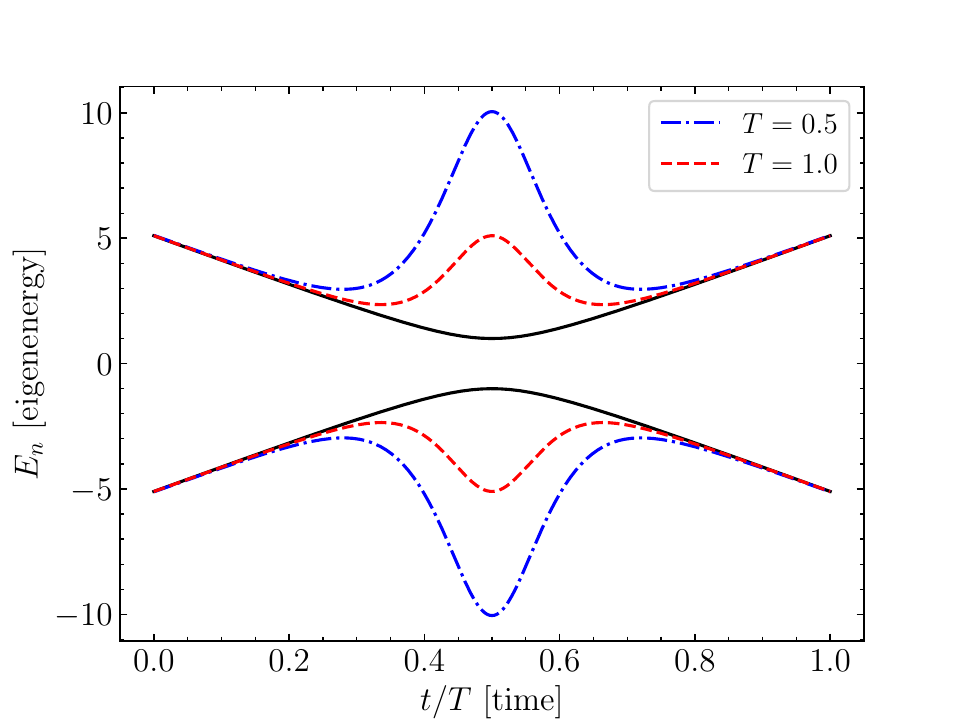}
\caption{Time dependence of the energy eigenvalues for the Landau-Zener model (solid, black) and for the controlled evolution, Eq.~\eqref{HCD_LZ} (dashed) for two different protocol durations assuming a linear ramp $v(t)=-5+10\tfrac{t}{T}$ and $\Delta\!=\!1$.}
\label{EnergiesHCD}
\end{figure}
For any choice of ramp, $\nu(t)$, evolving the system according to the Hamiltonian
\begin{equation}
\label{HCD_LZ}
H_\text{CD} = H + \mathcal{A} = \Delta \sigma^x + \nu(t)\sigma^z - \frac{\Delta~\partial_t \nu(t)}{2(\Delta^2 + \nu(t)^2)}\sigma^y,
\end{equation}
will ensure a perfectly adiabatic dynamics. As an example, if the system is initialised in the ground state for some initial value of the field, $\nu_0$, and we linearly ramp the system according to $\nu(t)=\nu_0+\nu_d \tfrac{t}{T}$, we can understand why this Hamiltonian achieves a perfectly adiabatic evolution by comparing the energy spectra of the original and the controlled Hamiltonians~\cite{Abah2019}. In Fig.~\ref{EnergiesHCD}, the solid curves show the behavior of the eigenenergies of the bare LZ model for a linear ramp and we clearly see the avoided crossing occurring for $t/T=0.5$. In contrast, the energy spectrum of the control Hamiltonian shows that adding the counterdiabatic term modifies the energy gap in such a way that, rather than exhibiting an avoided crossing, the energy levels are pushed apart in precisely the right manner to ensure that the rate at which the system is being driven is still slow enough, with respect to this new energy gap, that it continues to evolve adiabatically. Clearly the counterdiabatic term fundamentally changes the spectral properties of the system in order to achieve adiabaticity. Indeed, using the fact that $\mathcal{A}$ is a Hamiltonian term, several works use this as the basis for examining the thermodynamics of quantum control arguing that $\mathcal{A}$ can be thought of as a energetic toll that must be paid in order manipulate the system~\cite{ZhengPRA2016, SantosSciRep2015, CalzettaPRA2018, FunoPRL2017, AbahEPL2017, CampbellDeffnerPRL, CampbellEPL2023}. For simple models such as Eq.~\eqref{eq:LZham} it is intuitive to expect that $\mathcal{A}$ would modify the spectrum by simply opening an energy gap; however, this is not necessarily true, as has been shown by closing energy gaps in examples including non-interacting tight-binding models \cite{duncan2024exact} and many-body spin-1/2 lattice gauge models \cite{hartmann2019rapid}. In fact, the study of what constitutes the important characteristics that the counterdiabatic term induces in the spectrum of a complex many-body system is an open question. While interesting and insightful in its own right, understanding the control requirements for the LZ model is vital when moving to the genuine many-body Ising model. 

\subsubsection{Example: The Ising model in 1-dimension}

The transverse field nearest-neighbor Ising model describes a collection of interacting spin-$\frac{1}{2}$ degrees of freedom on a lattice under the action of a perpendicular magnetic field. This model is often taken as an example to first consider new control protocols due to its well-known solutions and relation to other higher-complexity cases that are studied \cite{SachdevQPTs,DziarmagaPRL,Damski2006}. We will consider the 1D Ising model for simplicity, which is given by 
\begin{equation}\label{eq:Ising}
H = - \sum_{j=1}^N \left( \sigma^x_j \sigma^x_{j+1} + g \sigma_j^z\right),
\end{equation}
$\sigma_j^a$, $a \in {x,y,z} $ representing the Pauli matrix acting on the $j$-th site and $g$ the relative strength of an applied external field. This model exhibits a quantum phase transition at $g=g_c=\pm 1$ between a paramagnetic ($|g|>1$) and ferromagnetic ($|g|<1$) phase. As it is the only parameter, we will consider the scenario where $g=g(t)$, i.e., we drive $g$ and are interested in implementing CD terms such that we negate any diabatic transitions so that, e.g., we can remain in the ground state of the system in a finite driving time.

\emph{The Jordan-Wigner transformation}. It is not always possible to derive the exact CD for arbitrary Hamiltonians, but luckily the Ising model is exactly solvable through the Jordan-Wigner transformation. This allows for the exact CD to be obtained, as was done in Refs.~\cite{delCampo2012Assisted,damskiJPA2013,damskiJSM2014}. First, we will assume periodic boundary conditions, i.e. $\sigma_{N+1}=\sigma_1$, for an $N$ site system with an even number of spins and $N\geq2$. It is simpler to consider the term across the periodic boundary separately and to write the Hamiltonian as
\begin{equation}\label{eq:Isingper}
H = - \sum_{j=1}^{N-1} \sigma^x_j \sigma^x_{j+1} - \sigma^x_N \sigma^x_{1} - g  \sum_{j=1}^{N} \sigma_j^z.
\end{equation}
The Jordan-Wigner transformation is used to transform the Hamiltonian from the interacting spin basis to a basis of non-interacting fermions, and a useful summary of its implementation can be found in Refs.~\cite{mbengArXiv2020,damskiJPA2013}. We can define a set of creation and annihilation fermionic operators from the spin basis as
\begin{equation}\label{eq:JW}
c_j^\dagger = K_j^\dagger \sigma_j^-, \qquad
c_j = K_j \sigma_j^+,
\end{equation}
with
\begin{equation}
\sigma_j^\pm = \frac{1}{2}\left(\sigma_j^x \pm i \sigma_j^y\right) \qquad K_j = e^{i\pi \sum_{k<j} \sigma_k^+ \sigma_k^-} = \prod_{k<j} \left( 1 - 2 n_j \right),
\end{equation}
where $n_j=c_j^\dagger c_j$ is the number operator and $K_j$, with eigenvalues $\pm 1$, is the non-local string operator which counts the parity of the number of fermions on the lattice before site $j$. The fermionic operators obey the anti-commutation relations
\begin{equation}\label{eq:anticomm}
\left\{ c_i, c_j^\dagger \right\} = \delta_{i,j} \qquad \left\{ c_i, c_j \right\} =  \left\{ c_i^\dagger, c_j^\dagger \right\} = 0.
\end{equation}
The one-body term is then relatively straightforward as we can write the $z$-Pauli matrix on site $j$ as 
\begin{eqnarray}
\sigma_j^z & = & 1 - 2 c_j^\dagger c_j, 
\end{eqnarray}
which can be seen from
\begin{eqnarray}
\sigma_j^z & = & 1 - 2 \sigma_j^- \sigma_j^+ = 1 - \frac{1}{2}\left( \sigma_j^x - i \sigma_j^y \right) \left( \sigma_j^x + i \sigma_j^y \right) = 1 - \frac{1}{2}\left( \sigma_j^x \sigma_j^x - i^2 \sigma_j^y \sigma_j^y - i \sigma_j^y \sigma_j^x + i \sigma_j^x \sigma_j^y \right) \\ \nonumber 
& = & 1 - \frac{1}{2}\left( 1 + 1 - i [\sigma_j^y, \sigma_j^x] \right) = 1 - \left( 1 - \sigma_j^z \right) =  \sigma_j^z , 
\end{eqnarray}
note that the ordering of the Pauli matrices is important as they do not commute and we have used the fact that $[\sigma_j^y, \sigma_j^x] = -2i\sigma_j^z$. We can also write out the $x$-Pauli matrix as,
\begin{eqnarray}
\sigma_j^x & = & K_j \left( c_j^\dagger + c_j \right) \\ \nonumber
& = & K_j \left(  K_j^\dagger \sigma_j^- + K_j \sigma_j^+ \right) = \sigma_j^- + K_j K_j \sigma_j^+ = \sigma_j^- + \sigma_j^+ = \sigma_j^x,
\end{eqnarray}
where we used the fact that $K_j K_j$ acting on any state is equivalent to $\mathds{1}_j$, i.e. the identity operator. With this we can write the Ising Hamiltonian given by Eq.~\eqref{eq:Isingper} as
\begin{eqnarray}\label{eq:intermediateH}
H & = & - \sum_{j=1}^{N-1} K_j \left( c_j^\dagger + c_j \right) K_{j+1} \left( c_{j+1}^\dagger + c_{j+1} \right) - K_N \left( c_N^\dagger + c_N \right) \left( c_{1}^\dagger + c_{1} \right) - g  \sum_{j=1}^{N} \left( 1 - 2 c_j^\dagger c_j \right) \nonumber \\ 
& = & - \sum_{j=1}^{N-1} K_j K_j \left( c_j^\dagger + c_j \right) \left(1-2n_j\right) \left( c_{j+1}^\dagger + c_{j+1} \right) - K_N \left( c_N^\dagger + c_N \right) \left( c_{1}^\dagger + c_{1} \right) - g  \sum_{j=1}^{N} \left( 1 - 2 c_j^\dagger c_j \right). 
\end{eqnarray}
We can simplify this further by considering each term in turn. The first term can be written as
\begin{eqnarray}
& & - \sum_{j=1}^{N-1} K_j K_j \left( c_j^\dagger + c_j \right)  \left(1-2n_j\right)\left( c_{j+1}^\dagger + c_{j+1} \right) = - \sum_{j=1}^{N-1} \left( c_j^\dagger + c_j \right) \left(1-2 c^\dagger_j c_j\right) \left( c_{j+1}^\dagger + c_{j+1} \right) \nonumber \\
& = & - \sum_{j=1}^{N-1} \left( c_j^\dagger + c_j - 2 c_j^\dagger c^\dagger_j c_j - 2 c_j c^\dagger_j c_j \right) \left( c_{j+1}^\dagger + c_{j+1} \right) = - \sum_{j=1}^{N-1} \left( c_j^\dagger + c_j - 2 c_j \right) \left( c_{j+1}^\dagger + c_{j+1} \right) \nonumber \\
& = & \sum_{j=1}^{N-1} \left( c_j - c_j^\dagger \right) \left( c_{j+1}^\dagger + c_{j+1} \right), 
\end{eqnarray}
the second term -- as
\begin{eqnarray}
& & - K_N \left( c_N^\dagger + c_N \right) \left( c_{1}^\dagger + c_{1} \right) = - K_{N+1} \left (1-2 n_N \right) \left( c_N^\dagger + c_N \right) \left( c_{1}^\dagger + c_{1} \right) \nonumber \\
& = & - P \left( c_N^\dagger + c_N - 2 c_N^\dagger c_N c_N^\dagger - 2 c_1^\dagger c_1 c_1 \right) \left( c_{1}^\dagger + c_{1} \right) = - P \left( c_N^\dagger + c_N - 2 c_N^\dagger \right) \left( c_{1}^\dagger + c_{1} \right) \nonumber \\ & = & - P \left( c_N - c_N^\dagger \right) \left( c_{1}^\dagger + c_{1} \right), 
\end{eqnarray}
where $P=K_{N+1}$ is the parity operator for the number of fermions in the full $N$-site system. The third term cannot be simplified further, though it is common to see it written in a different form given by
\begin{equation}
\sum_{j=1}^{N} \left( 1 - 2 c_j^\dagger c_j \right) = \sum_{j=1}^{N} \left( c_j^\dagger c_j + c_j c_j^\dagger - 2 c_j^\dagger c_j \right) = \sum_{j=1}^{N} \left( c_j c_j^\dagger - c_j^\dagger c_j \right),
\end{equation}
where we have used the anti-commutation relations given in Eq.~\eqref{eq:anticomm}. All this together means we can write Eq.~\eqref{eq:intermediateH} as
\begin{equation}
H = \sum_{j=1}^{N-1} \left( c_j - c_j^\dagger \right) \left( c_{j+1}^\dagger + c_{j+1} \right) - P \left( c_N - c_N^\dagger \right) \left( c_{1}^\dagger + c_{1} \right) -g \sum_{j=1}^{N} \left( c_j c_j^\dagger - c_j^\dagger c_j \right).
\end{equation}
This Hamiltonian is invariant under the parity operator and we can therefore split it into the even parity ($P=1$) and odd parity ($P=-1$) subspaces, and write the full Hamiltonian as
\begin{equation}
H = H^+ P^+ + H^- P^-,
\end{equation}
with
\begin{equation}
P^\pm = \frac{1}{2} \left( 1 \pm P \right),~~~~\text{and}~~~~ H^\pm = \sum_{j=1}^{N} \left( c_j - c_j^\dagger \right) \left( c_{j+1}^\dagger + c_{j+1} \right) - g \sum_{j=1}^{N} \left( c_j c_j^\dagger - c_j^\dagger c_j \right), 
\end{equation}
where $H^+$ has anti-periodic boundary conditions, i.e. $c_{N+1} \equiv - c_1$, and $H^-$ has periodic boundary conditions, i.e. $c_{N+1} \equiv c_1$.

\emph{Momentum space}. We will now focus our attention on the positive parity subspace, with the consideration of the negative parity subspace being similar to the discussion below. The dynamics of the Ising spin chain given by Hamiltonian~\eqref{eq:Ising} is then described by the Schr\"odinger equation and the fermionic Hamiltonian 
\begin{equation}
H^+ = \sum_{j=1}^{N} \left( c_j - c_j^\dagger \right) \left( c_{j+1}^\dagger + c_{j+1} \right) - g \sum_{j=1}^{N} \left( c_j c_j^\dagger - c_j^\dagger c_j \right),
\end{equation}
with anti-periodic boundary conditions, $c_{N+1}=-c_1$. 

It will now be convenient to write the model in (quasi)momentum $k$-space and to do this we can first write the fermion operators in $k$-space as
\begin{equation}\label{eq:FTJW}
c_k = \frac{e^{-i \phi}}{\sqrt{N}} \sum_{j=1}^N e^{-i k j} c_j, \qquad c_j = \frac{e^{i \phi}}{\sqrt{N}} \sum_{k} e^{i k j} c_k,
\end{equation}
with the overall phase given by $\phi$ changing the phase of the pairs that will be created in $k$-space as we will see below. Note, the choice of parity subspace informs the correct values of the quasimomentum $k$. For anti-periodic boundary conditions we must satisfy $c_{N+1} \equiv - c_1$, which after transforming to $k$-space will mean $e^{i k N} = -1$, which has solutions
\begin{equation}
k = \pm \frac{\left( 2 m - 1 \right)\pi}{N} \qquad \mathrm{with} \quad m = 1,2,\dots,\frac{N}{2}.
\end{equation}
First we can Fourier transform the local term
\begin{eqnarray}
\sum_{j=1}^{N} \left( c_j c_j^\dagger - c_j^\dagger c_j \right) & = &  \frac{1}{N} \sum_j \sum_{k,k'} \left( e^{-i\left( k' - k\right)j} c_k c_{k'}^\dagger - e^{-i\left( k' - k\right)j} c_{k'}^\dagger c_k \right) \nonumber \\
& = & \sum_{k,k'} \left( \delta_{k,k'} c_k c_{k'}^\dagger - \delta_{k,k'} c_{k'}^\dagger c_k \right) = \sum_k \left( c_k c_{k}^\dagger - c_{k}^\dagger c_k \right),
\end{eqnarray}
where we have used the fact that $\delta_{m,n}=\sum_j e^{-i(n-m)j}/N$. We can follow similar steps with the nearest-neighbor term to obtain
\begin{eqnarray}
\sum_{j=1}^{N} \left( c_j - c_j^\dagger \right) \left( c_{j+1}^\dagger + c_{j+1} \right) & = & \sum_{k,k'} \left( e^{2i\phi} e^{i k'} \delta_{-k,k'} c_k c_{k'} - e^{-2i\phi} e^{-i k'} \delta_{-k,k'} c_k^\dagger c_{k'}^\dagger - \delta_{k,k'} c_{k'}^\dagger c_k + \delta_{k,k'} c_{k'} c_k^\dagger \right) \nonumber \\ 
& = & \sum_k \left( 2\cos k \; c^\dagger_k c_k + e^{2i\phi} e^{i k} c_k c_{-k} - e^{-2i\phi} e^{-i k} c_k^\dagger c_{-k}^\dagger \right),
\end{eqnarray}
where in the final step we have neglected a constant energy offset. Combining these, and using the anticommutation relations as well as the fact that $\sum_k \cos k = 0$, we can obtain the $k$-space Hamiltonian to be
\begin{equation}
H = \sum_k \left[ \left( \cos k - g\right) \left( c_k c_{k}^\dagger - c_{k}^\dagger c_{k} \right) + e^{2i\phi} e^{ik} c_k c_{-k} - e^{-2i\phi} e^{-ik} c_k^\dagger c_{-k}^\dagger \right].
\end{equation}

All terms in the Hamiltonian now come in momentum pairs, $(k,-k)$, and we can define the full Hamiltonian by only considering the positive $k$ values. Let us now define a fermionic two-component spinor 
\begin{equation}
\Psi_k = \begin{pmatrix}
c_k \\ c_{-k}^\dagger
\end{pmatrix} \qquad \Psi_k^\dagger = \begin{pmatrix}
c_k^\dagger & c_{-k}
\end{pmatrix},
\end{equation}
which we can use to rewrite the Hamiltonian as
\begin{equation}
H = \sum_{k>0} H_k \qquad \mathrm{with} \quad H_k = 2 \Psi_k^\dagger \begin{pmatrix}
g - \cos k & - i e^{-2i\phi} \sin k \\
i e^{2i\phi} \sin k & - g + \cos k
\end{pmatrix} \Psi_k.
\end{equation}
We can write out the above $2\times2$ matrix in terms of a psuedo-spin as is shown in Eq.~(54) of Ref.~\cite{mbengArXiv2020}, the choice of $\phi$ then comes down to the direction of the effective magnetic field for this psuedo-spin, for the Ising model we have written the effective magnetic field of the psuedo-spin is in the $x\!-\!z$ plane which corresponds to $\phi=-\pi/4$. Taking this value of $\phi$ we can write the Hamiltonian in momentum space as
\begin{equation}
H_k = 2 \Psi_k^\dagger \left[ \left( g - \cos k \right) \sigma^z + \sin k \sigma^x \right] \Psi_k.
\end{equation}

\emph{Back to Landau-Zener}. We have now managed to decompose the 1D Ising model into a collection of independent two-level systems. Note that, under Schr\"odinger dynamics, each two-level system will undergo a Landau-Zener transition. We can now utilise the results of Sec.~\ref{sec:LandauZener}, identifying $\Delta = 2 \sin k$ and $v(t) = 2g - 2\cos k$, and the CD term can be written as
\begin{equation}
\mathcal{A} = - \sum_{k>0}  \frac{\dot{g}\sin k }{2\left( 1 + g^2 - 2g\cos k \right)} \Psi_k^\dagger \sigma^y \Psi_k.
\end{equation}
If one is only interested in being able to numerically calculate the outcome of a protocol with CD this is where we could stop. However, while we have the CD term in a nice form, expressing this in $k$-space may not be ideal for physical implementation. In addition, as the CD term is local in $k$-space, we should expect that it will be highly non-local in physical space; but this physical space alone is the space of the non-interacting fermions, and we must remember that the native basis for our initial Hamiltonian is in fact spin-$1/2$'s.

Having Fourier converted the problem of spin-$1/2$'s to non-interacting fermions and then transformed to $k$-space, we now find ourselves with the CD term but with the need to trek back along the path we have followed to this point. First we can write out the CD term in the fermion operators as
\begin{equation}
\mathcal{A} = - i \sum_{k>0} \frac{\dot{g}\sin k }{2\left( 1 + g^2 - 2g\cos k \right)} \left( c_k c_{-k} - c_k^\dagger c_{-k}^\dagger \right),
\end{equation}
then utilising the inverse Fourier transform defined in Eq.~\eqref{eq:FTJW} we can write this in position space as
\begin{equation}
\mathcal{A} = - \frac{i}{2N} \dot{g} \sum_{j,j'}^N \sum_{k>0} \frac{\sin k}{1 + g^2 - 2g\cos k} \sin \left( \left( j - j' \right) k \right) \left( c_j c_{j'} + c_j^\dagger c_{j'}^\dagger \right),
\end{equation}
remembering that we only need to consider positive $k$. We see that, as expected, the CD term is highly non-local in physical space as it is a sum over all pairs of sites in the 1D lattice. We can now write the CD term in terms of Pauli matrices, using Eq.~\eqref{eq:JW},
\begin{eqnarray}
\mathcal{A} & = & - \frac{i}{2} \dot{g} \sum_{j,j'}^N f\left( j - j' \right) \left( c_j c_{j'} + c_j^\dagger c_{j'}^\dagger \right) \nonumber \\
& = &  - \frac{i}{2} \dot{g} \sum_{j,j'}^N f\left( j - j' \right) \left[ K_j K_j \sigma_j^+ \left( \prod_{j \leq m < j'}  \sigma_m^z \right) \sigma_{j'}^+ +  K_j^\dagger K_j^\dagger \sigma_j^- \left( \prod_{j \leq m < j'}  \sigma_m^z \right) \sigma_{j'}^- \right] \nonumber \\
& = & - \dot{g} \sum_{j,j'}^N f\left( j - j' \right) \left[ \sigma_j^y \left( \prod_{j < m < j'}  \sigma_m^z \right) \sigma_{j'}^x + \sigma_j^x \left( \prod_{j < m < j'}  \sigma_m^z \right) \sigma_{j'}^y \right],
\label{eq:HCDIsing}
\end{eqnarray}
where in the last step we have removed a $\sigma_j^z$ from the product and used the fact that $\sigma^x_j \sigma_j^z = - i \sigma_j^x$ and $\sigma_j^y \sigma_j^z = i \sigma_j^x$. 

Achieving perfect control still requires knowledge of the coefficients, $f(j-j')$, the precise determination of which is highly non-trivial~\cite{delCampo2012Assisted, damskiJSM2014}, as we will discuss explicitly shortly. Nevertheless, Eq.~\eqref{eq:HCDIsing} is the exact CD term for the 1D Ising model in the native spin-$1/2$ basis. As anticipated, it is a highly non-local operator which couples spins across the full system and, in this regard, it is instructive to examine how the complexity of the CD term emerges as we scale up the size of the system. For $N\!=\!2$ we can directly compute the exact $\mathcal{A}$ using Eq.~\eqref{HCD} and find
\begin{equation}
\mathcal{A} \propto \left(\sigma^y_1\sigma_2^x + \sigma^x_1\sigma_2^y \right)
\end{equation}
similarly for $N=3$ 
\begin{equation}
\mathcal{A}\propto\left( \sigma^y_1\sigma_2^x + \sigma^x_1\sigma_2^y + \sigma^y_2\sigma_3^x + \sigma^x_2\sigma_3^y + \sigma^y_3\sigma_1^x + \sigma^x_3\sigma_1^y\right).
\end{equation}
where we see the operators entering the control terms are precisely those appearing in Eq.~\eqref{eq:HCDIsing}. Thus, while Eq.~\eqref{eq:HCDIsing} provides the overall form of the counterdiabatic term, it still remains to determine the correct form of the coefficients, $f(j-j')$. For $N=2$ it is easy to find the explicit form of the coefficient by direct calculation, resulting in the CD term taking the compact form
\begin{equation}
\mathcal{A} =-\frac{\dot{g}}{4(1+g^2)} \left(\sigma^y_1\sigma_2^x + \sigma^x_1\sigma_2^y \right).
\end{equation}
Already determining the precise coefficients even for $N=3$ becomes significantly more involved; however, if we restrict to only controlling the ground state we find by direct calculation
\begin{equation}
A_\text{GS}= -\dot{g}\frac{1 + g}{8(1+g^3)}\left( \sigma^y_1\sigma_2^x + \sigma^x_1\sigma_2^y + \sigma^y_2\sigma_3^x + \sigma^x_2\sigma_3^y + \sigma^y_3\sigma_1^x + \sigma^x_3\sigma_1^y\right),
\end{equation}
where the choice of symbol reflects the fact that, despite the operators appearing in $A_\text{GS}$ corresponding precisely to those that make up $\mathcal{A}$, the choice of coefficients means that this counterdiabatic term will not suppress excitations for all eigenstates of the Hamiltonian. In general, one can take a similar approach to Ref.~\cite{delCampo2012Assisted} and assume an ansatz form for the coefficients. If we restrict to control of the ground state for an even size chain with periodic boundary conditions, then we can use the results of Ref.~\cite{damskiJSM2014}, which provides an analytic form for the coefficients. For consistency with Refs.~\cite{delCampo2012Assisted, damskiJSM2014} we slightly rearrange Eq.~\eqref{eq:HCDIsing} into the form
\begin{equation}
\label{eq:HCDIsing2}
A_\text{GS}=-\dot{g} \left[\sum_{m=1}^{N/2-1} h_{m}(g) A_\text{GS}^{[m]} + \frac{1}{2}h_{N/2}(g) A_\text{GS}^{[N/2]} \right]    
\end{equation}
where the coefficients are now given precisely by
\begin{equation}
h_m(g) = \frac{g^{2m}+g^N}{8g^{m+1}(1+g^N)}.
\end{equation}
It should be clear that each $m$ term in Eq.~\eqref{eq:HCDIsing2} corresponds to a different ``range'' of interactions given by
\begin{equation}
A_\text{GS}^{[m]} = \sum_{n=1}^{N} \left[ \sigma^y_n \left( \prod_{j=n+1}^{n+m-1} \sigma^z_j \right) \sigma^x_{n+m} + \sigma^x_n \left( \prod_{j=n+1}^{n+m-1} \sigma^z_j \right) \sigma^y_{n+m} \right],
\end{equation}
with $m=1$ corresponding to nearest neighbor terms, i.e. no $\sigma^z_j$'s appearing as e.g. seen in the $N=2$ and 3 cases, and we extend up to a maximal range of $N/2$ due to the assumed periodic boundary conditions. Note that the factor of $1/2$ appearing in front of $A_\text{GS}^{[N/2]}$ is due to the even system size which results in there being only a single spin at the ``maximal" distance of $N/2$ from site 1. 
\begin{figure}[t]
\includegraphics[width=0.5\columnwidth]{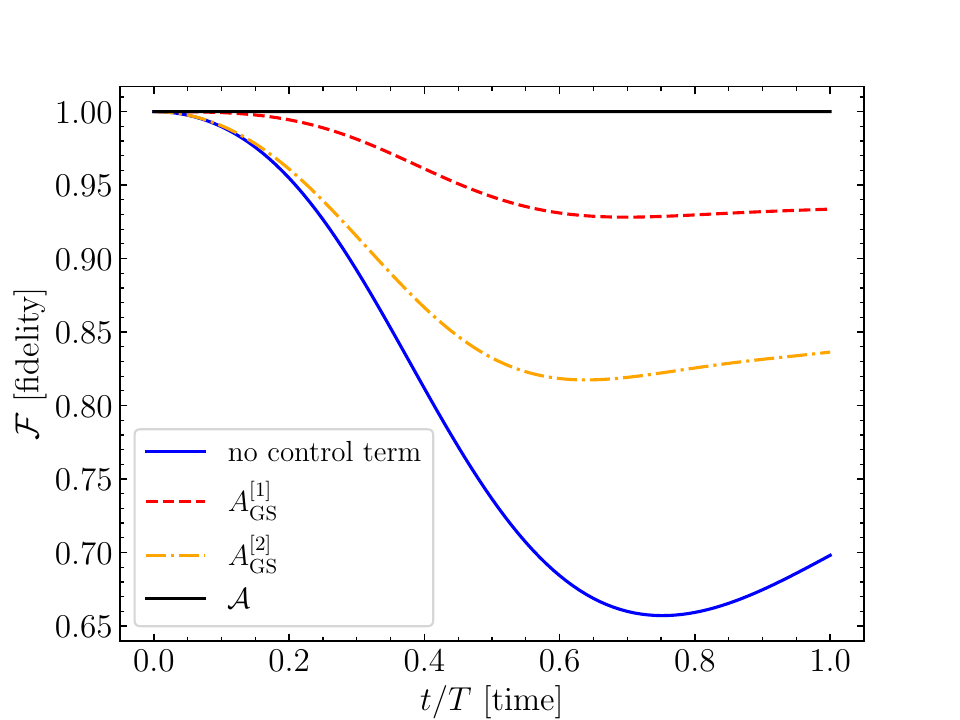}
\caption{Ising model for $N=4$ with $g(t)=0.1+1.9 t/T$ fixing $T \!=\!1$ and starting in the ground state. We show the fidelity with the instantaneous ground state for the full counterdiabtic term (black) involving all two and three body interactions i.e. Eq.~\eqref{eq:HCDIsingFull}, implementing only the two-body terms i.e. Eq.~\eqref{eq:HCDIsing2body} (dashed, red), implementing only the three-body terms i.e. Eq.~\eqref{eq:HCDIsing3body} (dot-dashed, orange). We also show the evolution for the bare Hamiltonian (blue).}
\label{IsingFig}
\end{figure}
This allows us to write out explicitly the control term that will maintain adiabatic dynamics for the ground state for $N=4$, which is the smallest system size such that there are contributions arising from both terms in Eq.~\eqref{eq:HCDIsing2} with 
\begin{eqnarray}
A_\text{GS} = A_\text{GS}^{(2)} &=& -\dot{g} \left[ h_1(g) A_\text{GS}^{[1]} + \frac{1}{2} h_2(g) A_\text{GS}^{[2]}  \right]~~\text{where} \label{eq:HCDIsingFull} \\
h_1(g) A_\text{GS}^{[1]} &=& \frac{1 + g^2}{8(1+g^4)} \left( \sigma^y_1\sigma_2^x + \sigma^x_1\sigma_2^y + \sigma^y_2\sigma_3^x + \sigma^x_2\sigma_3^y+ \sigma^y_3\sigma_4^x + \sigma^x_3\sigma_4^y + \sigma^y_4\sigma_1^x + \sigma^x_4\sigma_1^y \right) \label{eq:HCDIsing2body} \\ 
h_2(g) A_\text{GS}^{[2]} &=&  \frac{g}{4(1+g^4)} \left( 
\sigma^y_1\sigma_2^z\sigma_3^x + \sigma^x_1\sigma_2^z\sigma_3^y + 
\sigma^y_2\sigma_3^z\sigma_4^x + \sigma^x_2\sigma_3^z\sigma_4^y +
\sigma^y_3\sigma_4^z\sigma_1^x + \sigma^x_3\sigma_4^z\sigma_1^y +
\sigma^y_4\sigma_1^z\sigma_2^x + \sigma^x_4\sigma_1^z\sigma_2^y\right) \label{eq:HCDIsing3body}
\end{eqnarray}
where for consistency of notation with Sec.~\ref{subsec:varl_AGP} we introduce $A_\text{GS}^{(m)}$ to denote the cumulative control term utilising all interactions with ranges up to and including $m$-body terms, i.e. the sum of the individual terms with a specific range $A_\text{GS}^{[m]}$. In Fig.~\ref{IsingFig} we show the relevance of the different contributions in achieving control of the full system. We see that if we include only the two-body control terms, i.e. evolve according to $H-\dot{g}h_1(g)A_\text{GS}^{[1]}$, these terms provide the most significant improvement in the ability control of the system when compared to the performance if we include only the three body control terms, i.e. evolve according to $H-\dot{g}h_2(g)A_\text{GS}^{[2]}$. Clearly, given that we are considering only $N\!=\!4$, perfect control is achieved when both two- and three-body terms are present and the system is evolved under $H+A_\text{GS}^{(2)}$. 

\subsubsection{Example: The Ising model with all-to-all couplings (Lipkin-Meshkov-Glick model)}
\label{ex:LMGmodel}
While the previous section considered the solvable Ising model with nearest neighbor interactions, for which the exact CD term for any $N$ can be determined, we can also look at a complementary interaction setting, where all constituents mutually interact with one-another. This all-to-all coupling, or infinite range interaction, is captured by the Lipkin-Meshkov-Glick (LMG) Hamiltonian
\begin{equation}\label{eq:LMG}
H = - \frac{1}{N}\sum_{j<k}^N \sigma^x_j \sigma^x_{k} -\sum_j^N g(t) \sigma_j^z.
\end{equation}
Originally proposed as a model to study shape phase transitions in atomic nuclei, Eq.~\eqref{eq:LMG} has since come to encompass a wide range of physical phenomena including as a model of magnetism~\cite{LMG1, LMG2} and as a paradigmatic model for studying quantum phase transitions~\cite{ESQPTreview}. Notice that unless warranted for clarity, we use $g(t)\equiv g$ for brevity.

From the quantum many-body perspective, this model is interesting due to its very rich phase diagram, host to ground and excited state quantum phase transitions thanks to the high degree of symmetry the all-to-all interaction provides. In the thermodynamic limit, $N\to \infty$, the ground state phase transition is between symmetric ($0\!<\!g\!<\!1$) and symmetry broken ($g\!>\!1$) phases. In the symmetric phase, even and odd parity subspaces are degenerate (and quasi-degenerate for large but finite $N$ with the gap between even and odd sub-sector energies inversely scaling with system size), while in the symmetry broken phase this degeneracy is lifted~\cite{LMG1, FazioLMG}. Despite the similarities with the Ising model, while Eq.~\eqref{eq:Ising} can be solved for arbitrary values of $N$ the same is not true for Eq.~\eqref{eq:LMG}. However, due to the high degree of symmetry that this Hamiltonian exhibits, it is more amenable to a mean-field type approach, allowing for the model to be exactly solvable in the thermodynamic limit by mapping to a quantum harmonic oscillator. This raises interesting consequences when aiming to achieve control of the model: tools and techniques have been developed to control a harmonic oscillator~\cite{MugaJPB}; that said, clearly for any finite $N$ the mapping will only be an approximation and therefore leaves questions as to the efficacy of ``naively'' applying such control approaches. To examine this, however, we must first solve the model. In what follows we will focus on the symmetry broken phase only, i.e. $g>1$, carefully detailing the steps necessary to find the solution. An almost identical calculation can be performed for the symmetric phase, with some caveats that we will discuss later and details of which can be found in, e.g. Ref.~\cite{TakahashiPRE}. 

The first step is to exploit the symmetry induced by the homogeneous all-to-all interactions. This means that it is more convenient to write Eq.~\eqref{eq:LMG} using collective spin operators, $S_\alpha = \tfrac{1}{2}\sum_n \sigma^\alpha_n$ giving
\begin{equation}
\label{eq:LMGcoll}
H= -\frac{2}{N} S_x^2 -2g S_z + \frac{1}{2}\mathds{1}.
\end{equation}
Notice that the last term, proportional to $\mathds{1}$, will have no impact on the dynamics that we will study, it simply amounts to a global phase, and hence is often neglected. For pedagogical reasons of completeness we will keep this term. It is easy to confirm that $\mathbf{S}^2$ is a conserved quantity and we can view the $N$-particle system as an effective $(N\!+\!1)$-level spin. 
We can solve to find the ground state of the model by mapping to {\it bosonic} operators following the Holstein-Primakoff approximation, which is valid for $N\!\gg\! 1$. If we restrict to $g\!>\! 1$, the individual spins all tend to align along the $z$ direction, and therefore we can define our collective spin operators in reference to this axis,
\begin{eqnarray}
\label{eq:HPapprox}
S_+ = \sqrt{N} a,&&\qquad S_-=\sqrt{N}a^{\dagger},\nonumber\\ 
\implies~S_x=\frac{1}{2}(S_+ + S_-),~\qquad S_y&=&\frac{1}{2i}(S_+ - S_-),~\qquad S_z=\frac{N}{2}-a^\dagger a.
\end{eqnarray}
In this picture, a state with all spins pointing up along $z$ corresponds to the vacuum, i.e. $\langle a^\dagger a \rangle=0$. Plugging these into Eq.~\eqref{eq:LMGcoll} we find
\begin{eqnarray}
H&=& -\frac{2}{N} \left( \frac{1}{4}(S_+ + S_-)(S_+ + S_-) \right) - 2 g S_z + \frac{1}{2}\mathds{1} \nonumber \\
&=& -\frac{2}{N} \left( \frac{N}{4}(a + a^\dagger)(a + a^\dagger) \right) - 2 g \left(\frac{N}{2} - a^{\dagger}a\right) + \frac{1}{2} \\
&=& -\frac{1}{2} \left( a^2 + (a^\dagger)^2 + a^\dagger a + a a^\dagger \right) - 2g \left(\frac{N}{2} - a^\dagger a \right) + \frac{1}{2},  \nonumber
\end{eqnarray}
and collecting terms together and using the canonical commutation relation, $a a^\dagger = 1+a^\dagger a$, we finally arrive at our Hamiltonian written only in terms of bosonic annihilation and creation operators
\begin{eqnarray}
H&=& -\frac{1}{2} \left(a^2 + (a^\dagger)^2 \right) - \frac{1}{2}\left( a^\dagger a + a a^\dagger \right) + 2 g a^\dagger a + \frac{1-2g N}{2} \nonumber \\
&=& -\frac{1}{2}\left(a^2 + (a^\dagger)^2 \right) + a^\dagger a \left( 2 g -1 \right) - g N. \label{eq:LMGmapped}
\end{eqnarray}
We now want to diagonalise this Hamiltonian in order to remove the difficult terms in $a^2$ and $(a^\dagger)^2$. This can be achieved using a Bogoliubov transformation. For the symmetry broken phase ($g>1$) that we are focusing on, we start by making the following transformation
\begin{eqnarray}
a &=& \sinh\left( \frac{\alpha}{2} \right) b^\dagger + \cosh\left( \frac{\alpha}{2} \right) b, \nonumber \\
a^\dagger &=& \sinh\left( \frac{\alpha}{2} \right) b + \cosh\left( \frac{\alpha}{2} \right) b^\dagger. \label{eq:BogTrans}
\end{eqnarray}
While at first sight this would seem to complicate things, we will see that it is through a judicious choice of the argument, $\alpha$, that is determined by a suitable quotient of the coefficients in Eq.~\eqref{eq:LMGmapped}, which allows us to diagonalise the model. To proceed, we require the expressions $\left(a^2+(a^\dagger)^2\right)$ and $a^\dagger a$ in terms of the new bosonic operators $b$ and $b^\dagger$:
\begin{eqnarray}
a^2&=& \left( \sinh\left( \frac{\alpha}{2} \right) b + \cosh\left( \frac{\alpha}{2} \right) b^\dagger \right)^2 \nonumber \\
&=& \sinh^2\left( \frac{\alpha}{2} \right) (b^\dagger)^2 + \cosh^2\left( \frac{\alpha}{2} \right) b^2 + \cosh\left( \frac{\alpha}{2} \right)\sinh\left( \frac{\alpha}{2} \right) \left(1+2b^\dagger b \right), \label{eq:asq}\\
(a^\dagger)^2&=& \left( \sinh\left( \frac{\alpha}{2} \right) b^\dagger + \cosh\left( \frac{\alpha}{2} \right) b \right)^2 \nonumber \\
&=& \sinh^2\left( \frac{\alpha}{2} \right) b^2 + \cosh^2\left( \frac{\alpha}{2} \right) (b^\dagger)^2 + \cosh\left( \frac{\alpha}{2} \right)\sinh\left( \frac{\alpha}{2} \right) \left(1+2b^\dagger b \right), \label{eq:adagsq}\\
\implies a^2+(a^\dagger)^2 &= & \left( \cosh^2\left( \frac{\alpha}{2} \right)+ \sinh^2\left( \frac{\alpha}{2} \right) \right)\left( b^2 + \left(b^\dagger\right)^2 \right) +2 \cosh\left( \frac{\alpha}{2} \right)\sinh\left( \frac{\alpha}{2} \right) \left(1+2b^\dagger b \right) \nonumber \\
&=& \cosh\left( \alpha \right) \left( b^2 + \left(b^\dagger\right)^2 \right) + \sinh\left(\alpha \right) \left(1+2b^\dagger b \right), \\
a^\dagger a & = & \left( \sinh\left( \frac{\alpha}{2} \right) b + \cosh\left( \frac{\alpha}{2} \right) b^\dagger \right)\left( \sinh\left( \frac{\alpha}{2} \right) b^\dagger + \cosh\left( \frac{\alpha}{2} \right) b \right) \nonumber \\
& = &  \sinh\left( \frac{\alpha}{2} \right)  \cosh\left( \frac{\alpha}{2} \right) (b^\dagger)^2 + \sinh\left( \frac{\alpha}{2} \right)  \cosh\left( \frac{\alpha}{2} \right) b^2 +\cosh^2\left( \frac{\alpha}{2} \right) b^\dagger b + \sinh^2\left( \frac{\alpha}{2} \right) \left(1+b^\dagger b\right) \nonumber \\
&= & \frac{1}{2}\sinh\left(\alpha\right) \left(b^2 + (b^\dagger)^2\right) + \cosh\left(\alpha\right) b^\dagger b + \sinh^2\left(\frac{\alpha}{2}\right),
\end{eqnarray}
where we have used the relations $\sinh(2x)=2\sinh(x)\cosh(x)$ and $ \cosh^2\left( \frac{x}{2} \right)+ \sinh^2\left( \frac{x}{2} \right) = \cosh(x) $. We therefore find the Hamiltonian, Eq.~\eqref{eq:LMGmapped}, transformed according to Eq.~\eqref{eq:BogTrans} is
\begin{equation}
\label{eq:MappedMess}
H=-\frac{1}{2} \left( \cosh\left( \alpha \right) \left( b^2 + \left(b^\dagger\right)^2 \right) + \sinh\left(\alpha \right) \left(1+2b^\dagger b \right) \right) + \left(2 g -1 \right) \left( \frac{1}{2}\sinh\left(\alpha\right) \left(b^2 + (b^\dagger)^2\right) + \cosh\left(\alpha\right) b^\dagger b + \sinh^2\left(\frac{\alpha}{2}\right) \right) - g N.
\end{equation}
While indeed apparently more complicated than our original Hamiltonian, we now make use of the special choice for the argument
\begin{equation}
\label{eq:BogArgument}
\tanh \left(\alpha\right) = \frac{1}{2g-1},
\end{equation}
which we see is related simply to the ratio of the relevant coefficients appearing in Eq.~\eqref{eq:LMGmapped}. This then allows us to write Eq.~\eqref{eq:MappedMess} solely in terms of $\sinh\left(\alpha\right)$ since 
$$\cosh \left(\alpha\right)= \frac{\sinh \left(\alpha\right)}{\tanh \left(\alpha\right)} = (2 g -1)\sinh \left(\alpha\right).$$ 
Thus, we have
\begin{eqnarray}
H&=&-\frac{1}{2} \frac{\sinh\left( \alpha \right)}{\tanh\left( \alpha \right)} \left( b^2 + \left(b^\dagger\right)^2 \right) -\frac{1}{2} \sinh\left(\alpha \right) \left(1+2b^\dagger b \right) + \frac{1}{2}\left(2 g -1 \right) \sinh\left(\alpha\right) \left(b^2 + (b^\dagger)^2\right) \nonumber \\
  &&~~~~ + \left(2 g -1 \right) \frac{\sinh\left( \alpha \right)}{\tanh\left( \alpha \right)} b^\dagger b + \left(2 g -1 \right) \sinh^2\left(\frac{\alpha}{2}\right) - g N  \nonumber \\
&=& -\frac{1}{2} \left(2 g -1\right)\sinh \left(\alpha\right) \left( b^2 + \left(b^\dagger\right)^2 \right) + \frac{1}{2}\left(2 g -1 \right) \sinh\left(\alpha\right) \left(b^2 + (b^\dagger)^2\right)  -\frac{1}{2} \sinh\left(\alpha \right) \left(1+2b^\dagger b \right) \nonumber \\
        &&~~~~     + \left(2 g -1 \right)  (2 g -1)\sinh \left(\alpha\right) b^\dagger b + \left(2 g -1 \right) \sinh^2\left(\frac{\alpha}{2}\right) - g N,
\end{eqnarray}
where we now see that the terms involving $b^2$ and $(b^\dagger)^2$ cancel and, after a little algebra to collect together terms such that we can express $H$ in the standard harmonic oscillator form, we are left with
\begin{equation}
H=\left( 4g^2 -4g \right) \sinh \left(\alpha\right) \left(b^\dagger b + \frac{1}{2}\right) - g\left(N+1\right) +\frac{1}{2}.
\end{equation}
From Eq.~\eqref{eq:BogArgument} it follows that $\sinh \left(\alpha\right) = \left( 4 g^2 - 4g \right)^{-1/2}$, finally giving us our desired mapped Hamiltonian
\begin{equation}
\label{eq:LMGho}
H = 2\sqrt{g(g-1)} \left(b^\dagger b + \frac{1}{2} \right) - g(N+1) + \frac{1}{2}.
\end{equation}

\begin{figure}[t]
\includegraphics[width=0.49\columnwidth]{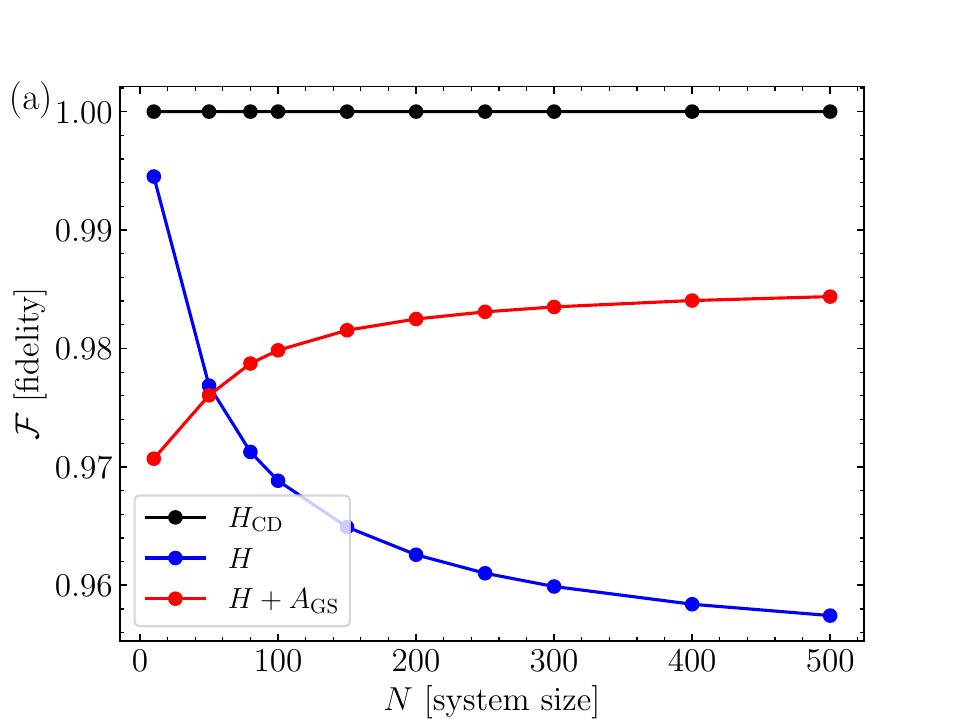}
\includegraphics[width=0.49\columnwidth]{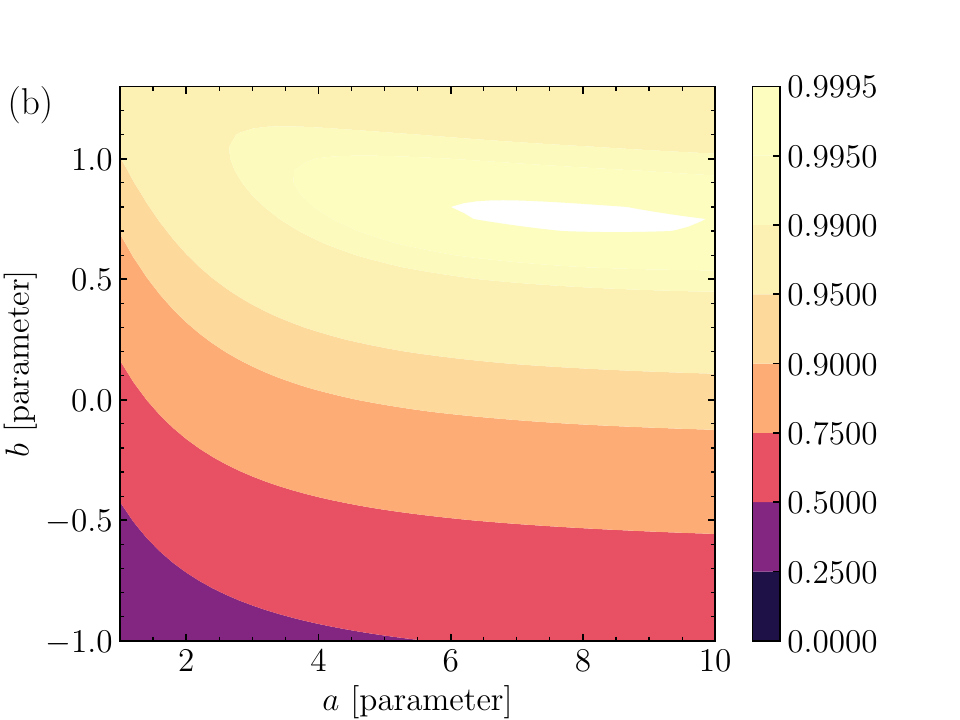}
\caption{(a) Ramping within the symmetry broken phase of the all-to-all model, with $g(t)=2-0.9 t$ where the aim is to connect the ground state of $H$ at $t=0$ with the ground state at $t=1$. We show the final target state fidelity at $t=1$ as a function of system size for the exact, numerically evaluated CD (black), bare evolution (blue), and using the QHO approximation, Eq.~\eqref{eq:LMGCDterm}, (red). (b) We fix $N=50$ and use the guess pulse Eq.~\eqref{eq:GuessPulseLMG} for the control given by the QHO approximation and sweep over the achievable fidelities.}
\label{LMGfig}
\end{figure}

The reason for doing this is that the precise form for the counterdiabatic term is known in the case of a quantum harmonic oscillator~\cite{MugaJPB} with a time dependent frequency, $\omega$, and is given by
\begin{equation}
\label{eq:CDqho}
\mathcal{A} = i \frac{\dot \omega}{4\omega} \left( b^2 - (b^\dagger)^2 \right).
\end{equation}
From Eqs.~\eqref{eq:asq} and \eqref{eq:adagsq} we see that
\begin{eqnarray}
a^2-a^{\dagger2} &=&  \left(\cosh^2\left(\frac{\alpha}{2}\right) - \sinh^2\left(\frac{\alpha}{2}\right) \right) b^2 + \left(\sinh^2\left(\frac{\alpha}{2}\right) - \cosh^2\left(\frac{\alpha}{2}\right) \right) b^{\dagger2} \nonumber \\
&=& b^2-b^{\dagger2}.
\end{eqnarray}
Therefore we can readily determine the operators that the counterdiabatic term depends on in the collective spin picture by simply reverting using Eq.~\eqref{eq:HPapprox}
\begin{eqnarray}
A_\text{GS} &\propto& \frac{1}{N} \left(S_+^2 - S_-^2 \right) \label{eq:propto} \\
 &=& \frac{1}{N}\left( \left(S_x+i S_y\right)^2 -  \left(S_x+i S_y\right)^2 \right) \nonumber \\
 &=& \frac{2i}{N}\left( S_xS_y + S_yS_x \right). \nonumber
\end{eqnarray}
The counterdiabatic term for the ground state of LMG model then follows by identifying the frequency, $\omega=2\sqrt{g(g-1)}$, where we remind that $g$ is the time-dependent parameter. We therefore finally find that the control term for $g>1$ is given by
\begin{equation}
\label{eq:LMGCDterm}
A_\text{GS} = -\frac{\left(2 g -1 \right) \dot g}{4N g\left(g-1 \right)} \left( S_xS_y + S_yS_x \right).
\end{equation}
We again stress that due to the regime of validity of the Holstein-Primakoff approximation, this term is an approximation for any finite $N$ and only strictly applies to the ground state; nevertheless, it provides useful insight insofar as it indicates the operators that are likely to be highly relevant in achieving control over the system. As it is an approximation, it is prudent to test its effectiveness. For a ramp restricting to the symmetry broken phase, taking $g(t)=g_0+g_d \frac{t}{T}$ we show in Fig.~\ref{LMGfig}(a) the performance, as determined by the final target state fidelity, of this ramp as we increase system size. For comparison we show unitary dynamics with no counterdiabatic driving term (blue) and the exact, numerically calculated counterdiabatic term (black), i.e. Eq.~\eqref{HCD} evaluated explicitly for the finite sized system. We clearly see that since Eq.~\eqref{eq:LMGCDterm} is an approximation it does not perfectly suppress the non-adiabatic transitions. However, crucially we see its performance improve as we increase the system size (notice that the apparent plateauing behavior is discussed by Takahashi~\cite{TakahashiPRE}). 

Given that Eq.~\eqref{eq:LMGCDterm} is an approximation, it leaves open the question as to whether we can do better by leveraging the information we have learned from the mapping, specifically the choice of operators that enter the control term, and examine whether there is a better choice for the driving profile. 
As a simple demonstration of this concept, somewhat inline with Ref.~\cite{CampbellPRL}, we consider the functional form of the coefficient in front of the operators in Eq.~\eqref{eq:LMGCDterm}. We can replace this with a simple pulse with a similar profile
\begin{equation}
\label{eq:GuessPulseLMG}
f(t) = \frac{\exp(t^a)-b}{N}.
\end{equation}
The task is then to determine what combination of the parameters $a$ and $b$ will give the best final state fidelity while still assuming the same ramping profile for $g(t)$, i.e. the linear protocol. For this simple case, we show the landscape in Fig.~\ref{LMGfig}(b) where we consider a fixed system size, $N=50$. We see that remarkably high target fidelities can be achieved. While the choice of pulse here was for ease of demonstration, it is clear that this hybrid approach is one which optimal control techniques are particularly suited to, especially if we allow for a more general pulse profile. 

While the above analysis holds for the symmetry broken phase, $g>1$, a qualitatively similar calculation can be done for the symmetric phase. This results in a control Hamiltonian with the same general form as Eq.~\eqref{eq:LMGCDterm} but with a different coefficient. In doing the mapping, the key difference is that the axis around which the Holstein-Primakoff approximation, Eq.~\eqref{eq:HPapprox}, (or equivalently the Bogoliubov transformation) is performed depends on the exact value of $g$. We refer to Refs.~\cite{TakahashiPRE, CampbellPRL, LMG1, LMG2} for more details. A remark regarding the simulation of the model in the symmetric phase: due to the quasi-degeneracy of the different parity sub-sectors, care must be taken when numerically implementing the control since the small energy gaps can lead to numerical instabilities in the dynamics. 

The above example demonstrates a key message of the present tutorial: hybrid approaches to control offer a powerful framework to coherently manipulate complex quantum systems. In particular, we see that the $N\to \infty$ limit serves as a good approximation for sufficiently large systems that we can use the functional form that the exact control term predicts and supplement it with optimal control techniques to achieve almost perfect performance. Naturally, one could have directly examined the problem from an optimal control stand point following e.g. Ref.~\cite{KochEPJQT2015}, dictating from the outset what operators to include in the control Hamiltonian and then seeking the best pulse, and for a model as relatively simple as Eq.~\eqref{eq:LMG} this would be feasible. The advantage of the above approach however is that it allows us to immediately see why a particular subset of operators are important in suppressing non-adiabatic transitions, a point that will be again explored in the next example.

The Ising and LMG models demonstrate that the exact CD terms for many-body systems necessitate non-local interactions. While some non-local interactions can be tuned in some systems, e.g. atoms coupled to cavity modes~\cite{Vaidya2018Tunable}, clearly there is an extremely high overhead in terms of the control needed over various ranges of interactions to implement the counterdiabatic term, cf. Eq.~\eqref{eq:HCDIsing} for the Ising model. In general, we find that interaction ranges necessary to achieve exact CD driving are not fixed and can be of higher order than the terms in the bare Hamiltonian. Indeed, this is the case with the Ising model in Eq.~\eqref{eq:HCDIsing}: when we map the non-local two-body terms obtained for spinless fermions back to the spin-1/2 basis, we find that we generate $N/2$-body terms with periodic boundaries and $N$ being the system size~\cite{delCampo2012Assisted}. While one approach to tackle this issue is to come up with new and novel ways in which to realise exact CD, e.g. by shaking the system in a particular way~\cite{Claeys2019Floquet}, another is to ask how important such long-range terms truly are, cfr. the discussion around Fig.~\ref{IsingFig}. In most scenarios, especially when keeping away from critical points, to obtain a substantial speed-up through a CD control protocol \steve{it} may be sufficient to consider only local approximations of the exact CD~\cite{COLD_PRXQ, Saberi2014}. In scenarios where it is possible to derive the exact counterdiabatic Hamiltonian then it is possible to take truncations of this in a sensible manner by considering dominant terms or taking a finite cut-off in the range of the exactly derived counterdiabatic terms, as demonstrated in Ref.~\cite{COLD_PRXQ} for the Ising model wherein it is shown that the variational adiabatic gauge potential provides the most insightful way to construct the control term. We next demonstrate this while also showing how the variational approach can be utilised for non-integrable models.

\subsubsection{Example: Variational counterdiabatic driving}
\label{subsec:varl_AGP}

The exact CD driving approach is powerful when feasible but limited in two central ways: {\it (i)} it requires the instantaneous Schr\"odinger equation to be solvable at all times of a given protocol, and {\it (ii)} it often results in a highly demanding non-local driving term to be implemented which can be beyond the capabilities of the hardware. The approach of variational CD driving and the use of the AGP, as detailed in Sec.~\ref{sec:varcd} was introduced in order to tackle both of these limitations at the cost of constructing only an approximate CD term.

We will take as an example the 1D Ising model, but will also include terms to make it non-integrable so that the Hamiltonian is given by
\begin{eqnarray}\label{eq:IsingNonIntegrable}
    H_0(\lambda) = -J(\lambda) \sum_j \sigma^z_j \sigma^z_{j+1} - Z(\lambda) \sum_j \sigma^z_j + X(\lambda) \sum_j \sigma^x_j,
\end{eqnarray}
with $\lambda$ being the time-dependent driving field. In this example, we will consider in detail the case of periodic boundary conditions for $N > 2$, however, we will outline where the boundary conditions become important and what happens in the case of $N=2$ for which we will derive the exact CD utilising the variational approach. In the limit of $N=1$ this model becomes equivalent to the Landau-Zener model considered in Sec. \ref{sec:LandauZener}.

From the commutator ansatz of Eq.~\eqref{eq:commansatz} we can find all possible operators on which the AGP can have non-zero support. However, this can be an exponentially large number of operators for an arbitrary $N$ size system. In this example, we will be interested in applying approximate CD driving up to two-body terms and we can turn to the commutator ansatz to obtain the correct set of operators for which the AGP can have support for up to this limit. Before moving further forward it is useful to write out the commutation rules 
\begin{eqnarray}
    \left[d A, B \right] & = & d \left[A, B \right] \nonumber\\
    \left[A, B + C\right] & = & \left[A, B \right] + \left[A, C\right] \nonumber\\
    \left[ A, BC \right] & = & B \left[ A, C \right] +  \left[ A, B \right] C \nonumber\\
    \left[ AB, C \right] & = & A \left[ B, C \right] +  \left[ A, B \right] B
\end{eqnarray}
with $A,B,C$ being operators and $d$ a scalar. Since they will be heavily used in the following, we explicitly state the commutator relations for the Pauli matrices
\begin{eqnarray}
    \left[ \sigma_j^x, \sigma_k^y \right] & = & 2 i \sigma_j^z \delta_{j,k} \nonumber \\
    \left[ \sigma_j^y, \sigma_k^z \right] & = & 2 i \sigma_j^x \delta_{j,k}  \nonumber \\
    \left[ \sigma_j^z, \sigma_k^x \right] & = & 2 i \sigma_j^y \delta_{j,k} \nonumber \\
    \left[ \sigma_j^z, \sigma_k^z \right] = \left[ \sigma_j^x, \sigma_k^x \right] & = & \left[ \sigma_j^y, \sigma_k^y \right] = 0
\end{eqnarray}
with $\delta_{j,k}$ the Kronecker delta, both of which we will use frequently in this example. We start by considering the first term of the commutator expansion, where we will drop the $\lambda$-dependence of each parameter for ease of reading,
\begin{eqnarray}
    A_\lambda^{(1)} = i \alpha_1 \left[ H_0(\lambda), \partial_\lambda H_0(\lambda) \right] & = & i \alpha_1 \sum_{j,k} \left( - J \dot{X} \left[ \sigma_j^z \sigma_{j+1}^z , \sigma_k^x \right] - Z \dot{X} \left[ \sigma_j^z, \sigma_k^x \right] - X \dot{J} \left[ \sigma_j^x, \sigma_k^z \sigma_{k+1}^z \right] - X \dot{Z} \left[ \sigma_j^x, \sigma_k^z \right] \right) \nonumber \\
    & = & i \alpha_1 \sum_{j,k} \left( \left( X \dot{J} - J \dot{X} \right) \left( \sigma_j^z \left[ \sigma_{j+1}^z, \sigma_k^x  \right] + \left[ \sigma_{j}^z, \sigma_k^x  \right] \sigma_{j+1}^z  \right) + \left( X \dot{Z} - Z \dot{X}  \right) \left[ \sigma_j^z, \sigma_k^x  \right]  \right) \nonumber \\
    & = & -2 \alpha_1 \sum_{j} \left( \left( X \dot{J} - J \dot{X} \right) \left( \sigma_j^z \sigma_{j+1}^y + \sigma_j^y \sigma_{j+1}^z  \right) + \left( X \dot{Z} - Z \dot{X}  \right) \sigma_j^y  \right) \ . \label{eq:CommutatorIsing}
\end{eqnarray}
Now we want to utilise only the information about what operators can be generated by the commutator ansatz up to two body terms. As a result, we will not utilise the coefficients derived in front of said operators and from Eq.~\eqref{eq:CommutatorIsing} we would write the ansatz for the AGP as
\begin{equation}
    A_\lambda^{(1)} = \alpha \sum_j \sigma_j^y + \gamma \sum_j \left( \sigma_j^z \sigma_{j+1}^y + \sigma_j^y \sigma_{j+1}^z  \right) \ ,
\end{equation}
with $\alpha$ and $\gamma$ being the coefficients we aim to determine. The next step would be to derive the second term of the commutator expansion, but since we are only interested in up to two body terms and the calculation of the second term is tedious, we will simply quote the result by stating that the full ansatz for the AGP including up to two-body terms is given by
\begin{equation}
    A_\lambda^{(2)} = \alpha \sum_j \sigma_j^y + \beta \sum_j \left( \sigma_j^x \sigma_{j+1}^y + \sigma_j^y \sigma_{j+1}^x  \right) + \gamma \sum_j \left( \sigma_j^z \sigma_{j+1}^y + \sigma_j^y \sigma_{j+1}^z  \right) \ ,
\end{equation}
where we also now need to find $\beta$. For comparison, we will write the ansatz for the AGP up to one-body terms as
\begin{equation}
    A_\lambda^{(1)} = \alpha \sum_j \sigma_j^y \ ,
\end{equation}
and note that the $\alpha$ coefficients for the one-body and two-body terms are not necessarily equivalent.

We now need to derive the distance we will minimise according to Sec.~\ref{sec:varcd}, first, we need to obtain $G_\lambda(A_\lambda^{(2)})$
\begin{eqnarray} \label{eq:GlambdaIsingStart}
    G_\lambda(A_\lambda^{(2)}) & = & \partial_\lambda H_0 + i \left[ A_\lambda^{(2)}, H_0 \right] \nonumber \\  & = & - \dot{J} \sum_j \sigma_j^z \sigma_{j+1}^z - \dot{Z} \sum_j \sigma_j^z + \dot{X} \sum_j \sigma_j^x + i \alpha \sum_{j,k} \left[ \sigma_j^y, - \dot{J} \sigma_k^z \sigma_{k+1}^z - \dot{Z} \sigma_k^z + \dot{X}  \sigma_k^x  \right] \nonumber \\ & & + i \beta \sum_{j,k} \left[ \sigma_j^x \sigma_{j+1}^y + \sigma_j^y \sigma_{j+1}^x, - \dot{J} \sigma_k^z \sigma_{k+1}^z - \dot{Z} \sigma_k^z + \dot{X}  \sigma_k^x  \right] + i \gamma \sum_{j,k} \left[ \sigma_j^z \sigma_{j+1}^y + \sigma_j^y \sigma_{j+1}^z, - \dot{J} \sigma_k^z \sigma_{k+1}^z - \dot{Z} \sigma_k^z + \dot{X}  \sigma_k^x  \right] \ .
\end{eqnarray}
At this point, it is useful to consider each one of the commutators in turn and we will look at each of these in the order they are presented above. Starting with the one corresponding to $\sigma^y_j$,
\begin{eqnarray}
    i \alpha \sum_{j,k} \left[ \sigma_j^y, - \dot{J} \sigma_k^z \sigma_{k+1}^z - \dot{Z} \sigma_k^z + \dot{X}  \sigma_k^x  \right] & = & i \alpha \sum_j \left( -J \sum_k \left( \sigma_k^z \left[ \sigma_j^y, \sigma_{k+1}^z \right] + \left[ \sigma_j^y, \sigma_{k}^z \right] \sigma_{k+1}^z \right) - Z \sum_k \left[ \sigma_j^y, \sigma_k^z \right] + X \sum_k \left[ \sigma_j^y, \sigma_k^x \right] \right) \nonumber \\ & = & i \alpha \sum_j \Bigg( -J \sum_k \left( \sigma_k^z \sigma_j^x \delta_{j,k+1} \left(2i\right) + \sigma_j^x \delta_{j,k} \left(2i\right)  \sigma_{k+1}^z \right) - Z \sum_k \left(2i\right) \delta_j,k \sigma_j^x + X \sum_k \left(-2i\right) \delta_j,k \sigma_j^z \Bigg) \nonumber \\ & = & 2 J \alpha \sum_j \left( \sigma_k^z \sigma_j^x + \sigma_j^x \sigma_{k+1}^z \right) + 2Z \alpha \sum_j \sigma_j^x + 2X \alpha \sum_j \sigma_j^z \ . \label{eq:Y}
\end{eqnarray}
Note that above, and in other places in the derivation of this example we use a subtle trick of labelling, the use of which can be frustrating for those who are not aware of it. For two operators labelled by $j$ and $k$, where $j$ and $k$ are summed independently over the same values, then we can always swap $j$ and $k$. This can be seen in a simple example
\begin{equation}
    \sum_{j,k} \left( A_j B_k +  A_k B_j \right) = \sum_{j,k} \left( A_j B_k +  A_j B_k \right) = 2 \sum_{j,k} A_j B_k \ ,
\end{equation}
while this trick is rather obvious when stated so bluntly, it is surprising how amid long analytical derivations such a trick can be missed, and how software, e.g., Mathematica and Maple, can not easily implement this in general. 

We now look at the commutator corresponding to $\sigma_j^x \sigma_{j+1}^y$ terms, which is rather involved, and requires the application of the commutation relation rules many times over. To start, we expand the commutator to get every term into the form of the commutation of two Pauli operators
\begin{eqnarray}
    i \beta &\sum_{j,k}& \left[ \sigma_j^x \sigma_{j+1}^y + \sigma_j^y \sigma_{j+1}^x, - \dot{J} \sigma_k^z \sigma_{k+1}^z - \dot{Z} \sigma_k^z + \dot{X}  \sigma_k^x  \right]  = \nonumber \\ 
    & = & i \beta \sum_{j} \Bigg( - J \sum_k \Big( \sigma_j^x \sigma_k^z \left[ \sigma_{j+1}^y, \sigma_{k+1}^z \right] + \sigma_j^x \left[ \sigma_{j+1}^y, \sigma_k^z \right] \sigma_{k+1}^z + \sigma_k^z \left[ \sigma_{j}^x, \sigma_{k+1}^z \right] \sigma_{j+1}^y + \left[ \sigma_{j}^x, \sigma_k^z \right] \sigma_{k+1}^z \sigma_{j+1}^y + \sigma_j^y \sigma_k^z \left[ \sigma_{j+1}^x, \sigma_{k+1}^z \right] \nonumber \\ 
    & &  + \sigma_j^y \left[ \sigma_{j+1}^x, \sigma_k^z \right] \sigma_{k+1}^z +  \sigma_k^z \left[ \sigma_{j}^y, \sigma_{k+1}^z \right] \sigma_{j+1}^x + \left[ \sigma_{j}^y, \sigma_k^z \right] \sigma_{k+1}^z \sigma_{j+1}^x \Big) - Z \sum_k \Big( \sigma_j^x \left[ \sigma_{j+1}^y, \sigma_k^z \right] + \left[ \sigma_{j}^x, \sigma_k^z \right] \sigma_{j+1}^y \nonumber \\ 
    & & + \sigma_j^y \left[ \sigma_{j+1}^x, \sigma_k^z \right] + \left[ \sigma_{j}^y, \sigma_k^z \right] \sigma_{j+1}^x \Big) + X \sum_k \left( \sigma_j^x \left[ \sigma_{j+1}^y, \sigma_k^x \right] + \left[ \sigma_{j}^y, \sigma_k^x \right] \sigma_{j+1}^x \right) \Bigg) .
\end{eqnarray}
We can then insert the known commutation relations and gather like terms
\begin{eqnarray}
    i \beta &\sum_{j,k}& \left[ \sigma_j^x \sigma_{j+1}^y + \sigma_j^y \sigma_{j+1}^x, - \dot{J} \sigma_k^z \sigma_{k+1}^z - \dot{Z} \sigma_k^z + \dot{X}  \sigma_k^x  \right]  = \nonumber \\ 
    & =  & i \beta \sum_{j} \Bigg( - J \sum_k \Big( \sigma_j^x \sigma_k^z \left(2i\delta_{j,k}\right) \sigma_{j+1}^x + \sigma_j^x \left(2i\delta_{j+1,k}\right) \sigma_{j+1}^x \sigma_{k+1}^z + \sigma_k^z \left(-2i\delta_{j,k+1}\right) \sigma_{j}^y \sigma_{j+1}^y + \left(-2i\delta_{j,k}\right) \sigma_{j}^y \sigma_{k+1}^z \sigma_{j+1}^y \nonumber \\ 
    & &+ \sigma_j^y \sigma_k^z \left(-2i\delta_{j,k}\right) \sigma_{j+1}^y + \sigma_j^y \left(-2i\delta_{j+1,k}\right) \sigma_{j+1}^y \sigma_{k+1}^z +  \sigma_k^z \left(2i\delta_{j,k+1}\right) \sigma_{j}^x \sigma_{j+1}^x + \left(2i\delta_{j,k}\right) \sigma_{j}^x \sigma_{k+1}^z \sigma_{j+1}^x \Big) - Z \sum_k \Big( \sigma_j^x \left(2i\delta_{j+1,k}\right) \sigma_{j+1}^x \nonumber \\ 
    & & + \left(-2i\delta_{j,k}\right) \sigma_{j}^y \sigma_{j+1}^y + \sigma_j^y \left(-2i\delta_{j+1,k}\right) \sigma_{j+1}^y + \left(2i\delta_{j,k}\right) \sigma_{j}^x \sigma_{j+1}^x \Big) + X \sum_k \left( \sigma_j^x \left(-2i\delta_{j+1,k}\right) \sigma_{j+1}^z + \left(-2i\delta_{j,k}\right) \sigma_{j}^z \sigma_{j+1}^x \right) \Bigg) \nonumber \\
    & = & 2J\beta \sum_j \left( \sigma_j^x \sigma_j^z \sigma_{j+1}^x + \sigma_j^x \sigma_{j+1}^z \sigma_{j+1}^x - \sigma_j^y \sigma_j^z \sigma_{j+1}^y - \sigma_j^y \sigma_{j+1}^z \sigma_{j+1}^y \right) + 2J\beta \sum_j \left( \sigma_j^x \sigma_{j+1}^x \sigma_{j+2}^z + \sigma_j^z \sigma_{j+1}^x \sigma_{j+2}^x - \sigma_j^z \sigma_{j+1}^y \sigma_{j+2}^y - \sigma_j^y \sigma_{j+1}^z \sigma_{j+2}^z  \right) \nonumber \\ 
    & & + 4Z\beta \sum_j \left( \sigma_j^x \sigma_{j+1}^x - \sigma_j^y \sigma_{j+1}^y \right) + 2X\beta \sum_j \left( \sigma_j^x \sigma_{j+1}^z + \sigma_j^z \sigma_{j+1}^x \right) .
\end{eqnarray}
Now, we can use the fact that we can always combine Pauli operators acting on the same site, as is present above, in this case we use that $\sigma_j^x \sigma_j^z = i \sigma_j^y$ and $\sigma_j^z \sigma_j^y = i \sigma_j^x$ to obtain
\begin{eqnarray}
     i \beta &\sum_{j,k}& \left[ \sigma_j^x \sigma_{j+1}^y + \sigma_j^y \sigma_{j+1}^x, - \dot{J} \sigma_k^z \sigma_{k+1}^z - \dot{Z} \sigma_k^z + \dot{X}  \sigma_k^x  \right]  = \nonumber \\ 
    & = & 2iJ\beta \sum_j \left( \sigma_j^y \sigma_{j+1}^x + \sigma_j^x \sigma_{j+1}^y - \sigma_j^x \sigma_{j+1}^y - \sigma_j^y \sigma_{j+1}^x \right) + 2J\beta \sum_j \left( \sigma_j^x \sigma_{j+1}^x \sigma_{j+2}^z + \sigma_j^z \sigma_{j+1}^x \sigma_{j+2}^x - \sigma_j^z \sigma_{j+1}^y \sigma_{j+2}^y - \sigma_j^y \sigma_{j+1}^z \sigma_{j+2}^z  \right) \nonumber \\ 
    & & + 4Z\beta \sum_j \left( \sigma_j^x \sigma_{j+1}^x - \sigma_j^y \sigma_{j+1}^y \right) + 2X\beta \sum_j \left( \sigma_j^x \sigma_{j+1}^z + \sigma_j^z \sigma_{j+1}^x \right) \label{eq:XY} \\
    & = & 2J\beta \sum_j \left( \sigma_j^x \sigma_{j+1}^x \sigma_{j+2}^z + \sigma_j^z \sigma_{j+1}^x \sigma_{j+2}^x - \sigma_j^z \sigma_{j+1}^y \sigma_{j+2}^y - \sigma_j^y \sigma_{j+1}^z \sigma_{j+2}^z  \right) + 4Z\beta \sum_j \left( \sigma_j^x \sigma_{j+1}^x - \sigma_j^y \sigma_{j+1}^y \right) + 2X\beta \sum_j \left( \sigma_j^x \sigma_{j+1}^z + \sigma_j^z \sigma_{j+1}^x \right) \ . \nonumber 
\end{eqnarray}

In the final stretch, we can turn to the commutator corresponding to $\sigma_j^z \sigma_{j+1}^y$ terms, following a similar procedure to above we first expand the commutator
\begin{eqnarray}
    & & i \gamma \sum_{j,k} \left[ \sigma_j^z \sigma_{j+1}^y + \sigma_j^y \sigma_{j+1}^z, - \dot{J} \sigma_k^z \sigma_{k+1}^z - \dot{Z} \sigma_k^z + \dot{X}  \sigma_k^x  \right] = \nonumber
    \\ & = & i \gamma \sum_j \Bigg( -J \sum_k \Big( \sigma_j^z \left[ \sigma_{j+1}^y, \sigma_k^z \sigma_{k+1}^z \right] + \left[ \sigma_{j}^y, \sigma_k^z \sigma_{k+1}^z \right]  \sigma_{j+1}^z \Big) \nonumber -Z \sum_k \Big( \sigma_j^z \left[ \sigma_{j+1}^y, \sigma_k^z \right] + \left[ \sigma_{j}^y, \sigma_k^z \right]  \sigma_{j+1}^z \Big) + X \sum_k \Big( \sigma_j^z \left[ \sigma_{j+1}^y, \sigma_k^x \right] \nonumber \\ & & + \left[ \sigma_{j}^y, \sigma_k^x \right]  \sigma_{j+1}^z + \sigma_j^y \left[ \sigma_{j+1}^z, \sigma_k^x \right] + \left[ \sigma_{j}^z, \sigma_k^x \right]  \sigma_{j+1}^y  \Big) \Bigg) \nonumber \\
    & = & i \gamma \sum_j \Bigg( -J \sum_k \Big( \sigma_j^z \sigma_k^z \left[ \sigma_{j+1}^y, \sigma_{k+1}^z \right] + \sigma_j^z \left[ \sigma_{j+1}^y, \sigma_k^z \right] \sigma_{k+1}^z + \sigma_k^z \left[ \sigma_{j}^y, \sigma_{k+1}^z \right] \sigma_{j+1}^z + \left[ \sigma_{j}^y, \sigma_k^z \right] \sigma_{k+1}^z \sigma_{j+1}^z \Big) \nonumber \\ & & -Z \sum_k \Big( \sigma_j^z \left[ \sigma_{j+1}^y, \sigma_k^z \right] + \left[ \sigma_{j}^y, \sigma_k^z \right]  \sigma_{j+1}^z \Big) + X \sum_k \Big( \sigma_j^z \left[ \sigma_{j+1}^y, \sigma_k^x \right] + \left[ \sigma_{j}^y, \sigma_k^x \right]  \sigma_{j+1}^z + \sigma_j^y \left[ \sigma_{j+1}^z, \sigma_k^x \right] + \left[ \sigma_{j}^z, \sigma_k^x \right]  \sigma_{j+1}^y  \Big) \Bigg) .
\end{eqnarray}
We can then insert the known Pauli commutation relations
\begin{eqnarray}
    & & i \gamma \sum_{j,k} \left[ \sigma_j^z \sigma_{j+1}^y + \sigma_j^y \sigma_{j+1}^z, - \dot{J} \sigma_k^z \sigma_{k+1}^z - \dot{Z} \sigma_k^z + \dot{X}  \sigma_k^x  \right] = \nonumber \\
    & = & i \gamma \sum_j \Bigg( -J \sum_k \Big( \sigma_j^z \sigma_k^z \left(2i\delta_{j,k}\right) \sigma_{j+1}^x + \sigma_j^z \left(2i\delta_{j+1,k}\right) \sigma_{j+1}^x \sigma_{k+1}^z + \sigma_k^z \left(2i\delta_{j,k+1}\right) \sigma_{j}^x \sigma_{j+1}^z + \left(2i\delta_{j,k}\right) \sigma_{j}^x \sigma_{k+1}^z \sigma_{j+1}^z \Big) -Z \sum_k \Big( \sigma_j^z \left(2i\delta_{j+1,k}\right) \sigma_{j+1}^x \nonumber \\ & & + \left(2i\delta_{j,k}\right) \sigma_{j}^x  \sigma_{j+1}^z \Big) + X \sum_k \Big( \sigma_j^z \left(-2i\delta_{j+1,k}\right) \sigma_{j+1}^z + \left(-2i\delta_{j,k}\right) \sigma_{j}^z \sigma_{j+1}^z + \sigma_j^y \left(2i\delta_{j+1,k}\right) \sigma_{j+1}^y + \left(2i\delta_{j,k}\right) \sigma_{j}^y \sigma_{j+1}^y  \Big) \Bigg) .
\end{eqnarray}
It is at this point that a peculiar thing happens, as the square of any Pauli matrix if the identity,  $\mathds{1}_j$, we get the following
\begin{eqnarray}
    & & i \gamma \sum_{j,k} \left[ \sigma_j^z \sigma_{j+1}^y + \sigma_j^y \sigma_{j+1}^z, - \dot{J} \sigma_k^z \sigma_{k+1}^z - \dot{Z} \sigma_k^z + \dot{X}  \sigma_k^x  \right] \nonumber \\
    & = & 4 J \gamma \sum_j  \sigma_j^z \sigma_{j+1}^x \sigma_{j+2}^z + 2 J \gamma \sum_k \left( \sigma_j^x \mathds{1}_{j+1} + \mathds{1}_j \sigma_{j+1}^x \right) + 2 Z \gamma \sum_j \left( \sigma_j^z \sigma_{j+1}^x + \sigma_j^x \sigma_{j+1}^z \right) + 4 X \gamma \sum_j \left( \sigma_j^z \sigma_{j+1}^z - \sigma_j^y \sigma_{j+1}^y \right) \nonumber \\
    & = & 4 J \gamma \sum_j  \sigma_j^z \sigma_{j+1}^x \sigma_{j+2}^z + 4 J \gamma \sum_k \sigma_j^x + 2 Z \gamma \sum_j \left( \sigma_j^z \sigma_{j+1}^x + \sigma_j^x \sigma_{j+1}^z \right) + 4 X \gamma \sum_j \left( \sigma_j^z \sigma_{j+1}^z - \sigma_j^y \sigma_{j+1}^y \right) \ . \label{eq:ZY}
\end{eqnarray}
Note, we highlight this step involving the $\mathds{1}_j$ terms, as this single-body operator in the second term is truly being obtained from a two body-operator. This has consequences dependent on the boundary conditions, and if $N=2$ then the resulting one-body term in the final line should have a coefficient of $2J$ instead of $4J$. We can now return to Eq.~\eqref{eq:GlambdaIsingStart} and, gathering like with like terms from Eqs.~\eqref{eq:Y}, \eqref{eq:XY} and \eqref{eq:ZY}, we obtain
\begin{eqnarray}
    G_\lambda\left(A_\lambda^{(2)}\right) & = &  \left(4 X \gamma - \dot{J}\right) \sum_j \sigma_j^z \sigma_{j+1}^z + \left( 2J\alpha + 2X\beta + 2Z\gamma \right) \sum_j \left( \sigma_j^x \sigma_{j+1}^z + \sigma_j^z \sigma_{j+1}^x \right) + \left( 2Z\alpha + 4J\gamma + \dot{X} \right) \sum_j \sigma_j^x + \left( 2X\alpha + \dot{Z} \right) \sum_j \sigma_j^z \nonumber \\ & & + 2J\beta \sum_j \left( \sigma_j^x \sigma_{j+1}^x \sigma_{j+2}^z + \sigma_j^z \sigma_{j+1}^x \sigma_{j+2}^x \right) - 2J\beta \sum_j \left( \sigma_j^y \sigma_{j+1}^y \sigma_{j+2}^z + \sigma_j^z \sigma_{j+1}^y \sigma_{j+2}^y \right) + 4Z \beta \sum_j \sigma_j^x \sigma_{j+1}^x \nonumber \\ & & - \left( 4Z \beta + 4X \gamma \right) \sum_j \sigma_j^y \sigma_{j+1}^y + 4J \gamma \sum_j \sigma_j^z \sigma_{j+1}^x + \sigma_j^z \ .
\end{eqnarray}
Note, we just subjected ourselves and the reader to a lot of algebra with the tracking of seemingly countless commutation relations and operators. This is done in part to, of course, educate the reader on these techniques, but also as a cautionary tale that this approach is not scalable and \steve{is therefore prone to human error}. We will present current leading alternatives to this pen and paper method outlined here in Sec.~\ref{sec:Outlook}.

Following the approach of Sec.~\ref{sec:varcd}, we now want to minimise $\Tr\left[G_\lambda^2\left(A_\lambda^{(2)}\right) \right]$ which at first can appear quite daunting as we now need to square the long expression given above. However, Pauli matrices are traceless, i.e.
\begin{equation}
    \Tr\left[\sigma_j^x\right] = \Tr\left[\sigma_j^y\right] = \Tr\left[\sigma_j^z\right] = 0 \nonumber \ ,
\end{equation}
which means that any term in $G_\lambda^2\left(A_\lambda^{(2)}\right)$ with a single Pauli matrix will not contribute to $\Tr\left[G_\lambda^2\left(A_\lambda^{(2)}\right)\right]$ and that only terms with Pauli matrices squared will be present, and that these will have a trace of $2$. Therefore, to write the $\Tr\left[G_\lambda^2\left(A_\lambda^{(2)}\right)\right]$ we skip calculating $G_\lambda^2\left(A_\lambda^{(2)}\right)$ and instead square the coefficients in front of each orthogonal operator giving
\begin{eqnarray}
   N 2^{-N} \Tr\left[G_\lambda^2\right] & = & \left(4 X \gamma - \dot{J}\right)^2 + \left( 2J\alpha + 2X\beta + 2Z\gamma \right)^2 + \left( 2Z\alpha + 4J\gamma + \dot{X} \right)^2 + \left( 2X\alpha + \dot{Z} \right)^2 + 8J^2\beta^2 \nonumber \\ & & + 16Z^2 \beta^2 - \left( 4Z \beta + 4X \gamma \right)^2 + 16J^2 \gamma^2 \ .
\end{eqnarray}
Note, we have dropped the $A_\lambda^{(2)}$ dependence of $G_\lambda$ for ease of reading. However, it is also at this step that boundary conditions become important, as each of the two- and three-body terms will be present a different number of times and will have coefficients of, e.g., $N-1$ for two body terms with open boundary conditions, instead of the $N$ for the case of periodic boundary conditions. It is simpler to think of the $N=2$ case as explicitly different, as there will be no three-body terms and any sum over two-body terms is irrelevant.

The final step in obtaining the coupled set of equations that we must solve to obtain ${\alpha,\beta,\gamma}$ is to differentiate and minimise with respect to each coefficient of our ansatz, i.e., consider 
\begin{equation}
    \frac{\partial \Tr\left[G_\lambda^2\right]}{\partial\alpha} = 0 \ , \: \frac{\partial \Tr\left[G_\lambda^2\right]}{\partial\beta} = 0 \ , \: \: \mathrm{and} \: \: \frac{\partial \Tr\left[G_\lambda^2\right]}{\partial\gamma} = 0 \ . \nonumber
\end{equation}
We can then write the obtained coupled set of linear equations as
\begin{equation}
    \begin{pmatrix}
        4J^2 + 2X^2 + 2Z^2 & 4JX & 8JZ \\
        JX & 2J^2 + X^2 + 4Z^2 & 3XZ \\
        4JZ & 6XZ & 8J^2 + 8X^2 + 2Z^2
    \end{pmatrix} \begin{pmatrix}
        \alpha \\
        \beta \\
        \gamma
    \end{pmatrix} = \begin{pmatrix}
        \dot{X}Z - X\dot{Z} \\
        0 \\
        \dot{X}J-X\dot{J}
    \end{pmatrix}\ .
\end{equation}
For completeness, we will also state the coupled set of equations for the case of $N=2$
\begin{equation}
    \begin{pmatrix}
        2J^2 + 2X^2 + 2Z^2 & 2JX & 4JZ \\
        2JX & 2X^2 + 4Z^2 & 6XZ \\
        4JZ & 6XZ & 2J^2 + 8X^2 + 2Z^2
    \end{pmatrix} \begin{pmatrix}
        \alpha \\
        \beta \\
        \gamma
    \end{pmatrix} = \begin{pmatrix}
        \dot{X}Z - X\dot{Z} \\
        0 \\
        \dot{X}J-X\dot{J}
    \end{pmatrix}\ ,
\end{equation}
as we will utilise these in the numerical example below, and it serves as a good test of the approach as we expect the variational CD with up to two-body terms to be the exact CD as there are only two bodies.

\begin{figure}[t!]
\includegraphics[width=0.49\columnwidth]{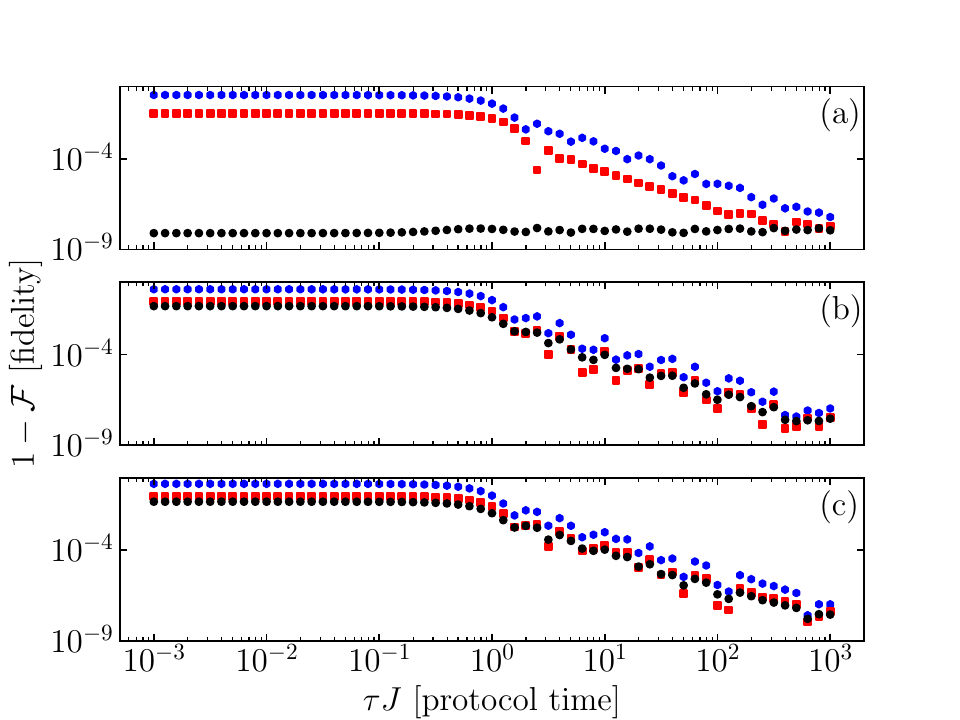}
\includegraphics[width=0.49\columnwidth]{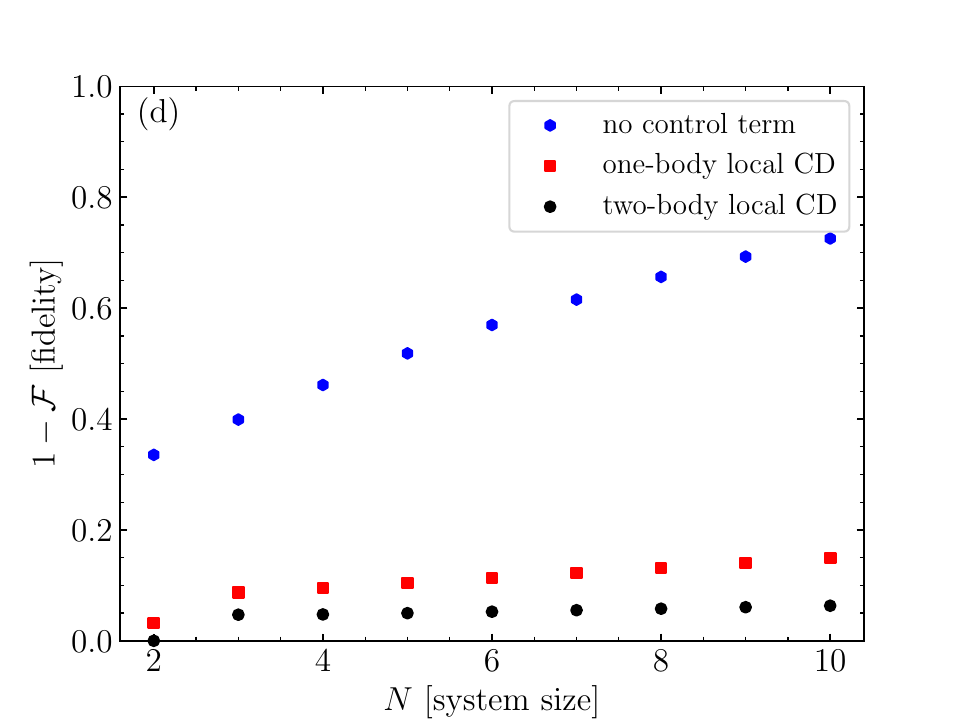}
\caption{Implementation of local CD in the Ising spin model. The case of \steve{no control term(s) being} implemented (blue) is compared to implementing one-body local CD (red) and two-body local CD (black). (a-c) The achieved fidelity ($1-\mathcal{F}$) for different total protocol times for (a) $N=2$, (b) $N=3$, and (c) $N=4$ spins. Note, for the case of (a) $N=2$ the two-body local CD gives unit fidelity to below single precision. (d) The achieved final fidelity for a fixed total time, $T=0.01 J^{-1}$, for various system sizes showing the saturation of the fidelity for the one and two-body implementations.}
\label{fig:figvarcd}
\end{figure}

We now turn to a concrete example of the implementation of variational CD in the Ising model. We will consider a system described by Hamiltonian~\eqref{eq:IsingNonIntegrable} with the following parameters
\begin{equation}
    J(\lambda) = J \ , \: Z(\lambda) = J \ , \: X(\lambda) = 2J\lambda \ , \nonumber
\end{equation}
with $\lambda$ being the time-dependent linear function
\begin{equation}
    \lambda(t) = \frac{t}{T} \,
\end{equation}
with $t$ going from $0$ to $T$ where $T$ is the total driving time. This allows us to explore both the adiabatic $T \gg J^{-1}$ and diabatic $T \ll J^{-1}$ regimes. First, we look to confirm that the variational CD terms are correctly derived by considering the case of $N=2$ in Fig.~\ref{fig:figvarcd}(a). Indeed, we observe that for all total driving times, $T$, the fidelity is perfect to single precision, whereas the one-body corrections only compensate some transitions, bringing the fidelity to $\sim\!0.9$. We note here, that there are limited other states for any excited dynamical state to populate with the Hilbert space consisting of only four states, as a result, the bare dynamical protocol with no variational CD still has an overlap of $\sim\!0.6$ with the ground state. Moving to larger systems of $N=3$ and $N=4$ sites in Fig.~\ref{fig:figvarcd}(b) and (c) we observe a clear decrease in the impact of variational CD to only two-body terms. This should be expected as the exact CD will include up to three- and four-body terms in the respective cases. The improvement that two-body variational CD can deliver in these larger systems is still substantial compared to the bare dynamical protocol, showing the utility of low-order variational CD when possible. In Fig.~\ref{fig:figvarcd}(d) we investigate the scaling of the fidelity for the bare and variational CD protocols for a fixed time $T=0.01J^{-1}$. We also briefly note that this approach of variational CD can be enhanced via the introduction of control fields \cite{COLD_PRXQ, morawetz2024efficient}, thus combining the methodologies described in this section with that of Sec.~\ref{sec:QOC}.

\subsection{Experimental implementations of counterdiabatic driving}
The first experimental demonstrations of such shortcuts to adiabaticity (STA) were performed using trapped Bose-Einstein condensates (BECs)~\cite{STAExp2011, CDExp2012}. In particular, control of the LZ model was achieved using a BEC trapped in an optical lattice and an effective two level system corresponding to Eq.~\eqref{eq:LZham} was accessed, in quasi-momentum space, by accelerating the gas~\cite{CDExp2012}. In this setting the exact CD dynamics, Eq~\eqref{HCD_LZ}, can be achieved in two ways: either directly introducing a second optical lattice that induces the required control field or by making a suitable transformation to the Hamiltonian, in effect absorbing the control field into modulations of the bare Hamiltonian parameters. The latter approach being clearly more experimentally preferable as it requires the manipulation of only a single optical potential. Indeed performing such a transformation to achieve the control is characteristic of many experimental realisation of STA protocols~\cite{CDExp2016a, CDExp2016b, CDExp2017}. The realised protocol was shown to be remarkably robust. Subsequently control of the LZ model was demonstrated in another experimental platform consisting of a single nitrogen-vacancy (NV) centre in diamond, where the effective two-level system was achieved by driving a particular subspace spanned by states of the electron spin~\cite{CDExp2013}. This implementation allowed to demonstrate the remarkable speed up that can be attained with CD driving, achieving an improvement of between 2 and 3 orders of magnitude. 

Counterdiabatic techniques have since been applied to more complex settings starting with three-level systems where CD control can significantly speed up stimulated Raman adiabatic passage protocols~\cite{CDExp2016a, CDExp2017, CDExp2022}, continuous variable control of a trapped ion~\cite{CDExp2016b}, and superconducting Xmon set ups to achieve high fidelity quantum gates~\cite{CDExp2018a} and characterising the thermodynamics of the control protocol~\cite{CDExp2018b}. The energetics of STA protocols has been recently explored in the context of the Landau-Zener model, Eq.~\eqref{HCD_LZ}, using a single trapped $^{40}$Ca$^+$ ion~\cite{STAExp2024} in order to experimentally demonstrate the tradeoff between resource consumption and achievable control~\cite{CampbellDeffnerPRL}.

Approximate CD driving has been demonstrated in controlling transport of population in a synthetic lattice of momentum states in ultracold ${}^{87}$Rb with a two orders of magnitude improvement in population transfer \cite{meier2020counterdiabatic} and in ground state preparation in a nuclear-magnetic-resonance setup \cite{zhou2020experimental}. The commutator exapnsion form of the AGP has led to the developlment of a new family of variaitional quantum algorithms, so-called CD Quantum Approximate Optimisation Algorithm (QAOA) \cite{wurtz2022counterdiabaticity,Chandarana2022digitized} which has been implemented on current hardware, including examining potential applications in protein folding~\cite{Chandarana2023digitized} and finance~\cite{Hegade2022portfolio}.
\clearpage


\section{Quantum Optimal Control} \label{sec:QOC}

In the previous section we discussed the basics of adiabatic control and the toolbox of shortcuts to adiabaticity. That framework assumed that the initial and target states of a control problem are given in terms of eigenstates of our Hamiltonian for different values of the control parameter. In some scenarios, however, this might not be the case. In this section we introduce the tools of quantum optimal control, which will allow us to tackle arbitrary quantum control problems by systematically searching for the optimal form of the control fields that drive a quantum system to given target configuration. \steve{Typical optimal control implementations require repeated numerical evaluations of the full dynamics of a quantum system, and thus are often, in practice, restricted to small systems. Nevertheless, the success of optimal control techniques in solving control problems in few-body settings is a testament to its importance in the field. In the following we introduce and illustrate these tools by revisiting some of the models introduced in Sec. \ref{sec:STA} and analyzing them through the lens of optimal control. First, we will study optimal state control for a single qubit system obeying  the Landau-Zener Hamiltonian (c.f. Sec.~\ref{sec:LandauZener}), and then its extension to unitary gate preparation with a model of phase-only control. Finally, we will apply optimal control to the problem of generating entangled states in an extended version of the all-to-all Ising model introduced in Sec.~\ref{ex:LMGmodel}.}


\subsection{Basic statement of an optimal control problem}
\label{subsec:QOC_intro}

In any quantum control problem, the goal is to maximize the success of a control protocol in driving a quantum system towards a predefined target configuration. For problems of state control, like the ones discussed in the previous section, this can be theoretically quantified by the quantum state fidelity, which measures how accurately the state at the final time $\ket{\psi(T)}$ approaches the target state $\ket{\psi_\ast}$ (typically, up to a global phase). This reads
\begin{equation}
    \mathcal{F}_S = \lvert \langle \psi_\ast |\psi(T)\rangle|^2,
\end{equation}
where $i\hbar \frac{d}{dt}\ket{\psi(t)}=H(t)\ket{\psi(t)}$. A more demanding control problem is that of unitary control, where we aim for the complete unitary evolution $U(T)$ to approach a target transformation $U_\ast$ (again up to a global phase). This will be the case, for instance, when the goal is to implement a quantum gate.  In such cases, the success of a protocol can be measured by 
\begin{equation}
    \mathcal{F}_U = \frac{1}{d^2}\left\lvert \Tr\left(U_\ast^\dagger U(T)\right)\right\rvert^2,
\end{equation}
 where now $i\hbar \frac{d}{dt}U(t)=H(t)U(t)$ and $d$ is the Hilbert space dimension. In this section, we introduce the tools of \textit{quantum optimal control} (QOC), which give a systematic approach to engineer $H(t)$ such that the target is approximately achieved, i.e., $\mathcal{F}\simeq 1$, where $\mathcal{F}$ symbolizes the state or unitary fidelities, depending on the problem at hand. Note that, depending on the specific control problem, alternative measures of protocol success can be defined. For instance, one could be interested in minimizing the energy of a system, or in achieving a specific expectation value of an observable. Here, for concreteness, we will focus on the use of the fidelity as such a measure.

We start by considering the system Hamiltonian $H(t)$ to be a function of a set of real \textit{control fields} $\bm{a}(t)=\{a_k(t)\}$, $j=1,\ldots,K$. For instance, we could consider the Hamiltonian to be written as
\begin{equation}
    H(t) = H_0 + H_C(t) = H_0 + \sum\limits_{k=1}^K a_k(t) H_{C,k},
    \label{eq:QOC_hami}
\end{equation}
which makes explicit the existence of a drift (or free) Hamiltonian $H_0$, which we cannot manipulate directly, and the control Hamiltonian $H_C(t)$, which depends on time through the control fields $\bm{a}(t)$ modulating the individual operators $H_{C,k}$. The properties of the drift and control Hamiltonians determine which states and transformations are in principle reachable to the system. Well-established methods to characterize this aspect exist \cite{dalessandro_book,poggi2019geometric}, and use tools rooted in Lie group theory to determine the degree of controllability of a given model.

Here we take a pragmatic approach, and establish a systematic way to find the control functions $\bm{a}(t)$ that minimize a \textit{cost functional}
\begin{equation}
    J[\bm{a}(t)] = 1-\mathcal{F}[\bm{a}(t)]
    \label{eq:QOC_functional}
\end{equation}
where $\mathcal{F}$ refers to either the state or unitary fidelity, and we have indicated the functional dependence of the fidelity explicitly. Note that here we follow the convention of defining a functional to be \textit{minimized}; one could similarly define a functional to be maximized instead. We will focus on solving this problem via numerical optimization, for which we introduce a parametrization of the control fields in terms of real numbers
\begin{equation}
    \bm{\alpha}=(\alpha_1,\alpha_2,\ldots,\alpha_M) \rightarrow \bm{a}(t),
    \label{eq:QOC_parametrization}
\end{equation}
which serve as a our control variables, and will allow us to treat the problem of minimizing Eq. (\ref{eq:QOC_functional}) with standard numerical optimization tools. Note that, while we do not treat them in this tutorial, there are approaches in optimal control theory which rely on functional optimization (see Krotov's method \cite{krotov1999,reich2012} and the \steve{Pontryagin} Maximum Principle~\cite{PRXQtutorial}). 

Generically, a numerical minimization routine involves initializing the controls with a guess $\bm{\alpha}_{0}$, and then iteratively updating it following a set of rules that depend on the optimization algorithm. These can be broadly classified as gradient-based and gradient-free depending on whether they compute derivatives of the cost function $J(\bm{\alpha})$ at various points to decide the most appropriate way to update the control field $\alpha(t)$ at each iteration. The choice of parametrization in Eq.~\eqref{eq:QOC_parametrization} is in principle arbitrary, but can be tailored to fit computational or experimental constraints. For instance, a Fourier parametrization would take the form
\begin{equation}
    \bm{\alpha} = (A_1,\ldots,A_{n_F},B_1,\ldots,B_{n_F})\rightarrow a(t) = \sum\limits_{m=1}^{n_F}\left[A_m \cos(\omega_m t) + B_m\sin(\omega_m t)\right], 
    \label{eq:QOC_fourier}
\end{equation}
where $n_F$ is the number of Fourier components and $\omega_m$ could correspond to the harmonics of some base frequency, or be chosen randomly (as in the chopped randomised basis, CRAB, algorithm \cite{doria2011_crab}). For concreteness, here we focus on parametrizing the field as a piecewise-constant function, 
\begin{equation}
\bm{\alpha} = (\alpha_1,\alpha_2,\ldots,\alpha_M)\rightarrow a(t) = \alpha_j\ \text{if}\ (j-1)\Delta t \leq t < j\Delta t,\ \text{where}\  j=1,\ldots,M.
\label{eq:QOC_pwc}
\end{equation}
where we have chosen a uniform time step $\Delta t$ for convenience. This choice is conceptually simple and allows us to analytically compute the gradient of the cost function of Eq. (\ref{eq:QOC_functional}) with respect to the control variables, which greatly improves the efficiency of the numerical search. We will work out this derivation explicitly in the next section.

\begin{table}[b]
\centering
\begin{tabular}{|p{12.5cm}|c|}
\hline
Object & Proposed Symbol \\
\hline\hline
Target state & $\ket{\psi_\ast}$ \\
\hline
Fidelity (state) & $\mathcal{F}_S$ \\
\hline
Target unitary & $\ket{U_\ast}$ \\
\hline
Fidelity (unitary) & $\mathcal{F}_U$ \\
\hline
Control field & $a(t)$ \\
\hline
Cost functional & $J[a(t)]$ \\
\hline
Control parameters & $\bm{\alpha}={\alpha_1,\ldots,\alpha_M}$  \\
\hline
Number of time steps & $M$\\
\hline 
Time step length & $\Delta t$ \\
\hline
Total evolution time & $T$ \\
\hline
Quantum speed limit time & $\tau_{\mathrm{QSL}}^\ast$ \\
\hline
\end{tabular}
\caption{Summary of notation for the most common quantities used in quantum optimal control.}
\label{table:QOCSymbols}
\end{table}

\begin{figure}[t]
\centering
\includegraphics[width=0.55\columnwidth]{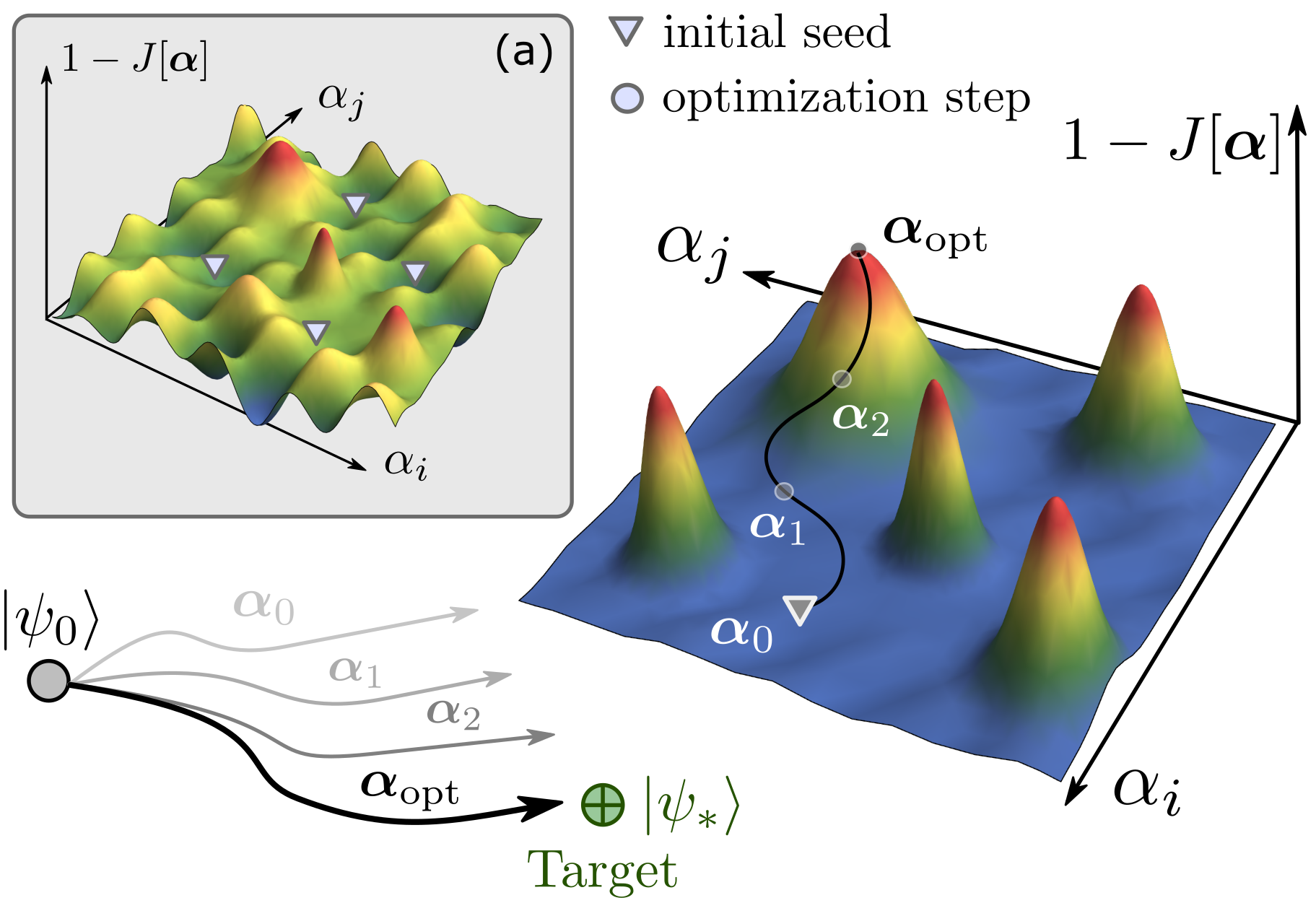}
\caption{Quantum control landscapes are determined by the dependence of the cost function $J(\bm{\alpha})$ on the control variables $\bm{\alpha}$. Main figure depicts how the optimal solutions, initial seeds, and optimization steps would appear in a two-dimensional landscape (note that real control problems typically have much more than $M=2$ dimensions). An optimization routine starts with an initial guess for the control variable $\bm{\alpha}_0$ which typically lands far from an extrema of the landscape, meaning that the evolution of the quantum state is far from the target configuration at the final time. The optimization consists in updating the control variables sequentially until an optimum is achieved $\bm{\alpha}_{\text{opt}}$. In the presence of constraints, quantum control problems can have rugged landscapes with suboptimal extrema, as depicted in subfigure (a). In such cases, using many different initial seeds helps explore various parameter regions of the landscape.}
\label{fig:QOC_landscapes}
\end{figure}

The definition of the control Hamiltonian, e.g. Eq. (\ref{eq:QOC_hami}), and parametrization, c.f. Eq. (\ref{eq:QOC_parametrization}), together with the choice of initial and target states ($\ket{\psi_0}$ and $\ket{\psi_\ast}$ for state control, $\mathds{1}$ and $U_\ast$ for unitary control), completely determine the cost function $J(\bm{\alpha})$, which can be seen as an optimization \textit{landscape}. Each point on the landscape corresponds to a different control field, with optimal protocols being global minima. The goal of any optimization algorithm is to traverse this landscape, starting from the initial guess $\bm{\alpha}_0$, to eventually reach the global extremum via a series of updates of the control parameters $\bm{\alpha}_0\rightarrow\bm{\alpha}_1\rightarrow \ldots \rightarrow \bm{\alpha}_{\text{opt}}$. Here, we assume that such updates are done via a local search (instead of a global one), where only properties about the current point and its surrounding region informs the update rule. This process is depicted in Fig. \ref{fig:QOC_landscapes}. The properties of the landscape determine the hardness of finding a solution to the control problem. In a generic optimization problem, the landscape is rugged and filled with suboptimal minima or traps, as can be seen in Fig.~\ref{fig:QOC_landscapes}(a), which often increases the computational expense of finding a near-optimal solution. In this scenario, it is typically convenient to balance the computational cost of the optimization between the exploration and the exploitation of the landscape. An optimization landscape can be explored by starting the search multiple times with different initial guesses (or ``seeds'') which aim to cover different regions of parameter space. Each run is then exploited by allowing the search a sufficient number of iterations to converge to a good solution.  For quantum control problems, remarkably, it has been shown that optimization landscapes tend to have benign features \cite{rabitz2004}; in particular, provided there are no constraints to the control \steve{problem} \cite{zhdanov2015}, quantum control landscapes can be shown to be generically devoid of traps \steve{in fully controllable systems} (see Fig. \ref{fig:QOC_landscapes}). \steve{In practice, however, constraints on control problems (which we discuss below) will typically affect the topology of control landscapes}. 
Interestingly, recent developments in quantum computing, particularly variational quantum algorithms, have revealed new properties of quantum control landscapes, like the phenomenon of barren plateaus, which we discuss briefly in Sec. \ref{outlook_sec_qc}.

Often, constraints need to be taken into account when solving practical quantum control problems, and thus understanding their effect on the optimization search is an important task. Constraints can be broadly divided into two categories: 
\begin{enumerate}
    \item Constraints which are \textbf{built into} the control model. For instance, a fixed set of control Hamiltonians in the expansion of Eq. (\ref{eq:QOC_hami}), a limited bandwidth set by the number of Fourier components in Eq. (\ref{eq:QOC_fourier}), or a limited evolution time $T = M\Delta t$ in Eq. (\ref{eq:QOC_pwc});
    \item Constraints which are imposed \textbf{a posteriori} to the problem as additional terms to the cost function. For instance, bounding the amplitude of driving field to limit its energetic cost \cite{reich2012} or demanding the control processes to be robust to certain external perturbations \cite{kosut2022,poggi2023_urc}.
\end{enumerate}

We highlight here the role of one of these constraints: limiting the total evolution time $T$. Quantum mechanics imposes fundamental limitations to the rate at which one state can evolve into a different state. These are formalized as quantum speed limits (QSLs)~\cite{Deffner2017}, which provide lower bounds on the total evolution time, c.f. $T\geq \tau_{\text{QSL}}$. For unitary dynamics, QSLs can be traced back to time-energy uncertainty relations, and take several forms \cite{taddei2013, pires2016, mondal2016,marvian2016}. In particular, we consider the bound derived from the work of Mandeltstam and Tamm \cite{mandelstam_45,bhattacharyya1983}:
\begin{equation}
    T \geq \frac{\hbar}{\overline{\Delta E}}\arccos\left(\lvert \langle \psi_\ast|\psi_0\rangle\rvert\right) \equiv \tau_{\text{QSL}},
    \label{eq:Tqsl}
\end{equation}
 where 
\begin{equation}
    \overline{\Delta E} = \frac{1}{T}\int\limits_0^T dt \Delta E(t)
\end{equation}
and $\Delta E(t)$ is the variance of the Hamiltonian $H(t)$,
\begin{equation}
    \Delta E(t)^2 = \bra{\psi(t)}H(t)^2\ket{\psi(t)}-\bra{\psi(t)}H(t)\ket{\psi(t)}^2.
\end{equation}
with $H(t)$ the total Hamiltonian, e.g. as defined in Eq.~\eqref{eq:QOC_hami}. The existence of QSLs fundamentally impacts quantum control problems as any optimization is destined to fail if one fixes the evolution time $T$ to be below $\tau_{\text{QSL}}$. Furthermore, it is usually desirable to derive control processes which are as fast as possible since long evolution times allow for more sources of errors and decoherence to affect the dynamics. While in some cases optimal control solutions exist exactly at the fundamental bound $T=\tau_\text{QSL}$ \cite{caneva2009}, typically constraints mean we can achieve  solutions only at longer times $\tau_\text{QSL}^\ast>\tau_\text{QSL}$ \cite{hegerfeldt2013,poggi2019geometric}. By definition, the time $\tau_\text{QSL}^\ast$ is the minimum possible evolution time for which a given optimal control problem has a solution (i.e., one that achieves a fidelity $\mathcal{F}=1$). In Sec. \ref{subsec:QOC_example_state} we will illustrate how to use optimal control methods to systematically search for the shortest evolution time $\tau_{\text{QSL}}^\ast$ for simple control problems, and explore how constraining the evolution time affects the optimization landscape. \steve{We point that out that geometric QSL bounds like Eq. (\ref{eq:Tqsl}), while universal, tend to become loose as the size of the system increases (see \cite{bukov2019geometric} for an exception) This is because the energy variance in the denominator is typically an extensive quantity.}

\subsection{Gradient-based optimal control using GRAPE} \label{subsec:QOC_grape}
In this section we discuss in more detail how to approach an optimal control problem numerically. We focus on the simplest case of having a single control field $a(t)$ parametrized with the piecewise constant ansatz of Eq. (\ref{eq:QOC_pwc}), leading to control variables grouped in a vector $\bm{\alpha}=\{\alpha_1,\alpha_2,\ldots,\alpha_M\}$. For a given choice of number of steps $M$ and total evolution time of $T$, the associated time step is $\Delta t=T/M$. This choice allows us to write the final-time evolution operator in closed form
\begin{equation}
    U(T) = U_M U_{M-1} \ldots U_2 U_1,\ \text{where}\  U_j = \exp\left(-iH_j\Delta t\right)\ \text{and}\ H_j =H(\alpha_j)
\end{equation}

To search for a choice of $\bm{\alpha}$ that minimizes the cost function $J(\bm{\alpha})$, one can resort to a variety of numerical optimization algorithms. Because we expect the cost function to have a smooth dependence on the control variables in general, it is convenient to use gradient-based methods, which use information about the derivatives $\partial J/\partial \alpha_k$ to inform the search at each step. Examples of these include gradient descent, Newton - Raphson and the Broyden-Fletcher-Goldfarb-Shannon (BFGS) algorithms. These are widely used optimization methods which are easily available in Python, Matlab, and other languages using standard packages (note that some quantum optimal control approaches also benefit from using gradient-free methods, like Nelder-Mead or Powell \cite{doria2011_crab,COLD_PRXQ}). For these methods, the computation of the gradient can be done numerically (e.g., via finite-difference approximations, which are typically carried out automatically by these optimization routines) or it can be provided explicitly such that it only has to be numerically evaluated. The piecewise-constant parametrization of the control field we are using here actually allows us to compute the gradient of $J(\bm{\alpha})$ analytically in a relatively straightforward way. The use of this parametrization together with the analytical form of the cost gradient in a quantum control problem constitutes the GRAPE (gradient-ascent pulse engineering) algorithm, first proposed by Khaneja \textit{et al.} \cite{khaneja2005}. 

To compute the gradient, we first note that for both state and unitary control problems the cost function can be written as $J(\bm{\alpha})=1-\mathcal{F}(\bm{\alpha})=1-|z(\bm{\alpha})|^2$ where
\begin{align}
\text{state control:} & \ z(\bm{\alpha})=\bra{\psi_\ast}U(T)\ket{\psi_0} \\
\text{unitary control:} & \ z(\bm{\alpha})=\frac{1}{d}\Tr\left(U_\ast^\dagger U(T)\right), 
\end{align}
and thus
\begin{equation}
    \frac{\partial J}{\partial \alpha_j} = \frac{\partial}{\partial \alpha_j}(z^\ast z) =  -2\Re{z^\ast \frac{\partial z}{\partial \alpha_j}}.
    \label{eq:QOC_gradZ}
\end{equation}

As the control dependence is entirely contained within $U(T)$, to obtain $\frac{\partial z}{\partial \alpha_j}$ we require to calculate
\begin{equation}
      \frac{\partial U(T)}{\partial \alpha_j} =U_M U_{M-1}\ldots \frac{\partial U_j}{\partial \alpha_j} U_{j-1}\ldots U_1
    \label{eq:QOC_gradient0}
\end{equation}
and thus we need to compute $\frac{\partial U_j}{\partial \alpha_j}$. Here, one should proceed with care because $U_j=\exp(-i H_j \Delta t)$ does not necessarily commute with $\partial H_j/\partial \alpha_j$. In general, we have

\begin{align}
    \frac{\partial U_j}{\partial \alpha_j} & = \frac{\partial}{\partial \alpha_j}\sum\limits_{k=0}^\infty \frac{(-i)^k}{k!}\Delta t^k H_j^k \\
    & = \sum\limits_{k=0}^\infty \frac{(-i)^k}{k!}\Delta t^k \sum\limits_{i=1}^k H_j^{i-1} \frac{\partial H_j}{\partial \alpha_j} H_j^{k-i}
    \label{eq:QOC_derUj_aj}
\end{align}

In the limit where the time step $\Delta t$ is small, and thus the piecewise constant field resembles a continuous function, the expression in Eq. (\ref{eq:QOC_derUj_aj}) can be simplified, see for instance Ref. \cite{khaneja2005,ansel2024_arxiv}. For that limit to be relevant, however, we need to deal with a large number $M$ of control variables, which can be undesirable. Here we focus on obtaining an exact closed form for $\partial U_j/\partial \alpha_j$, valid for arbitrary time step sizes $\Delta t$ \cite{machnes2011,motzoi2011}. For this, we introduce the spectral decomposition of $H_j = \sum_{l=1}^d e_l \ketbra{l}{l}$ (note that both the eigenvalues and eigenvectors will be different for each $j$, i.e. for each time step). Inserting this into Eq. (\ref{eq:QOC_derUj_aj}), we obtain

\begin{align}
    \frac{\partial U_j}{\partial \alpha_j} & = \sum\limits_{k=0}^\infty \frac{(-i)^k}{k!}\Delta t^k \sum\limits_{i=1}^k \sum\limits_{l,m} e_l^{i-1}e_m^{k-i} \ketbra{l}{m} \bra{l}\frac{\partial H_j}{\partial \alpha_j}\ket{m} \nonumber \\
    & = \sum\limits_{l,m} \ketbra{l}{m} \bra{l}\frac{\partial H_j}{\partial \alpha_j}\ket{m} \sum\limits_{k=0}^\infty \frac{(-i)^k}{k!}\Delta t^k \sum\limits_{i=1}^k e_m^{k-1} \left(e_l/e_m\right)^{i-1}.
    \label{eq:QOC_gradient1}
\end{align}

The infinite series appearing in the above expression can be evaluated exactly. For the case $e_l = e_m$, we obtain

\begin{equation}
    \sum\limits_{k=0}^\infty \frac{(-i)^k}{k!}\Delta t^k k e_m^{k-1} = -i\Delta t \sum\limits_{k=0}^\infty \frac{(-i)^{k-1}}{(k-1)!}\Delta t^{k-1} e_m^{k-1}=-i\Delta t \exp{-i e_m \Delta t}.
    \label{eq:QOC_gradient2}
\end{equation}

If $e_l\neq e_m$, we can proceed as follows
\begin{align}
    \sum\limits_{k=0}^\infty \frac{(-i)^k}{k!}\Delta t^k e_m^{k-1}\sum\limits_{i=1}^k \left(e_l/e_m\right)^{i-1} & = \sum\limits_{k=0}^\infty \frac{(-i)^k}{k!}\Delta t^k \frac{e_m^k}{e_m}\sum\limits_{n=0}^{k-1} \left(e_l/e_m\right)^{n} \nonumber \\
    & = \sum\limits_{k=0}^\infty \frac{(-i)^k}{k!}\Delta t^k \frac{e_k^k - e_m^k}{e_l-e_m} \nonumber \\
    & = \frac{1}{e_l-e_m}\left(e^{-i e_l \Delta t}-e^{-i e_m \Delta t}\right).
    \label{eq:QOC_gradient3}
\end{align}

Combining expressions (\ref{eq:QOC_gradient1}), (\ref{eq:QOC_gradient2}) and (\ref{eq:QOC_gradient3}) yields an analytical expression for all the matrix elements of $\partial U_j/\partial \alpha_j$. Notice that this requires us to diagonalize the operator $H_j$ for each $\alpha_j$, which will need to be done numerically in most cases. For a single qubit, the eigenvectors and eigenvalues of $H_j$ have a analytic form for an arbitrary $\alpha_j$. Here we do not use this form to illustrate the more general procedure, which applies to more complex systems. Once the factor from Eq. (\ref{eq:QOC_gradient1}) is computed, the gradient can be constructed from Eq. (\ref{eq:QOC_gradient0}),

\begin{align}
    \text{state control:} & \     \frac{\partial z}{\partial \alpha_j} = \bra{\psi_\ast}U_{j+1,M}\frac{\partial U_j}
    {\partial \alpha_j}U_{1,j-1}\ket{\psi_0}\\
    \text{unitary control:} & \ \frac{\partial z}{\partial \alpha_j} = \frac{1}{d}\Tr\left(U_\ast^\dagger U_{j+1,M}\frac{\partial U_j}
    {\partial \alpha_j}U_{1,j-1}\right)
\end{align}
where we introduced the notation $U_{k,l}=U_l U_{l-1}\ldots U_{k+1}U_k$. Note that in the case of state control, the cost function gradient at step $j$ depends on the initial state $\ket{\psi_0}$ forward-evolved until time-step $j-1$, and on the target state $\ket{\psi_\ast}$ backwards-evolved until time-step $j+1$.\\

We can now construct a basic optimal control routine by following these steps:

\begin{enumerate}
    \item Set the final evolution time $T$ and the number of timesteps $M$ (which defines the number of control variables). This suffices to define a discretized time variable $t\rightarrow \{t_k = k\Delta t\}$, with $k=0,1,\ldots,M$, and $M\Delta t=T$.
    \item Choose an initial guess or ansatz $\bm{\alpha}_0$ for the control variable. 
    \item Implement a numerical minimization routine. Typical inputs for such routine are
    \begin{itemize}
        \item The cost function $J(\bm{\alpha})$.
        \item The particular minimization method to be used (BFGS, gradient descent, ...)
        \item The associated gradient function $g(\bm{\alpha})=\frac{\partial}{\partial \bm{\alpha}}J(\bm{\alpha})$, Eq. (\ref{eq:QOC_gradZ}).
        \item The initial guess $\bm{\alpha}_0$, defined in step 2.
        \item Bounds on the control variables such that the search is restricted to $\alpha_{min}\leq \alpha_j \leq \alpha_{max}$ (optional)
        \item Numerical tolerances or thresholds on the cost and/or the gradient, which determine at what level of precision the search is allowed to stop.
    \end{itemize}
\end{enumerate}

In the remainder of this section, we consider two examples in which we illustrate this procedure for both state and unitary control in a two-level system.


\subsection{Examples}
\subsubsection{Example: optimal control in the Landau-Zener model}
\label{subsec:QOC_example_state}
We begin by considering control of the Landau-Zener Hamiltonian, already introduced in Sec. \ref{sec:LandauZener}, which has the form 
\begin{equation}
    H(t) = H_0 + \nu(t) H_C = \Delta \sigma^x + \nu(t) \sigma^z.
    \label{eq:QOC_hami_LZ}
\end{equation}
We also recall here the form of the ground state of $H(t)$ as a function of instantaneous value of $\nu(t)$:
\begin{equation}
    \ket{\phi_g(\nu)} \equiv \cos\left(\frac{\theta}{2}\right) \ket{0}+\sin\left(\frac{\theta}{2}\right) \ket{1},\ \text{with}\ \tan\theta = \frac{\Delta}{\nu}.
\end{equation}
For this model we are interested in the problem of driving the system from the initial state 
\begin{equation}
    \ket{\psi_0}=\ket{\phi_g(-\nu_0)}
\end{equation}
to the target state
\begin{equation}
    \ket{\psi_\ast}=\ket{\phi_g(+\nu_0)}
\end{equation}
for some $\nu_0>0$. Note that, as $\nu_0\rightarrow \infty$, we have $\theta\rightarrow 0 $ and the desired evolution is the one that connects $\ket{\psi_0}=\ket{0}$ to $\ket{\psi_\ast}=\ket{1}$. We consider the field $\nu(t)$ to be piecewise-constant with $M=10$ steps and denote its values $\{\alpha_1,\ldots,\alpha_M\}$ to make notation consistent with the previous section. Note that this example obeys the decomposition of Eq. (\ref{eq:QOC_hami}). Even in this very simple case, one finds that $H_j = H_0+\alpha_j H_C$ and $\partial H_j/\partial \alpha_j=H_C=\sigma^z$ do not commute, which motivates the need for the calculation of the gradient shown in Sec. \ref{subsec:QOC_grape}.\\

\begin{figure}[t]
\includegraphics[width=0.55\columnwidth]{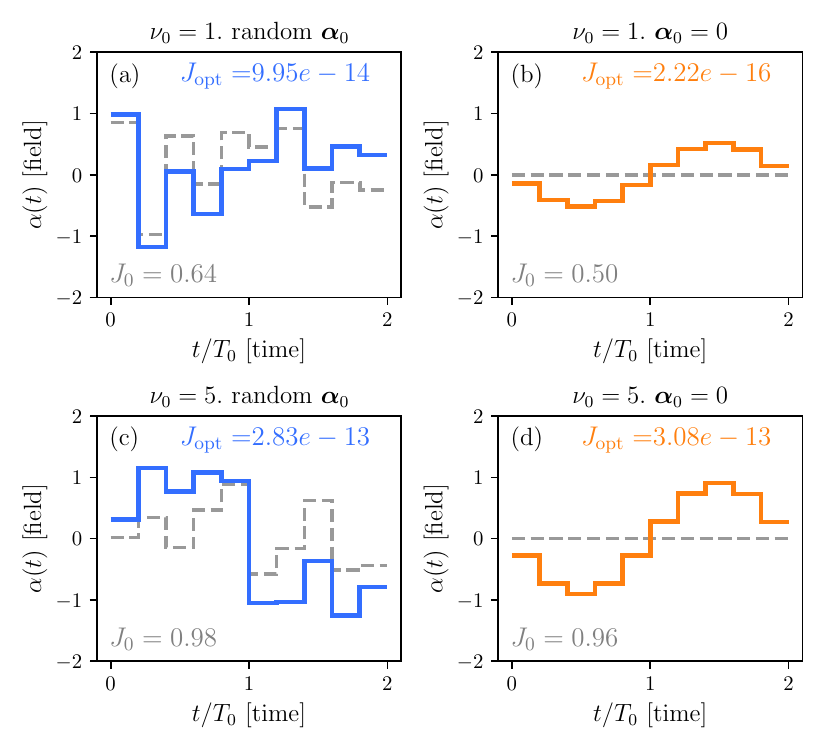}
\includegraphics[width=0.43\columnwidth]{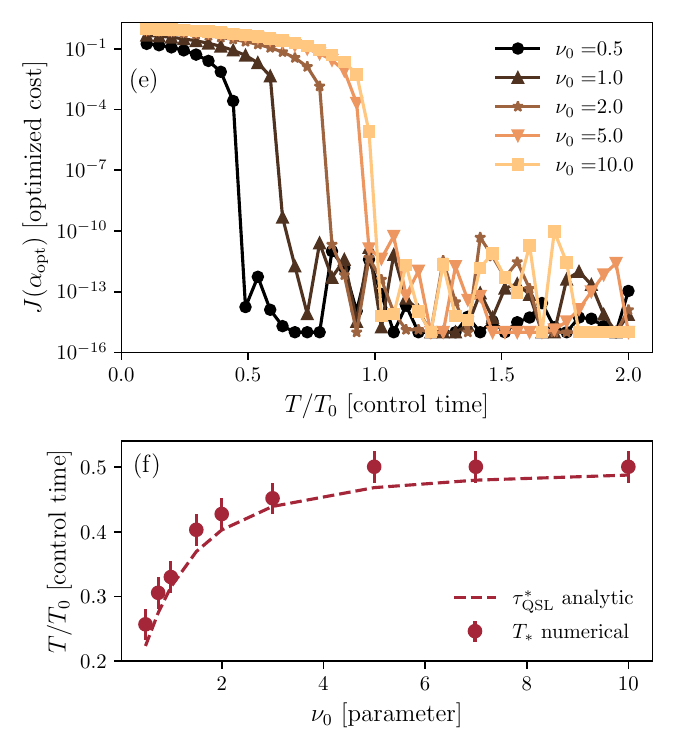}
\caption{Quantum optimal state control for the Landau-Zener model. (a)-(d) Show various control fields related to realizing a state transfer process between ground states of the Landau-Zener Hamiltonian, c.f. Eq. (\ref{eq:QOC_hami_LZ}). Shown are the initial guesses (gray dashed lines) and optimized fields (thick colored lines) for four different cases. (a) Initial state given by $\nu_0=1$, and initial guess for control field $\bm{\alpha}_0$ chosen randomly. (b) $\nu_0=1$ and $\bm{\alpha}_0=0$. (c) $\nu_0=5$ and $\bm{\alpha_0}$ random. (d) $\nu_0=5$ and $\bm{\alpha}_0=0$. In all cases the initial $J_0\equiv J({\bm{\alpha}_0})$ and optimized $J_{\mathrm{opt}}\equiv J(\bm{\alpha}_{\mathrm{opt}})$ values of the cost function are displayed. Parameters used $M=10$, $T=2T_0$ where $T_0=\pi/(2\Delta)$. (e) Optimized cost functions achieved by the optimization as a function of total evolution time $T$, for various choices of the parameter $\nu_0$ determining the initial and final state. (f) Minimum control time for the Landau-Zener Hamiltonian, as a function of $\nu_0$. Circles indicate the numerical estimates obtained from the data shown in (e). Full line corresponds to the analytical solution derived in Ref. \cite{hegerfeldt2013}.}
\label{fig:QOC1}
\end{figure}

To test the optimization procedure of steps 1-3 above, we start by fixing $T=\pi/\Delta$, and using a limited-memory variant of the BFGS method (L-BFGS-B) as our optimization routine. We also restrict the search to $|\alpha_j|\leq c = 2\Delta$ to avoid control fields with large amplitudes. With these specifications, we run the optimal control routine for two choices of the parameter $\nu_0$ determining the initial and target states ($\nu_0=1$ and $\nu_0=5$), and two choices of initial ansatz; one where the $\alpha_j$'s are sampled from a uniform distribution between $[-1,1]$, and another where we choose $\alpha_j=0$ for all $j=1,\ldots,M$. Results are shown in Fig.~\ref{fig:QOC1}, where we display the initial guess for the control field (gray dashed lines) and the optimal field obtained with the optimization routine (colored full lines). In all cases we observe that the cost function is initially of order $J(\bm{\alpha}_0) \in (0.5 , 1)$, and that the optimization is able to achieve $J(\bm{\alpha}_{\text{opt}})\in \left( 10^{-16},10^{-13}\right)$ in any standard run. Comparison of Figs.~\ref{fig:QOC1} (a)-(b) and (c)-(d) reveals that for a given control problem (i.e., for a fixed $\nu_0$), different initial guesses perform similarly. The comparison also illustrates two generic facts of quantum optimal control. \steve{When the system is controllable and given enough control resources,} control problems typically have multiple possible solutions which are (roughly) equally good. Even for fixed, finite values of $T$ and $M$ (which effectively impose constraints on the control problem) these multitude of solutions can be accessed by exploration of the control landscape with different initial guesses $\bm{\alpha_0}$ (see illustration in Fig. \ref{fig:QOC_landscapes}). \steve{In addition}, the optimal fields often inherit properties of the initial guesses. This is because in presence of multiple minima,  a local optimization routine like the one considered here will lead to a solution close to the initial point. In our example, we see that the optimal fields obtained following the random ansatz look random themselves, while the optimal fields obtained from the constant initial guess are reveal to be structured and symmetric. For simple systems like the one studied here, these features might be easily traced back to symmetries of the model \cite{larocca2018} but, in more complex cases, the structure of the optimal field can be useful to identify nontrivial properties of the system \cite{rabitz2006_topology,poggi2015,bukov2018_symmetry}. \\

{\it Identifying quantum speed limits using optimal control.---} Following on from previous discussions on quantum speed limits, it is natural to ask what is the minimum evolution time $\tau_{\text{QSL}}^\ast$ required to achieve a good solution for this control problem. To explore this question, we set out to solve the optimization problem for different values of the total evolution time $\{T_1,T_2,\ldots,T_R\}$, arranged in increasing order. In principle, each choice of $T$ leads to an independent optimization procedure, given by the steps 1-3 above. Here we describe a slightly more elaborate approach, which has been routinely used in previous studies \cite{caneva2009,murphy2010,goerz2011,poggi2015_qsl,goerz2017, ref2, ref3, ref4}. We start by solving the optimal control problem for the largest value of evolution time $T_R$ with a given choice of initial ansatz $\bm{\alpha}_0^{(R)}$ (say, a uniformly random one). If $T_R>\tau_{\text{QSL}}^\ast$, then we expect to be able to obtain an optimal field $\bm{\alpha}_{\text{opt}}^{(R)}$ yielding a very small cost function. We then move on to solve the optimization for $T_{R-1}$, but now we choose as an initial guess the optimal field from the previous run, i.e.
\begin{equation}
    \text{initial guess for $T=T_{j-1}$}\rightarrow \bm{\alpha}_0^{(j-1)} =  \bm{\alpha_{\text{opt}}^{(j)}}\leftarrow \text{optimal field from $T=T_j$}.
\end{equation}
When repeating this procedure for every choice of $T$ in our list, we are ``helping'' each new optimization by seeding a field which we expect to lead to a small cost (as long as $T_j-T_{j-1}$ is sufficiently small). In Fig.~\ref{fig:QOC1}(e)  we plot the optimized cost as a function of the evolution time $T$ for this procedure, for various choices of $\nu_0$ (i.e. for various choices of initial and target states). Analyzing the curves starting from the right (large values of $T$), in all cases we observe nearly optimal fidelities, corresponding to costs of orders between $10^{-12}$ and $10^{-15}$. As the allowed evolution time decreases, the optimization is able to find good solutions up until a critical time, after which the optimized cost grows rapidly and takes values $\sim 10^{-1}$. This critical time $T_*$ divides the controllable regime where $T>T_*$ and the uncontrollable regime where $T<T_*$. Because $T_*$ depends on details of the optimization procedure (for instance, on the number of time steps $M$), it can only provide  an upper bound to the true minimum control time $\tau_{\text{QSL}}^\ast$. Nevertheless, in many situations of interest, we can take $T_*$ as a good approximation of $\tau_{\text{QSL}}^\ast$. For the problem treated here, the actual optimal time is known analytically \cite{hegerfeldt2013,poggi2013}. In Fig. \ref{fig:QOC1}(f) we compare the numerically obtained $T_*$ and the analytical $\tau_{\text{QSL}}^\ast$ as a function of $\nu_0$. The results indicate that the numerical optimal control procedure is able to obtain good estimates of the fundamental optimal control time for this problem. In particular, we see how as $\nu_0\rightarrow \infty$ and thus $\ket{\psi_0}\rightarrow \ket{0}$ and $\ket{\psi_\ast}\rightarrow\ket{1}$, the control time reaches $\tau_{\mathrm{QSL}}^\ast=\frac{\pi}{2\Delta}$. This is the time required by $H=\Delta \sigma_x$ to rotate $\ket{0}$ to $\ket{1}$. 

\subsubsection{Example: optimal control of a single-qubit gate}
\label{subsec:QOC_example_gate}
\begin{figure}[t]
\includegraphics[width=0.55\columnwidth]{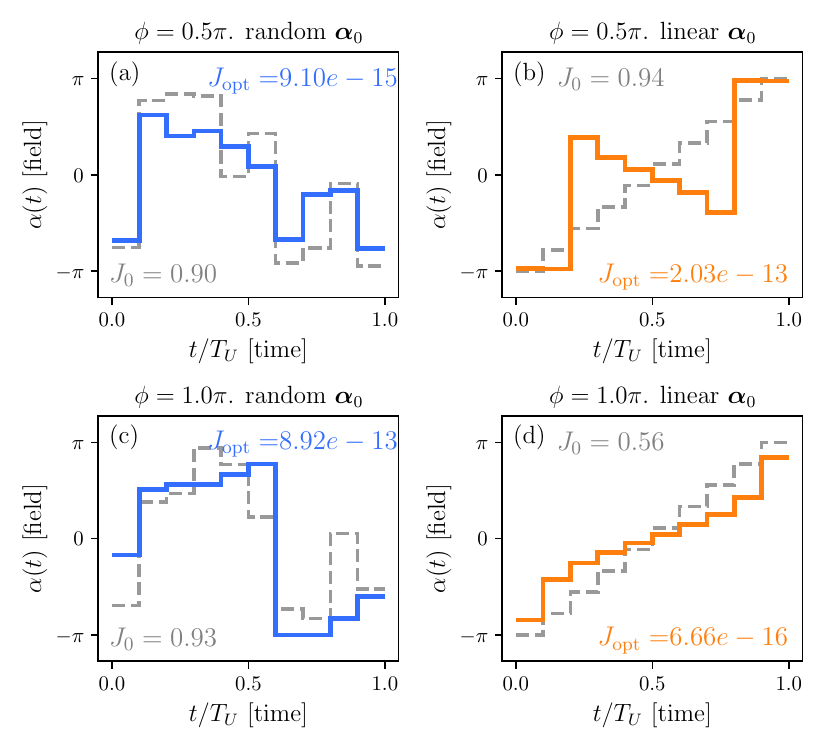}
\includegraphics[width=0.43\columnwidth]{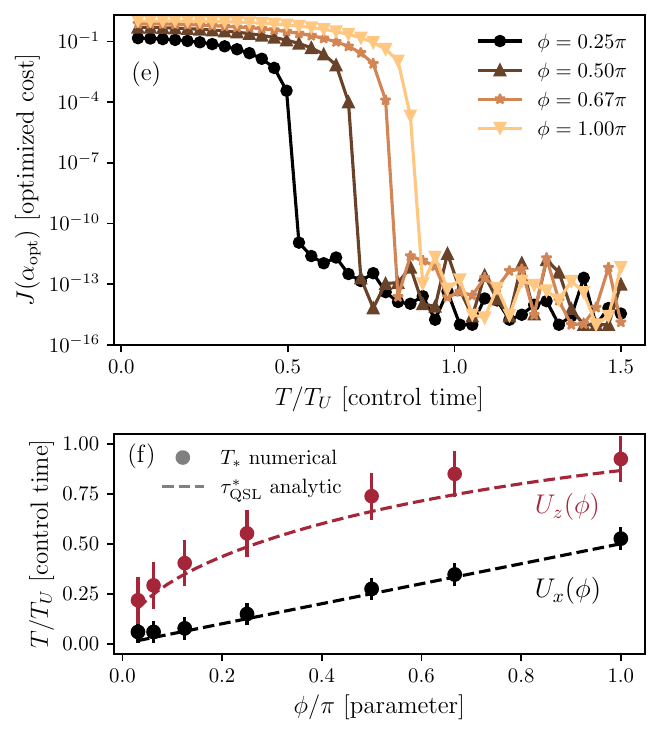}
\caption{Quantum optimal control for a single-qubit gate. (a)-(d) Show control fields related to implementing the target gates $U_z(\phi)$ in Eq. (\ref{eq:QOC_gate_target_z}) using the control Hamiltonian in Eq. (\ref{eq:QOC_gate_hami}). We display
initial guesses (gray dashed lines) and optimized fields (thick colored lines) for four different cases. (a) Target gate given by $\phi=\pi/2$, and initial guess for control field $\bm{\alpha}_0$ chosen randomly. (b) $\phi=\pi/2$ and $\bm{\alpha}_0$ linear. (c) $\phi=\pi/2$ and $\bm{\alpha}_0$ random. (d) $\phi=\pi$ and $\bm{\alpha}_0$ linear. In all cases the initial $J_0\equiv J({\bm{\alpha}_0})$ and optimized $J_{\mathrm{opt}}\equiv J(\bm{\alpha}_{\mathrm{opt}})$ values of the cost function are displayed. Parameters used $M=10$, $T=T_U$, where $T_U=2\pi/\Omega$. (e) Analysis of optimal cost functions achieved by the optimization as a function of total evolution time $T$, for various choices of the target $U_z(\phi)$.
(f) Minimum control time as a function of $\phi$ for the two families of target gates $U_x(\phi)$ and $U_z(\phi)$. Circles indicate the numerical estimates obtained from the optimal control procedure. Dashed lines correspond to analytical solutions. For $U_x$, $\tau_{\mathrm{QSL}}^\ast/T_U=\phi/(2\pi)$. For $U_z$ the optimal evolution time has been derived in Ref. \cite{boozer2012}.}
\label{fig:QOC2}
\end{figure}

Here we consider a different model of a two-level system where we aim at generating unitary transformations, or gates, using optimal control. To this effect, we will study the Hamiltonian
\begin{equation}
    H(t) = \frac{\Omega}{2}\left( \cos(\alpha(t)) \sigma_x + \sin(\alpha(t)) \sigma_y\right).
    \label{eq:QOC_gate_hami}
\end{equation}
This model corresponds to a single spin-$1/2$ particle being driven by a field with constant amplitude $\Omega$ (i.e. Rabi frequency) but variable direction in the $x\!-\!y$ plane, determined by $\alpha(t)$. When compared to the Landau-Zener Hamiltonian, c.f. Eq. (\ref{eq:QOC_hami_LZ}), this model has two important features. First, the energy of the system is bounded by $\Omega/2$ irrespective of the control field. Thus, increasing the amplitude or frequency of the field will not necessarily make the system evolve faster in time. Second, it does not follow the decomposition of Eq.~(\ref{eq:QOC_hami}). In particular, one finds that $\partial H_j/\partial \alpha_j$ is now a function of $\alpha_j$ due to the nonlinear dependence on the control field. Nonetheless, this case is naturally contemplated in the general calculation of the gradient of Sec. \ref{subsec:QOC_grape}. 

In a unitary control problem, we need to define the target transformation. We consider two families of target gates,
\begin{align}
    U_x(\phi) &= \exp\left(-i \sigma_x \phi/2\right), \label{eq:QOC_gate_target_x}\\
    U_z(\phi) &= \exp\left(-i \sigma_z \phi/2\right). \label{eq:QOC_gate_target_z}
\end{align}
Note that for the targets $U_x(\phi)$, the problem has a straightforward solution, obtained by setting $\alpha(t)=0$ and evolving the system for a time $T=\phi/\Omega$. On the other hand, for the case $U_z(\phi)$ the control field needs to play a nontrivial role to generate the desired gate. 

We analyze this problem numerically by choosing the field $\alpha(t)$ to be piecewise-constant with $M=10$, as in the previous example. We start considering an evolution time of $T=2\pi/\Omega$ and two instances of the target gates $U_z(\phi)$ with $\phi=\pi/2$ and $\phi = \pi$. To illustrate the results of the search, we again consider two different types of initial guess: control parameters randomly sampled from a uniform distribution in $[-\pi,\pi]$, or increasing linearly in time from $-\pi$ at $t=0$ to $\pi$ at $t=T$. Results are shown in Fig. \ref{fig:QOC2}. Similarly to the results displayed in the previous section, we observe how the optimization routine is able to decrease the value of the cost by over 10 orders of magnitude and achieve very high fidelities. We also observe how different initializations of the search lead to different optimal fields, which inherit properties of the initial seed.\\

Finally, we can use our optimization procedure to explore the quantum speed limit to achieve the indicated target gates. We do this in exactly the same way as described for the state control problem in the previous section. In Fig.~\ref{fig:QOC2}(e) we show the optimized cost values as a function of evolution time $T$ for the gates $U_z(\phi)$ and various choices of $\phi$. We observe a clear separation between the controllable regime for $T>\tau_{\text{QSL}}^\ast$ where very low cost values are achievable, and the uncontrollable regime where $T<\tau_{\text{QSL}}^\ast$ and the optimal cost is highly constrained. The critical evolution time $T_*$ increases with $\phi$, which is intuitive as gates with larger $\phi$ are further away from the identity, i.e. the initial state for unitary control. A deeper analysis of this feature is presented in Fig.~\ref{fig:QOC2}(f), revealing a  square-root-like growth of $T_*$ with $\phi$. This is consistent with the actual minimum evolution time $\tau_{\text{QSL}}^{*(z)}$ which for this problem has been analytically obtained in Ref.~\cite{boozer2012}. For completeness, we include results corresponding to the family of gates $U_x(\phi)$ for which, as mentioned before, optimization was not really necessary. Nonetheless, we verify with numerical optimization that it is not possible to break the minimum evolution time given by $\tau_{\text{QSL}}^{*(x)} = \phi/\Omega$, which scales as $\phi$ instead of $\sqrt{\phi}$. Similarly to what happens with state control problems and the quantum speed limit, lower bounds to the minimum control times can be obtained via geometric arguments~\cite{poggi2019geometric}.

\subsubsection{Example: generation of many-body entanglement}

The examples shown in the Secs. \ref{subsec:QOC_example_state} and \ref{subsec:QOC_example_gate} showcase the application of QOC to the simplest model of a two-level system or qubit.  One obvious extension of these cases is the generation of two-qubit gates, which has been thoroughly studied for various platforms, including superconducting qubits, trapped ions, and neutral atoms. Nevertheless, the formalism described in Sec. \ref{subsec:QOC_grape} is general and can be directly applied to more complex scenarios.\\

Here we consider the problem of many-body state preparation, which is closely related to the control tasks treated in Sec. \ref{sec:RL_theory} with the STA formalism. While the direct application of QOC to many-body systems is naturally limited by the computational complexity of numerically simulating these systems, we will show that even for modest system sizes QOC can provide interesting insights into the resources needed to perform control on a high-dimensional Hilbert space. This, in turn, will allow us to address a key point raised in Sec. \ref{sec:Introduction}: the use of quantum control tools to bridge the gap between implementability and scalability of quantum operations. \\

We consider a model of $N$ qubits undergoing global control and all-to-all interactions according to the Hamiltonian

\begin{equation}
    H(t) = \frac{\Omega_x(t)}{2}\sum\limits_{i=1}^N \sigma_i^x + \frac{\Omega_y(t)}{2}\sum\limits_{i=1}^N \sigma_i^y + \frac{\beta}{4N}\sum\limits_{i,j=1}^N \sigma_i^z \sigma_j^z
    \label{eq:QOC_hami_many_body}
\end{equation}

Note that, as opposed to the single-qubit example, here the model has a drift term which is time-independent (thus, not directly controllable), which corresponds to the interaction between the qubits. The relevant time-scale of the problem is set by $T_\beta=2\pi/\beta$. We have also chosen to normalize the interaction term to ensure that the Hamiltonian remains extensive as $N$ increases. 

We consider the following control task. Starting from the state

\begin{equation}    \ket{\psi_0}=\ket{00\ldots 0},
\end{equation}
we want our system to reach an entangled state $\ket{D_k^{(N)}}$ characterized by the symmetric superposition of all possible $k$ bit-flips. For instance, if $N=4$ some of the target states we are interested in are 
\begin{align}
    \ket{D_1^{(N=4)}} &= \frac{1}{2}\left(\ket{1000}+\ket{0100}+\ket{0010}+\ket{0001}\right)\\
    \ket{D_2^{(N=4)}} &= \frac{1}{\sqrt{6}}\left(\ket{1100}+\ket{1010}+\ket{1001}+\ket{0110}+\ket{0101}+\ket{0011}\right)
\end{align}

These types of states display multipartite entanglement, and are regarded as a resource in quantum sensing \cite{toth2012} and cavity quantum optics \cite{pezze2018}, among others. \\

To tackle this problem, we first note that the Hamiltonian in Eq. (\ref{eq:QOC_hami_many_body}) is invariant under any permutation of the qubits. This allow us to write this operator elegantly in terms of collective spin operators
\begin{equation}
    J_\alpha = \frac{1}{2} \sum\limits_{i=1}^N \sigma_i^\alpha,
\end{equation}
leading to

\begin{equation}
    H(t) = \Omega_x(t) J_x + \Omega_y(t) J_y + \frac{\beta}{N} J_z^2.
\label{eq:QOC_hami_spinJ}
\end{equation}

In this form, it also becomes clear that at all times $[H(t),\mathbf{J}^2]=0$, where $\mathbf{J}^2=J_x^2+J_y^2+J_z^2$ is the total spin of the system. In addition, the set of initial and target states all correspond to joint eigenstates of $J_z$ and $\mathbf{J}^2$ (note that $\ket{\psi_0}=\ket{D_0^{(N)}}$). These are also called Dicke states, which obey \cite{dicke1954}
\begin{align}
    J_z \ket{D_k^{(N)}} &= \left(\frac{N}{2}-k\right)\ket{D_k^{(N)}} \\
    \mathbf{J}^2 \ket{D_k^{(N)}} &= \frac{N}{2}\left(\frac{N}{2}+1\right)\ket{D_k^{(N)}}.
\end{align}

Thus, our control problem takes place in a reduced subspace spanned by these states, which has a dimension $N+1$, as the value of excitations $k$ goes from 0 to $N$. This reduced subspace is actually exactly equivalent to the Hilbert space of a single spin-$J$ particle with $J=N/2$. In fact, the Hamiltonian in Eq. (\ref{eq:QOC_hami_spinJ}) accurately models the dynamics of electronic states in individual atoms, where the $J_z^2$ now corresponds to a non-linear light-shift induced by non-resonant optical driving \cite{deutsch2010}. These systems are now routinely being considered as qudits for quantum computing platforms, and the results shown here directly carry over to that application as well \cite{omanakuttan2021} (see also Sec. \ref{outlook_sec_qc} for further information).\\

With this model we can explore how control resources change as the complexity of the control task increases, both in terms of increasing the number of target excitations $k$ and the number of total particles $N$. To do this, we run the QOC procedure outlined in Sec. \ref{subsec:QOC_grape} to this problem for $N=6,8,10$ particles, and in each case to reach a target state with $k=1,2,..,N/2$ excitations. We note that the case $k>N/2$ turns out to be equivalent to the cases considered here, up to global rotations. \\

\begin{figure}[t]
\includegraphics[width=\columnwidth]{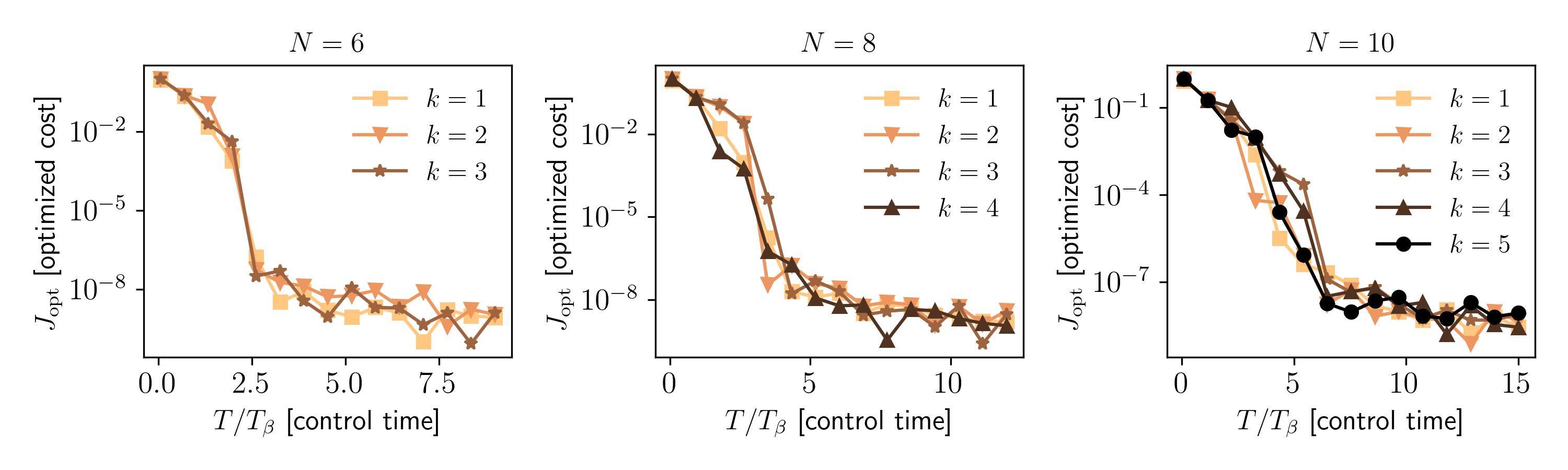}
\caption{Quantum optimal control for many-body entangled state generation. Panels show the optimized cost obtained from QOC on the protocol described in the main text, as a function of the allotted control time $T$. In all cases, both fields $\Omega_x(t)$ and $\Omega_y(t)$ are discretized with $M=15$ steps and constrained to maximum amplitude of $\Omega_{\mathrm{max}}=3\beta$.} 
\label{fig:QOC3}
\end{figure}

Similar to the analysis shown in, e.g., Fig. \ref{fig:QOC1} (e), we run the optimization for various choices of the total evolution time, in order to identify an approximate quantum speed limit required to produce these various transformations. Results are shown in Fig. \ref{fig:QOC3}. For each value of $N$ considered, we find that the time required for the control to find high-fidelity solutions does not depend significantly on the number of excitations $k$ in the target state. This indicates that even when states with higher $k$ require more bit-flips, the optimal solution is able to efficiently navigate Hilbert space and takes a shortcut to generate the target state directly. In turn, by comparing the curves for different values of $N$, we do observe that the total amount of time increases with system size in a seemingly linear fashion. This is expected given the linear increase of accessible Hilbert space. We note that this is also a consequence of the choice of normalization in Eq. (\ref{eq:QOC_hami_many_body}).

\subsection{Experimental implementations of quantum optimal control}
Quantum optimal control tools have been successfully applied to various experimental platforms involving manipulation of quantum systems. Interestingly, the flexibility of QOC has led each community to adapt these tools to tackle and mitigate errors and limitations which are native to each implementation. Following the discussion in Sec.~\ref{subsec:QOC_intro}, this translates into imposing different constraints to the base QOC problem. Examples of this include leaking reduction in superconducting qubits~\cite{motzoi2009} and inhomogeneous broadening mitigation in atomic and nuclear magnetic resonance (NMR) systems~\cite{ruths2012}.

Here we name some examples of successful implementations of QOC in different platforms, in a manifestly non-exhaustive way; the reader is referred to Reviews such as Refs.~\cite{Glaser2015,Koch2022} for further examples. In atomic systems, optimal control has been used to coherently control large manifolds of electronic states in individual neutral atoms. This includes both Zeeman \cite{smith2013,anderson2015,lysne2020} and Stark states \cite{larrouy2020}. In these cases both radiofrequency and microwave fields are modulated in time to achieve the desired control. Demonstrated tasks include the generation of circular Rydberg and nonclassical superposition states, as well the implementation of arbitrary unitary operations within the corresponding electronic state manifold.

Likewise, QOC has become a standard tool to design entangling gates between neutral atoms for quantum computing. In this case, time-modulation of laser intensity and detuning can be optimized to produce operations which are highly robust to imperfections like light shift broadening or spontaneous decay. Recent demonstrations which leverage QOC tools include Refs.~\cite{omran2019,evered2023,cao2024}. A similar approach has been applied to trapped-ion quantum processors, with entangling gates enabled by optimal design of frequency comb structure~\cite{choi2014}. These systems have also benefited from QOC in the task of transporting individual ions along a trap~\cite{sterk2022}.

Superconducting circuits have also successfully implemented optimal control tools for various tasks. In these systems, control is exerted by microwave signals; recently demonstrated tasks include the implementation of a universal gate set on a logical cat qubit~\cite{heeres2017} and high-fidelity qudit gates~\cite{seifert2023}. Since superconducting qubits are two-levels embedded in a larger Hilbert space, it is crucial to minimize population leaking outside of the relevant computational subspace. The DRAG (derivative removal by adiabatic gate) method is a QOC approach that has been proposed to address this issue~\cite{motzoi2009}, and which has been experimentally demonstrated in Ref.~\cite{werninghaus2021}. Following its demonstrated success and flexibility, quantum optimal control is now becoming an industrial standard for quantum computing, with many quantum hardware and software providers utilizing quantum control tools in various layers of the computing stack for tasks like circuit compilation, gate synthesis, state preparation and readout, and error mitigation.

Finally, we briefly mention that quantum optimal control tools have a long history beyond the quantum information processing platforms mentioned here. In particular, optimization is a standard tool in composite pulse design in NMR experiments~\cite{bonnard2012}. Also, the control of chemical reactions using shaped laser pulses was one of the first applications of quantum optimal control theory, see for instance Ref.~\cite{tannor1985}.

\clearpage


\section{\label{sec:RL_theory}Reinforcement learning for optimal quantum control}


In the previous sections, we introduced the concepts of counterdiabatic driving and optimal control. Here we take on a different approach to quantum control using reinforcement learning (RL). We also discuss how ideas from CD driving can be used in combination with RL to control quantum systems, leading to powerful hybrid approaches. While the previous sections would be useful to grasp the overarching theme, the material below can be considered a self-consistent standalone tutorial on reinforcement learning for quantum control. 
\steve{
We begin with a short overview of reinforcement learning, defining the basic notions and concepts, and elaborating on the details behind a simple RL algorithm, cf.~Sec.~\ref{subsec:OC_via_RL}. We then move on to discuss three examples that illustrate some of the features of RL algorithms applied to quantum control: in particular, in Sec.~\ref{sec:RL_1q}, we discuss how to train an RL agent to learn how to prepare a fixed target starting from an arbitrary initial state. Section~\ref{sec:RL_2q} combines ideas from shortcuts to adiabaticity with RL that give rise to hybrid RL/OC algorithms. Finally, in Sec.~\ref{subsec:RL_qubit_ancilla} we discuss how to extract partial information about the state of a controlled qubit with the help of an ancilla to train an RL agent using feedback control.
} 

\begin{figure}[h!]
\includegraphics[width=0.4\columnwidth]{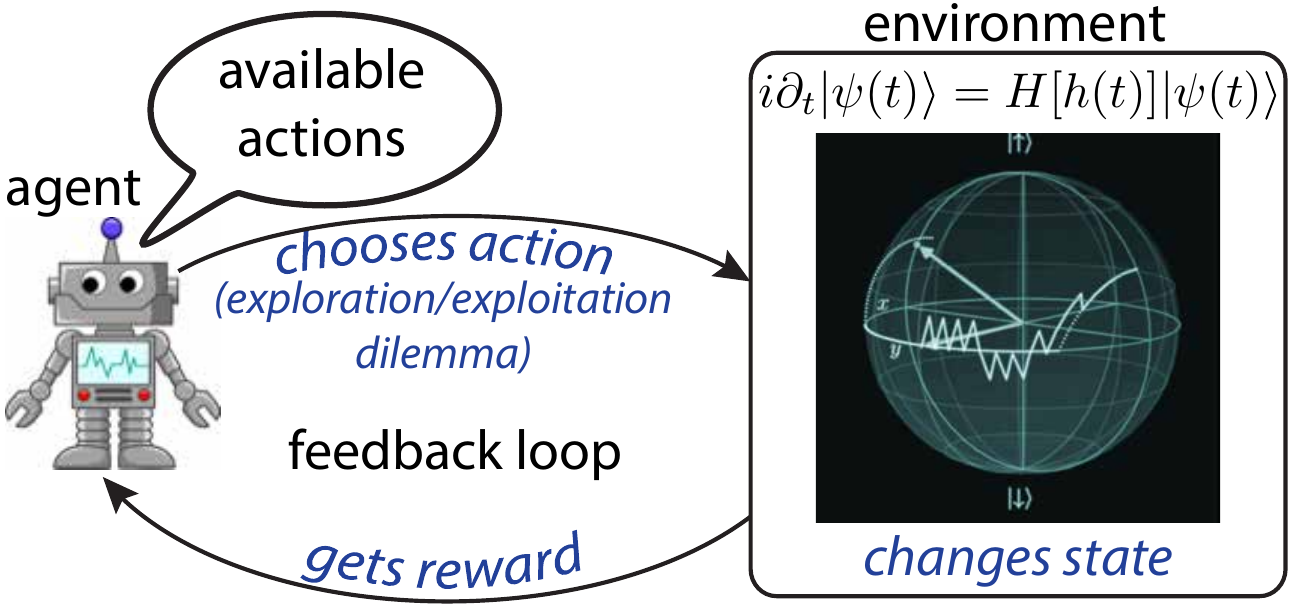}
\caption{
RL in a nutshell: the agent observes its environment in a given state, and is presented with a choice of actions to pick from, each of which changes the state of the environment. As a consequence of its decision, the agent is given a reward, depending on the new environment state. Similar to animals being rewarded for achieving a task during training, the goal in RL is to maximize the long-term expected return. This is accomplished through learning about those properties of the environment that are particularly relevant for maximizing the cumulative reward. During the training process, RL agents autonomously figure out an optimal policy to perform the task.
}
\label{fig:RL-scheme}
\end{figure}

Modern machine learning (ML) stands on three pillars~\cite{mehta2019high,carleo2019machine,carrasquilla2020machine,krenn2023artificial}:
(i) \textit{Supervised learning} deals with labeled data, and aims at finding the most probable label for new, unseen datapoints. Notable examples include image recognition or classification tasks based on linear and logistic regression. (ii) The goal of \textit{unsupervised learning}, on the other hand, is to produce an approximate model for the probability distribution that generates a given dataset; once such a model is trained, we can use it to produce more data with the desired properties. This includes, for instance, generative models that can compose music, draw paintings, or write text. (iii) Last, in \textit{reinforcement learning} (RL)~\cite{sutton_barto_rl}, an agent learns to execute a given task by interacting with its environment. 
Note that in RL the term `environment' refers to the physical system the agent interacts with [see Fig.~\ref{fig:RL-scheme}]; this is in contrast to the notion of environment in open systems.
RL is designed for problems that require an interactive feedback loop. The main objective is to learn a strategy (a.k.a.~policy) that, based on observations of the environment, helps the agent manipulate or steer the environment. Reinforcement learning and optimal control share many common features and can be thought of as the two sides of the same coin~\cite{todorov2006optimal}. Therefore, when leveraging ML for quantum control~\cite{ref5, ref6}, RL appears as a natural choice for many (though certainly not all) quantum control setups.

\begin{table}[!h]
\centering
\begin{tabular}{|p{12.5cm}|c|}
\hline
Object & Proposed Symbol \\
\hline\hline
states and state space & $s\in\mathbf{S}$ \\
\hline
actions and action space & $a\in\mathbf{A}$ \\
\hline
rewards and reward space & $r\in\mathbf{R}$ \\
\hline
trajectory & $\tau$ \\
\hline
expected return & $\mathbf{G}(\tau)$ \\
\hline
episode length (number of time steps) & $T$ \\
\hline
transition probability & $p(s'|s,a)$ \\
\hline
policy (strategy) & $\pi(a|s)$ \\
\hline
Q function & $Q_\pi(s,a)$ \\
\hline
probability for a trajectory & $P_\pi(\tau)$ \\
\hline
RL objective & $\mathbf{J}$ \\
\hline
variational (learning) parameters & $\theta$ \\
\hline
Monte Carlo samples & $N$ \\
\hline

\hline
\end{tabular}
\caption{Summary of notation for the most common quantities used in RL.}
\label{table:RLSymbols}
\end{table}

In RL, the environment is initialized in a certain state $s_0$, and the agent is presented with a set of actions $\mathbf{A}$ to choose from. Depending on the action $a_t$ selected, the environment reacts and changes state: $s_t\to s_{t+1}$. After each such step $t$, the agent observes the new state of the environment, and is given a reward $r_{t+1}$ that quantifies the quality of the action taken w.r.t.~the objective of the task. Based on this information, the agent then makes an educated guess to select the next action, and so forth, cf.~Fig.~\ref{fig:RL-scheme}. The goal of the agent is to bring the environment into a desired state. To do so, its objective is to maximize the cumulative expected reward $\mathbf{G}$, called return. The expectation is taken over the policy (i.e., the strategy) of the agent and the environment dynamics if the latter is stochastic. Such optimization in expectation brings out as a clear advantage of RL the ability to learn in uncertain or stochastic environments. The learning process in RL is often divided into episodes, which typically comprise a fixed number of steps; once the episode comes to an end, the environment is reset to its initial state $s_0$ and the agent starts over again. Importantly, however, the agent keeps building on the experience gained in previous episodes. 

To select an action, the agent uses a probability distribution $\pi$, called a policy (or strategy). Given a state of the environment, the policy $\pi(a|s)$ defines the probability of taking each of the available actions. If the policy is a delta-function over the actions, we call it deterministic; such a case corresponds to selecting specific actions with certainty, e.g., when applying a fixed external control protocol to the environment. One issue with deterministic policies is that they always produce the same output, and hence the RL agent cannot cause the environment to behave differently, to learn from. To explore new actions (and ultimately determine which one is optimal for the given state), the policy in RL is typically non-deterministic: for instance, if the action space is discrete, one can think of the policy as a normalized state-dependent histogram (distribution) over the available actions; for continuous action spaces, the policy is a continuous probability density, such as a Gaussian or a Lorentzian. However, if the agent explores too much (i.e., its policy is close to a uniform distribution), it cannot learn a better policy either. This tension between exploring new actions and exploiting the already gathered knowledge is known as the `exploration-exploitation dilemma'. 

RL algorithms iteratively improve the policy until it becomes (close to) optimal. Optimal policies, by definition, maximize the total expected return $\mathbf{G}$. In an alternative formulation of the learning procedure, we can look at the state of the environment, and try to predict what the expected return from this state onward (following a fixed policy) should be. This leads us to the concept of action-value (or Q-) functions $Q_\pi(s,a)$; they assign an expected return value to each state-action pair. Once the Q-function is known, one can extract a policy from it by, e.g., looking up the values of different actions for a fixed state, and taking the action that maximizes the action-value function. 

\begin{figure}[h!]
\includegraphics[width=0.3\columnwidth]{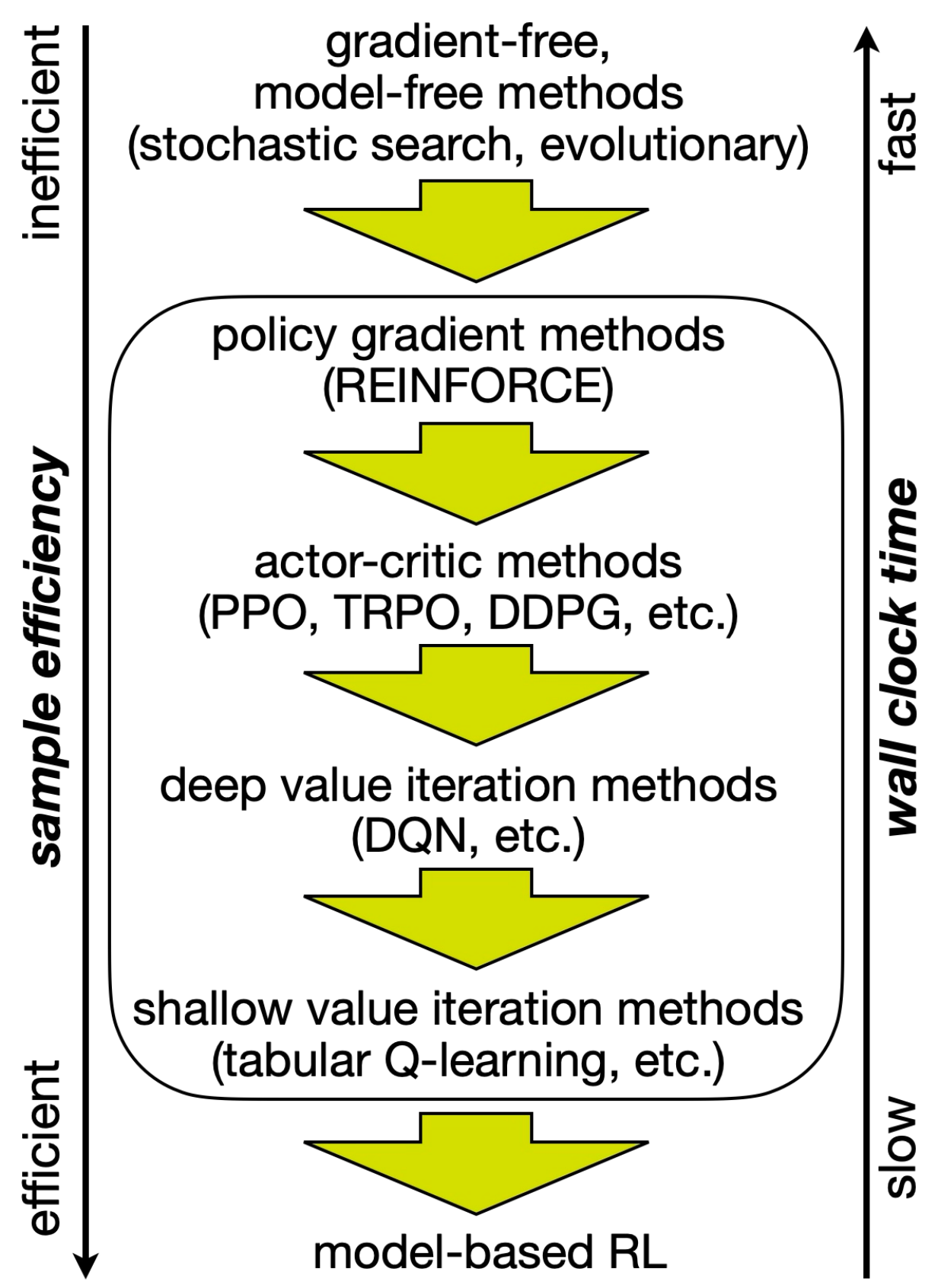}
\caption{
Heuristic schematic of sample efficiency and runtime of some RL algorithms.
}
\label{fig:RL_sample_eff}
\end{figure}

Reinforcement learning is not a single algorithm; instead, it should be viewed as a framework that comprises a collection of optimization algorithms, broadly divided into two classes: In policy gradient methods one approximates the policy $\pi\approx\pi_\theta$ with a set of variational parameters $\theta$. For instance, $\theta$ can be the weights and biases of a deep neural network, or $\theta=(\mu,\sigma^2)$ contain the mean and the variance of a Gaussian distribution in the case of continuous actions~\cite{sutton_barto_rl}, but other variational ans\"atze (e.g., a matrix product state or a tensor network more generally~\cite{metz2023self,rose2024combining}) might be better suited to the particular problem at hand. The parameters $\theta$ are then updated using the gradient of the total expected return, estimated, e.g., using Monte Carlo sampling. Examples of widely used algorithms include Policy Gradient (PG)~\cite{williams1992simple}, Proximal Policy Optimization (PPO)~\cite{schulman2017proximal}, Trust-Region Policy Optimization (TRPO)~\cite{schulman2015trust}, and Actor-Critic (AC) methods~\cite{konda1999actor}. Alternatively, Value-Function methods allow us to encode information about the policy; they aim at improving the $Q$-function itself which allows us to indirectly improve the policy~\cite{sutton_barto_rl}. Such methods include, among others, Q-learning~\cite{watkins1992q-learning} and its deep-learning version -- DQN~\cite{mnih2015human}, Deep Deterministic Policy Gradient (DDPG)~\cite{lillicrap2015continuous}, etc. Policy gradient methods are typically less sample-efficient, since updating the policy renders the old data obsolete; thus, new data have to be collected more frequently. Value-function methods, on the other hand, are often off-policy: this feature allows them to learn from old data and makes them more sample efficient. 
Sample efficiency determines the suitability of RL algorithms for quantum control. This is particularly relevant for experimental applications. Depending on the details, collecting rewards from the environment can bring a substantial overhead. In a simulator, if it is easier/faster to generate data than training the agent, Policy Gradient and Actor-Critic algorithms are typically a good starting point. In potential applications to experiments, Deep Q-Learning may be a good choice since it is off-policy. The relative sample efficiency is shown in Fig.~\ref{fig:RL_sample_eff}; the runtime (wallclock time) usually goes in the opposite direction. Deep value iteration methods align with the objective of developing algorithms for both theory and experiment.
Ultimately, which algorithm to use depends on the concrete problem and setup at hand: in some cases, it is easier to learn the policy directly, whereas in others -- the policy may be too complex, but the value function may be easier to approximate.

When it comes to applications of RL for the manipulation of quantum systems, we can broadly identify four mutually overlapping areas: quantum state manipulation, or preparation,~\cite{bukov2018reinforcement,yao2021reinforcement,sivak2022model-free,Porotti2022,metz2023self,reuer2023realizing, XiChenRL1, XiChenRL2, XiChenRL3}, quantum metrology~\cite{MarcoRL1}, quantum gate design~\cite{niu2019universal,dalgaard2020global,baum2021experimental,moro2021quantum,nguyen2023reinforcement}, quantum circuit design~\cite{foesel2021quantum,bolens2021reinforcement,he2021variational,herreramarti2022policy,patel2024curriculum}, and quantum error correction~\cite{foesel2018reinforcement, sweke2020reinforcement, andreasson2019quantum, fitzek2020deep,sivak2023real}. 
Out of these, by far the most common set of problems relates to quantum state manipulation. As exemplified in Sec.~\ref{subsec:OC_via_RL} below, the idea here is to use an RL agent to learn a protocol that transfers the population from an initial to a target state. This has been done extensively for both few-particle and many-body systems, in the context of bosonic, fermionic, and spin systems; RL agents have been shown to perform on par with, or better than, some optimal control algorithms in rugged optimization landscapes. An important concern when applying RL to quantum systems is the lack of accessibility to the exact quantum state (quantum states are unmeasurable mathematical constructs). While this is not an issue when training in simulators, it quickly poses a formidable challenge when applications to real experiments come into sight. For this reason, when applying RL to quantum systems one typically defines the RL state as a set of accessible observables, or a sequence of previously taken actions [cf.~Sec.~\ref{subsec:RL_qubit_ancilla}]. Whereas in many cases RL algorithms work using such partially observable environments, extra care needs to be taken to make sure the learning problem remains well-posed. In turn, the reward function in quantum control problems ranges from maximizing a fidelity (or the negative log-fidelity, in the case of many-body states) through minimizing the energy (as is common when targeting ground states), to optimizing the expectation value of a specific observable; RL agents can also be used to manipulate quantum entanglement~\cite{tashev2024reinforcement}. 



\subsection{\label{subsec:OC_via_RL}Optimal control via policy gradients}

In RL, an agent is confronted with the problem of learning to execute a task by interacting with its environment. The environment is described by a state $s$. The agent observes the state of the environment, and is given a choice of actions $a$ to select from, that alter the state of the environment. The environment can react deterministically or stochastically to these actions, following a set of prescribed rules, the so-called transition probabilities $p(s'|s,a)$, that contain the laws governing the physics of the system. The agent is then given a reward $r(s,a)$, which determines how well it performs on the task. Based on this information, the agent then selects a new action, in an attempt to maximize its total expected return $\mathbf{G}$.
The RL agent selects actions based on its policy $\pi(a|s)$, which defines the probability of choosing action $a$ from state $s$. This is repeated for a fixed number of time steps, $T$, that define a learning episode. The sequence of state-action-reward triples $(s,a,r)$ concatenated over the episode, is called a trajectory $\tau$.

In the following, we will take a look at some simple-to-implement reinforcement learning (RL) algorithms. Let us start by deriving REINFORCE -- the simplest policy gradient algorithm. The reinforcement learning objective $\mathbf{J}$ is to maximize the expected total return, starting from an initial state and taking actions according to the policy $\pi(a|s)$ thereafter. Let us denote the probability for the environment to transition from state $s$ to state $s'$ upon taking action $a$ by $p(s'|s,a)$; for an episode of $T$ steps, denoting the initial state distribution by $p(s_0)$, the probability for the trajectory $\tau = (s_0,a_0,r_1,s_1,a_1,\dots,s_{T-1},a_{T-1},r_T,s_T)$ to occur is given by
\begin{equation}
\label{eq:prob_trajs}
    P_\pi(\tau) = p(s_0)\prod_{t=1}^T \pi(a_t|s_t)p(s_{t+1}|s_t,a_t). 
\end{equation}
The RL objective can be defined formally as
\begin{equation}
\label{eq:RL_obj}
    \mathbf J = \mathbb{E}_{\tau\sim P_\pi} \left[ \mathbf G(\tau) | s_{t=0}=s_0 \right],\qquad \mathbf G(\tau)=\sum_{t=1}^T r(s_t,a_t),
\end{equation}
where $\mathbf G$ is the total return for a fixed trajectory $\tau$. The notation $\mathbb{E}_{\tau\sim P_\pi}$ means that we draw a trajectory $\tau$ according to the probability $P_\pi$; in practice, we do this by sampling the initial state distribution $p(s_0)$ and then the policy $\pi(a|s)$ for each episode step at a time (this is similar to generating runs when playing a video game). The expectation value of the return $\mathbf G(\tau)$ over $P_\pi$ requires a sum over all possible trajectories $\tau$, each weighted by its probability to occur $P_\pi(\tau)$. While this sum can formally be written down, it often contains either exponentially or even infinitely many trajectories. This is similar to the concept of partition functions of Feynman path integrals; we can thus borrow the notation:
\begin{equation}
    \mathbf J = \mathbb{E}_{\tau\sim P_\pi} \left[ \mathbf G(\tau) | s_{t=0}=s_0 \right]
    = \int\mathcal{D}\tau\; P_{\pi}(\tau)\; G(\tau),
\end{equation}
where $\mathcal{D}\tau$ is just a formal notation that allows us to do algebraic manipulations.
Notice that RL is designed to optimize in expectation, which allows it to handle non-differentiable or even discontinuous reward functions $r(s,a)$.  

Policy gradient methods in RL directly approximate the policy $\pi$ using a variational ansatz $\pi_\theta$, parametrized by the unknown parameters $\theta$, $\pi\approx\pi_\theta$. The goal is then to find those optimal parameters $\theta$, which optimize the RL objective $\mathbf J(\theta)$ (think of the agent \steve{that} needs to win the game by maximizing the score). To define an update rule for $\theta$, we may use gradient ascent (or any of its generalizations~\cite{mehta2019high}). This requires us to evaluate the gradient of the RL objective w.r.t.~the parameters $\theta$:
\begin{equation}
\label{eq:nabla_J}
\nabla_\theta \mathbf J(\theta) = \nabla_\theta \mathbb{E}_{\tau\sim P_\pi} \left[ \sum_{t=1}^T r(s_t,a_t) | s_{t=0}=s_0 \right] 
= \nabla_\theta \int\mathcal{D}\tau\; P_{\pi_\theta}(\tau)\; \mathbf G(\tau)
= \int\mathcal{D}\tau\; \nabla_\theta P_{\pi_\theta}(\tau)\; \mathbf G(\tau),
\end{equation}
where we brought the derivative inside the path integral; note that the return $\mathbf G(\tau)$ is independent of $\theta$ and hence it is not affected by the derivative. 

Computing the path integral in Eq.~\eqref{eq:nabla_J} exactly is out of reach because we cannot exhaust the space of all trajectories (similar to the difficulty with computing partition functions in statistical mechanics). In addition, in a model-free setting, we do not know the laws of physics governing the dynamics of the system; as a result, our agent does not have access to the transition probabilities $p(s'|s,a)$. Hence, computing the path integral exactly is not a viable approach. These issues require us to find alternative ways, for instance, by estimating the gradient $\nabla_\theta \mathbf J(\theta)$ from samples. This is feasible because our agent interacts with the environment, and hence we can sample trajectories from it. To this end, we need to write the expression for the gradient $\nabla_\theta \mathbf J(\theta)$ as an expectation value over $P_{\pi_\theta}(\tau)$.
This can be accomplished by using the logarithmic derivative rule $\nabla_\theta P_{\pi_\theta} = P_{\pi_\theta} \nabla_\theta \log P_{\pi_\theta}$ almost everywhere (i.e., up to a set of measure zero), in Eq.~\eqref{eq:nabla_J}:
\begin{equation}
\nabla_\theta \mathbf J(\theta) = \int\mathcal{D}\tau\; \nabla_\theta P_{\pi_\theta}(\tau) \mathbf G(\tau) = \int\mathcal{D}\tau\; P_{\pi_\theta}(\tau)\; \nabla_\theta \log P_{\pi_\theta}(\tau)\; \mathbf G(\tau) = \mathbb{E}_{\tau\sim P_\pi} \left[\nabla_\theta \log P_{\pi_\theta}(\tau)\; \mathbf G(\tau)\right].
\end{equation}

The next step is to evaluate $\nabla_\theta \log P_{\pi_\theta}(\tau)$. Since the initial state distribution and the transition probabilities are independent of $\theta$, using the definition of $P_{\pi_\theta}$ in Eq.~\eqref{eq:prob_trajs}, we have $\nabla_\theta \log P_{\pi_\theta}(\tau) = \nabla_\theta \log \pi_\theta(\tau)$ where $\pi_\theta(\tau) = \prod_{t=1}^T \pi_\theta(a_t|s_t)$. We can therefore use Monte Carlo sampling to estimate the gradients directly from a sample of trajectories $\{\tau_j\}_{j=1}^N$ using the policy $\pi_\theta(\tau)$ (together with the initial state and transition probabilities):
\begin{equation}
\label{eq:RL_grad_sampled}
    \nabla_\theta \mathbf J(\theta) = \mathbb{E}_{\tau\sim P_\pi} \left[\nabla_\theta \log \pi_\theta(\tau)\; \mathbf G(\tau)\right]
    \approx \frac{1}{N}\sum_{j=1}^N \nabla_\theta \log \pi_\theta(\tau_j)\; \mathbf G(\tau_j)
    = \frac{1}{N}\sum_{j=1}^N \left( \sum_{t=1}^T \nabla_\theta \log\pi_\theta(a^j_t|s^j_t) \sum_{t'=1}^T r(a^j_{t'},s^j_{t'}) \right).    
\end{equation}
\steve{Here, we first replace the expectation value over trajectories by its Monte Carlo approximation, $\mathbb{E}_{\tau\sim P_\pi}[\cdot]\approx N^{-1}\sum_{j=1}^N(\cdot)$, where each trajectory $\tau_j$ is drawn from the probability $P_\pi$. Then, in the last equality, we used $\log \pi_\theta(\tau) = \log \prod_{t=1}^T \pi_\theta(a_t|s_t) = \sum_{t=1}^T \log\pi_\theta(a^j_t|s^j_t)$, together with the definition of the total return $\mathbf G(\tau)$ from Eq.~\eqref{eq:RL_obj}.
}

Since, in practice, we always deal with a finite number $N$ of trajectories in our sample, the estimate of the gradient for some parameters $\theta_k$ may have a large variance. To alleviate this issue, we use a two-fold strategy: 
first, notice that the $t'$-sum in Eq.~\eqref{eq:RL_grad_sampled} runs over all time steps $t'=1,\dots,T$; however, only times $t'>t$ should contribute to the gradient since rewards obtained at earlier times $t'<t$ do not affect the action at step $t$. Keeping a smaller number of summands naturally reduces the variance of the estimate.
Second, we introduce a so-called baseline $b=N^{-1}\sum_j \sum_{t'=t}^T r(a^j_{t'}|s^j_{t'})$, defined as the average return over the sample. This so-called centering of the training data is a common trick used in ML.
The PG update then takes the form
\begin{equation}
    \nabla_\theta \mathbf J(\theta)
    \approx \frac{1}{N}\sum_{j=1}^N \sum_{t=1}^T \nabla_\theta \log \pi_\theta(a^j_t|s^j_t) \left[\sum_{t'=t}^T r(a^j_{t'},s^j_{t'}) - b\right].
\end{equation}
One can show that these tricks do not change the expectation value $\nabla_\theta \mathbf J(\theta) = \mathbb{E}_{\tau\sim P_\pi} \left[\nabla_\theta \log \pi_\theta(\tau)\; \mathbf G(\tau)\right]$. The corresponding gradient ascent update rule reads as
\begin{equation}
    \theta \leftarrow \theta + \alpha \nabla_\theta \mathbf J(\theta),
\end{equation}
for some step size (sometimes called a learning rate) $\alpha$.

\subsection{Examples}

Let us now apply the REINFORCE algorithm to the single-qubit quantum control problem. We will do this in a series of three examples of increasing complexity. 
In Example 1, we consider fully observable qubit control: while this is somewhat unrealistic (as we explain later on), it represents the simplest quantum environment, and we use it to gain intuition about how RL works in practice. 
Subsequently, in Example 2 we use RL in combination with insights from CD driving to design a simple algorithm for variational quantum control. Unlike CD driving, our agent aims at preparing the target only at the final time step. 
Finally, in Example 3 we consider a realistic quantum control problem; in contrast to Example 1, the environment here is stochastic and minimally observable.
\steve{Importantly, as will become clear below, the challenge of applying RL to quantum systems lies with the definition of the RL environment -- the choice of actions, states, and rewards -- rather than with the RL algorithm itself. We therefore place significant focus on discussing how to set up the environment, and refer the interested reader to the accompanying Jupyter notebooks for a practical implementation of the PG algorithm.}

\subsubsection{\label{sec:RL_1q}Example: Universal single-qubit state preparation}

Our goal in this example will be to train an RL agent to learn how to prepare the target state $\ket{\psi_\ast}=(1,0)^t$ starting from any single-qubit initial state $|\psi\rangle$ by using a set of predefined gates.

We begin by defining the qubit environment that models the action of gates on the state of the two-level system (2LS).
The state of a qubit $|\psi\rangle\in\mathbb{C}^2$ is modeled by a two-dimensional complex-valued vector with unit norm: $\langle\psi|\psi\rangle:=\sqrt{|\psi_1|^2+|\psi_2|^2}=1$. Every qubit state is uniquely described by two angles $\theta\in[0,\pi]$ and $\varphi\in[0,2\pi)$:
\begin{eqnarray}
\label{eq:blochsphere_state}
|\psi\rangle=
\begin{pmatrix}
\psi_1 \\ \psi_2
\end{pmatrix}=
\mathrm{e}^{i\alpha}
\begin{pmatrix}
\cos\frac{\theta}{2} \\
\mathrm{e}^{i\varphi}\sin\frac{\theta}{2}
\end{pmatrix}
\end{eqnarray}
The overall phase $\alpha$ of a single quantum state has no physical meaning.
Thus, any \steve{single} qubit state can be pictured as an arrow on the unit sphere (called the Bloch sphere) with spherical coordinates $(\theta,\varphi)$. 

To operate on qubits, we use quantum gates. Quantum gates are represented as unitary transformations $U\in \mathrm{U(2)}$, where $\mathrm{U(2)}$ is the unitary group in two dimensions. Gates act on qubit states by matrix multiplication transforming an input state $|\psi\rangle$ into an output state $|\psi'\rangle$: $|\psi'\rangle=U|\psi\rangle$. For this problem, we consider four gates and their inverses:
\begin{equation}
U_0=\mathds{1},\qquad 
U_x=\mathrm{exp}(-i\delta t \sigma^x/2),\qquad
U_y=\mathrm{exp}(-i\delta t \sigma^y/2),\qquad 
U_z=\mathrm{exp}(-i\delta t \sigma^z/2),
\end{equation}
where $\delta t$ is a fixed time step, $\mathrm{exp}(\cdot)$ is the matrix exponential, 
$\mathds{1}$ is the identity, and we remind that the Pauli matrices are defined
\begin{equation}
\mathds{1}=\begin{pmatrix}
1 & 0 \\ 0 & 1
\end{pmatrix}
,\qquad
\sigma^x=\begin{pmatrix}
0 & 1 \\ 1 & 0
\end{pmatrix}
,\qquad
\sigma^y=\begin{pmatrix}
0 & -i \\ i & 0
\end{pmatrix}
,\ \qquad
\sigma^z=\begin{pmatrix}
1 & 0 \\ 0 & -1
\end{pmatrix}.
\end{equation}
We also allow our RL agent to use the inverse operations for $U_x^\dagger, U_y^\dagger, U_z^\dagger$. 

To determine if a qubit, described by the state $|\psi\rangle$, is in a desired target state $|\psi_\ast\rangle$, we compute the fidelity

\begin{eqnarray}
\mathcal F=|\langle\psi_\ast|\psi\rangle|^2\in[0,1].
\end{eqnarray} 
Physically \steve{for a single qubit}, the fidelity corresponds to the angle between the arrows representing the state on the Bloch sphere (we want to maximize the fidelity but minimize the angle between the states).

Now, let us define an episodic RL environment, which contains the laws of physics that govern the dynamics of the qubit (i.e., the application of the gate operations to the qubit state). \steve{We recall that the environment in RL is different from the notion of environment in open systems.} Our RL agent will interact with this environment \steve{by taking actions and observing its states} to learn how to control the qubit to bring it from an initial state to a prescribed target state. To this end, we define the RL states $s=(\theta,\varphi)$ as an array containing the Bloch sphere angles of the quantum state. For each step within an episode, the agent can choose to apply one out of the actions, corresponding to the seven gates $(\mathds{1},U_x,U_y,U_z,U_x^\dagger,U_y^\dagger,U_z^\dagger)$. We use the instantaneous fidelity w.r.t.~the target state as a reward: $r_t=\mathcal F=|\langle\psi_\ast|\psi(t)\rangle|^2$. More formally, we have to define: 
\begin{itemize}

    \item \textbf{State space:} $\mathbf{S} = \{(\theta,\varphi)\;|\;\theta\in[0,\pi],\varphi\in[0,2\pi)\}$. There are no well-defined states to terminate episodes in this task. Instead, we consider a fixed number of time steps $T$, after which the episode terminates deterministically. The target state (i.e., the qubit state we want to prepare) is $|\psi_\ast\rangle=(1,0)^t$: it has the Bloch sphere coordinates $s_\ast=(0,0)$. 

    \item \textbf{Action space:} $\mathbf{A} = \{\boldsymbol{1},U_x,U_y,U_z,U_x^\dagger,U_y^\dagger,U_z^\dagger\}$. Actions \steve{change the state of the controlled system; they} act on RL states as follows: 
    \begin{enumerate}
        \item if the current state is $s=(\theta,\varphi)$, we first create the quantum state $|\psi(s)\rangle$; 
        \item we apply the gate $U_a$ corresponding to action $a$ to the quantum state, and obtain the new quantum state $|\psi(s')\rangle = U_a|\psi(s)\rangle$. 
        \item last, we compute the Bloch sphere coordinates which define the next state $s'=(\theta',\varphi')$, using the Bloch sphere parametrization for qubits given in Eq.~\eqref{eq:blochsphere_state}.
    Note that all actions are allowed from every state. 
    \end{enumerate}

    \item \textbf{Reward space:} $\mathbf{R}=[0,1]$. We use the fidelity between the next state $s'$ and the target state $s_\ast$ as a reward at every episode step: 
    \begin{equation}
        r(s,s',a)= \mathcal F = |\langle\psi_\ast|U_a|\psi(s)\rangle|^2=|\langle\psi_\ast|\psi(s')\rangle|^2
    \end{equation}
    for all states $s,s'\in\mathcal{S}$ and actions $a\in\mathbf{A}$. \steve{This choice is motivated by the fact that the fidelity of the instantaneous state with the target state is a measure of the distance between the states. Alternatively, one can use other reward functions, such as the ground state expectation of the parent Hamiltonian for the target state.}
\end{itemize}

\begin{figure}[t!]
\includegraphics[width=0.45\columnwidth]{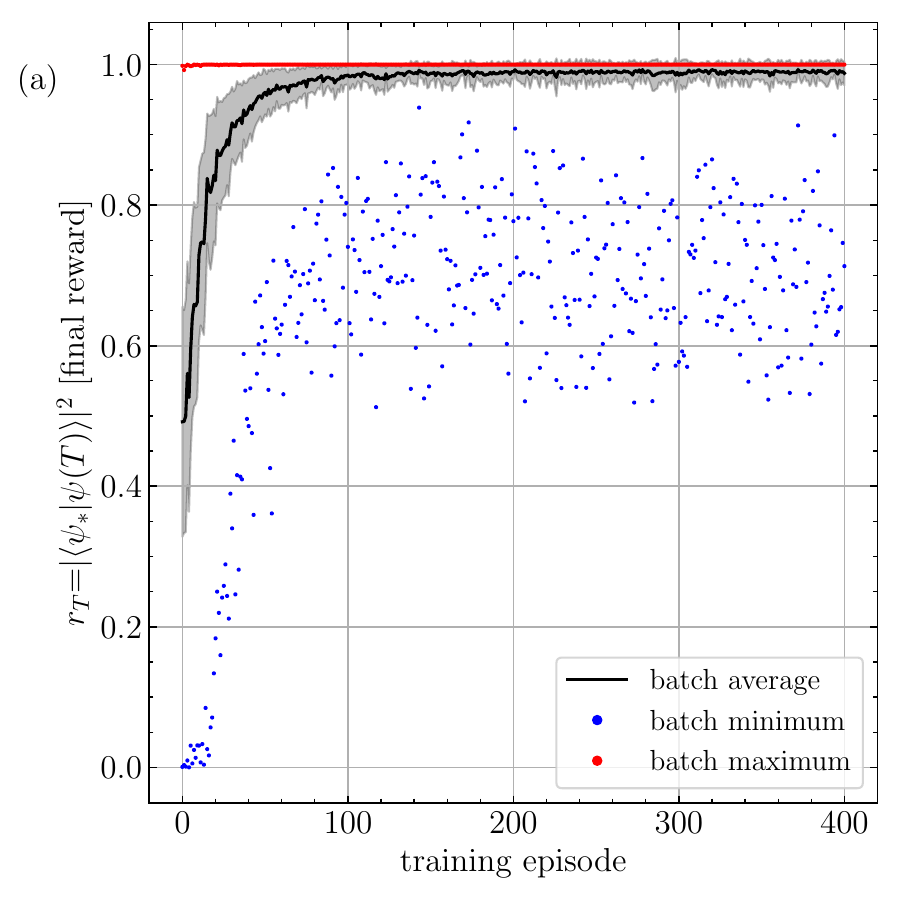}
\includegraphics[width=0.45\columnwidth]{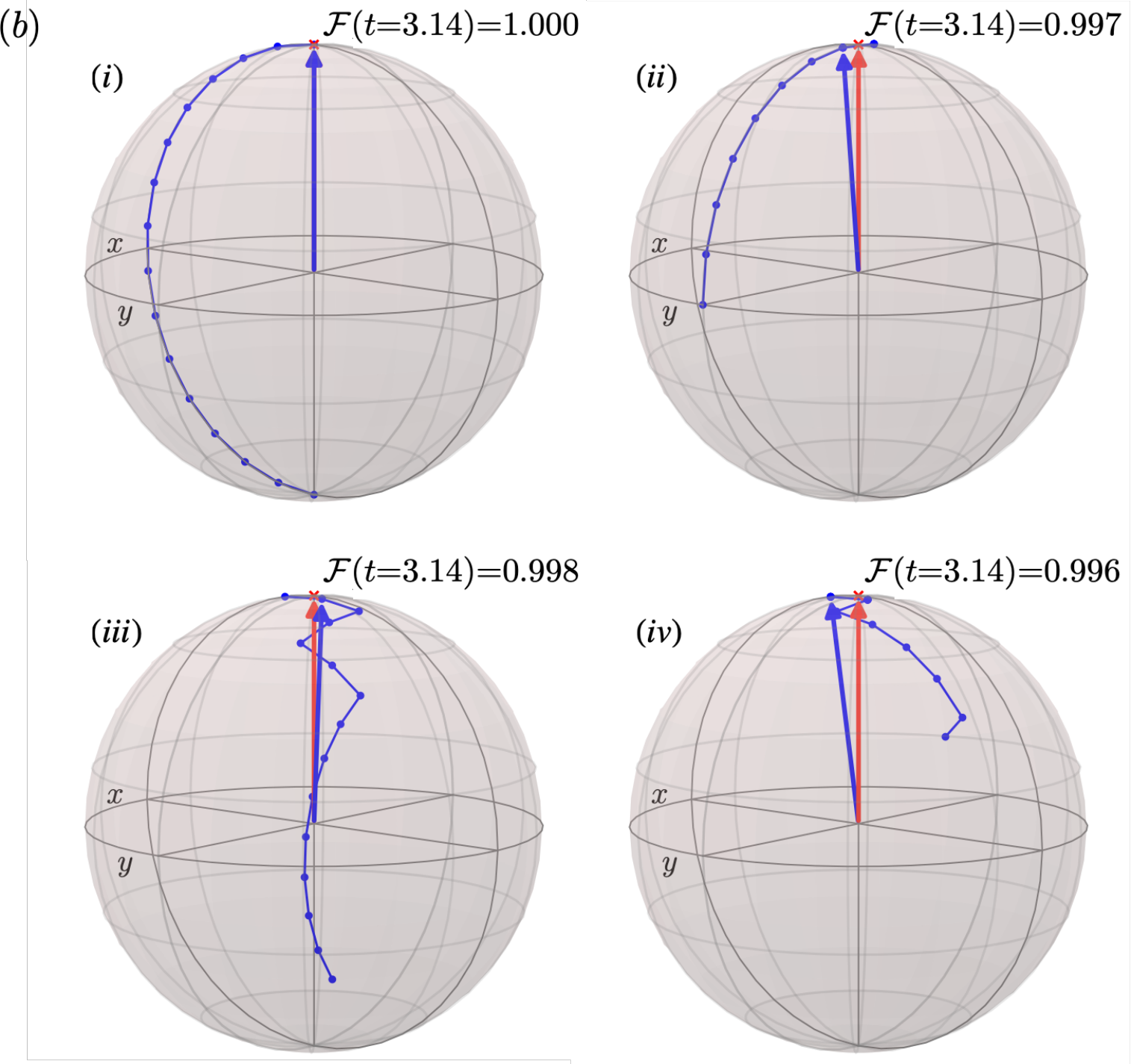}
\caption{
Universal single-qubit control using reinforcement learning:
\textbf{(a)} training curves for an RL agent learning to prepare the target $|\uparrow\rangle$ states (north pole of Bloch sphere) starting from any initial state. The final reward at the end of the protocol is shown, averaged over a batch of $N=256$ random initial states (black), as well as the minimum/worst-case (blue dashed) and the maximum (red dashed) values within the batch. 
\textbf{(b)} (i-iv) trajectories learned by the RL agent for four different initial states; the target (final) state is shown in red (blue). The agent achieves a fidelity $1-\mathcal{F}$ within the $1\%$ error range. 
We used a fully connected neural network with $(512, 256,7)$ neurons in each layer to learn the optimal strategy using the Policy Gradient algorithm.
The time step is $\delta t = T/\pi$ for a total of $T=15$ time steps.}
\label{fig:RL_qubit}
\end{figure}

Figure~\ref{fig:RL_qubit}(a) shows the result of the training procedure. We use a uniform initial state distribution $p(s_0)$ \steve{since this will expose the agent to a sample of the entire single-qubit Hilbert space; in turn, this will help it learn how to steer any initial state into the target}. To train the agent, we employ the REINFORCE algorithm with a batch of $N=256$ trajectories; \steve{we invite the interested reader to explore the accompanying Jupyter notebook. To this end, the agent runs a large number of episodes during which it explores the control protocol space by taking actions}. The reward at the end of the episode $r_T=|\langle\psi_\ast|\psi(T)\rangle|^2$ tells us how close the final state is to the target. Monitoring this quantity as a function of the training episodes (i.e., iterations of the algorithm) shows a steady increase of the batch average (black solid line) to a value close to unity. For comparison, we also show the best (red) and worst (blue) trajectories from the batch. The spread in the worst trajectories, as training progresses, comes from exploration: the RL agent takes actions probabilistically, and hence occasionally, suboptimal actions can be taken. 

To test the performance of the trained agent, we consider a greedy policy that takes the action with the highest probability deterministically. Figure~\ref{fig:RL_qubit}(b) shows the Bloch sphere trajectories of qubits controlled by the trained RL agent, starting from four different initial states (the time discretization is marked by the blue dots on the trajectory); the target state is shown in as a red arrow, and the final state in blue. \steve{Importantly, we emphasize that there is no further optimization required: once the training process has completed successfully, the same agent knows how to produce the optimal protocol starting from any initial state. This is a prime example of the learning (or generalization) capabilities of RL agents, which contrasts RL from optimal control.} In all cases, the agent manages to bring the state close to the target; the small leftover deviation is caused by the finite timestep $\delta t$. 
Supplementary movies show the trajectories in time.
We invite the interested reader to check out the accompanying Jupyter notebook and explore the behavior of the agent further~\cite{github_code}.

\steve{The example above serves as a stepping stone to learn how RL works in practice. In principle, all one has to do to control more complex quantum systems is redefine the environment, and the same implementation of the RL algorithm can be re-used. However, for quantum many-body systems, this is a tricky task: for instance, there is no convenient Bloch sphere representation for the quantum states. In this case, one can resort to using the bare wavefunction amplitudes, or other exact representations of the state (e.g., different momentum modes in cases, such as the transverse-field Ising model, where the problem can be mapped to a free system, or Clifford tableaus for Clifford circuit dynamics, etc.). One may also resort to suitable approximations, e.g., matrix product states~\cite{metz2023self} whenever appropriate.
We should also point out that the fidelity is typically exponentially small (in the number of qubits) for many-body states; in this case, one typically resorts to minimizing the energy of a parent Hamiltonian (i.e., a Hamiltonian whose ground state is the target state)~\cite{yao2021reinforcement}. It is important to keep in mind that, while RL provides a way to finding optimal protocols for quantum systems, it does not resolve any of the problems of many-body physics associated with representing quantum states or extracting information from them.  
}

\subsubsection{\label{sec:RL_2q}Example: RL vs.~counter-diabatic driving in the presence of trotterization errors}

We now build on the example above and apply RL to the problem of preparing the ground state of a target single-qubit Hamiltonian. We adopt a hybrid approach where the structure of the control terms in the Hamiltonian are designed using knowledge from CD driving (Sec.~\ref{sec:CD}).
In particular, we compare and contrast the behavior of the RL agent to counterdiabatic (CD) driving. Before we dive into the RL implementation, let us briefly mention that CD driving can be enhanced using other ML techniques, see, e.g., Ref.~\cite{ferrer2023physics}. 

As we have seen in Sec.~\ref{sec:STA}, CD driving introduces additional control terms to suppress excitations during the protocol: 
\begin{equation}
\label{eq:RL_HCD}
    H_\text{CD}(t) = \Delta\sigma^x + \nu(t)\sigma^z - \frac{1}{2}\frac{\Delta\; \dot\nu(t)}{\Delta^2 + \nu^2(t)}\sigma^y.
\end{equation}
As a result, evolving an eigenstate of the initial Hamiltonian $H(0)=\Delta\sigma^x+\nu(0)\sigma^z$ in time with $H_\text{CD}(t)$, the system follows the instantaneous eigenstates of the Hamiltonian $H(t)$ at all times. 

We now initialize the system in the ground state of the Hamiltonian $H(0)$ at $\nu(0)=\nu_i=+2\Delta$, and want to transfer the population in the ground state at $H(t_\ast)$ at $\nu(t_\ast)=\nu_\ast=-2\Delta$ in a finite amount of time $T$ (referred to as the duration of the protocol). During the evolution, the system is subject to the Hamitlnoan
\begin{equation}
\label{eq:Hg}
    H_g(t) = \Delta\sigma^x + \nu(t)\sigma^z - g(t)\sigma^y,
\end{equation}
where $\nu(t) = (\nu_\ast-\nu_i)t/T + \nu_i$ is a linear drift term that is always kept on (and the RL agent has no control over it), while $g(t)$ is the unknown counter protocol that we want our RL agent to find. The form of the Hamiltonian $H_g(t)$ is motivated by the CD Hamiltonian in Eq.~\eqref{eq:RL_HCD}; however, we do not make any assumptions on the structure of the drive protocol $g(t)$ whatsoever, and we leave it to the RL agent to find its shape.

The qubit evolves following $H_g(t)$, and its state at time $t$ is determined by the time evolution operator, which can be expressed in terms of the time-ordered exponential of the Hamiltonian $H_g(t)$ generating the dynamics:
\begin{equation}
    |\psi(t)\rangle =U(T,0)|\psi(0)\rangle = \mathcal{T}\exp\left(-i \int^t_0\mathrm ds\;  H_g(s)\right)|\psi(0)\rangle .
\end{equation}
Once again, we use the fidelity to determine if a qubit, described by the state $|\psi(t)\rangle$, is in a desired target state $|\psi_\ast\rangle$:
\begin{eqnarray}
    \mathcal F=|\langle\psi_\ast|\psi(t)\rangle|^2 ,\qquad \mathcal F\in[0,1]\; .
\end{eqnarray}

\steve{Now that we have defined the control problem, we discuss how to phrase it within the RL framework. Similar to the example in Sec.~\ref{sec:RL_1q}, we define an episodic RL environment, which contains the laws of physics that govern the dynamics of the qubit (i.e., the time evolution following the Hamiltonian $H_g(t)$). The environment will provide data for the RL agent to learn how to bring the qubit to the target ground state. 
To this end, we discretize the protocol into $N_T$ time steps of size $t$, so that the total duration is $T=N_T\delta t$. At each step, the RL agent constructs the value of the control protocol by selecting an action out of the available action set. The total number of timesteps in the protocol comprises a learning episode. Once an episode comes to an end, the environment is reset to its initial state, and the agent starts constructing the protocol again. In each subsequent episode, the agent uses the knowledge gained during previous protocols. 
}

Now, for each timestep of size $\delta t$ within the episode, the agent has to determine the optimal value of the protocol function $g(t)$; to this end, starting from some initial value $g_i$, we let the agent find the optimal relative change in the protocol (i.e., by how much the protocol value has to change at the given timestep). In practice, we define a minimum protocol size $\delta g$, and consider $2n+1$ steps of size $-n\delta g, -(n-1)\delta g, \dots, -\delta g, 0,\delta g,\dots, (n-1)\delta g, n\delta g$ that the agent has to choose from. We also set $g_i=\frac{1}{2}\frac{\Delta\; \dot\nu(0)}{\Delta^2 + \nu^2(0)}$, which fits our goal to directly compare the performance of the agent with CD driving, cf.~Eq.~\eqref{eq:Hg}.

\begin{itemize}
    \item \textbf{state space:} $\mathbf{S} = \{(\theta,\varphi)\;|\;\theta\in[0,\pi],\varphi\in[0,2\pi)\}$. There are no well-defined terminal states in this task. Instead, we consider a fixed number of time steps, after which the episode terminates deterministically. 

    \item \textbf{action space:} $\mathbf{A} = \{-n,-(n-1),\cdots,-1,0,1,\cdots,n-1,n\}$. Actions correspond to fixed discrete amounts by which the agent can change the protocol $g(t)$ (see above). Actions act on RL states as follows:
    \begin{enumerate}
        \item if the state at time step $n$ (corresponding to physical time is $t=n\delta t$) is $s=(\theta,\varphi)$, we first create the quantum state $|\psi(s)\rangle$; 

        \item we apply to the quantum state the unitary $U_a$ corresponding to action $a\in\mathbf{A}$, defined by:
            \begin{equation}
                 U_a = \exp(-i \delta t H_{a\times\delta g}(n\delta t)),\qquad 
                 H_{a\delta g}(n\delta t) = \Delta\sigma^x + \nu(n\delta t)\sigma^z - a\delta g\sigma^y,
            \end{equation}
        where $n$ labels the time steps, and $\delta g$ ($\delta t$) is a fixed minimum protocol (time) step size (parameters of our choice). 

        \item we obtain the new quantum state $|\psi(s')\rangle = U_a|\psi(s)\rangle$. 

        \item last, we compute the Bloch sphere coordinates which define the next state $s'=(\theta',\varphi')$, using the Bloch sphere parametrization for qubits given above.
        Note that all actions are allowed from every state. 
    \end{enumerate}

    \item \textbf{reward space:} $\mathbf{R}=[0,+\infty)$. 
    Since we want to incentivize the agent to prepare the target state at the end of the protocol, the reward is given only at the end of the episode. Note the difference to CD driving, where the system is in the instantaneous ground state at any given time during the protocol. In general, such reward functions are referred to as sparse rewards in RL. The return $\mathbf G=r_T$ is then equal to the final reward. 
    
    Thus, the reward is zero at each step during the protocol: $r_{t<T}=0$, except at the last step, where we use the fidelity between the final state $|\psi(T)\rangle$ and the target state $|\psi_\ast\rangle$ as a reward at the end of every protocol: 

    \begin{equation}
    r(s,s',a)= r_t=
    \left\{\begin{array}{lr}
        0, & \text{for}\qquad t<T\\
        -\log_{10}\left(1-\mathcal F\right), & \text{for}\quad t=T
        \end{array}\right.
    ,\qquad \mathcal F=|\langle\psi_\ast|\psi(T)\rangle|^2,\quad |\psi(T)\rangle = U(T,0)|\psi(0)\rangle,
    \end{equation}
    for all states $s,s'\in\mathbf{S}$ and actions $a\in\mathbf{A}$. 

    Since we are interested in high-fidelity protocols, it is important to build a high resolution into the reward signal. One way to do this is to use the logarithmic fidelity $1-\mathcal{F}$ w.r.t.~the target state at the end of the protocol as a reward. The logarithm allows us to learn on a logarithmic scale. Using a base-$10$ logarithm allows us to interpret the reward as the number of decimal places in the deviation of the fidelity from unity: e.g., $r_t=1$ corresponds to $\mathcal F=0.9$, while $r_t=2$ is equivalent to $\mathcal F=0.99$, etc.

\end{itemize}

\begin{figure}[t]
\includegraphics[width=0.33\columnwidth]{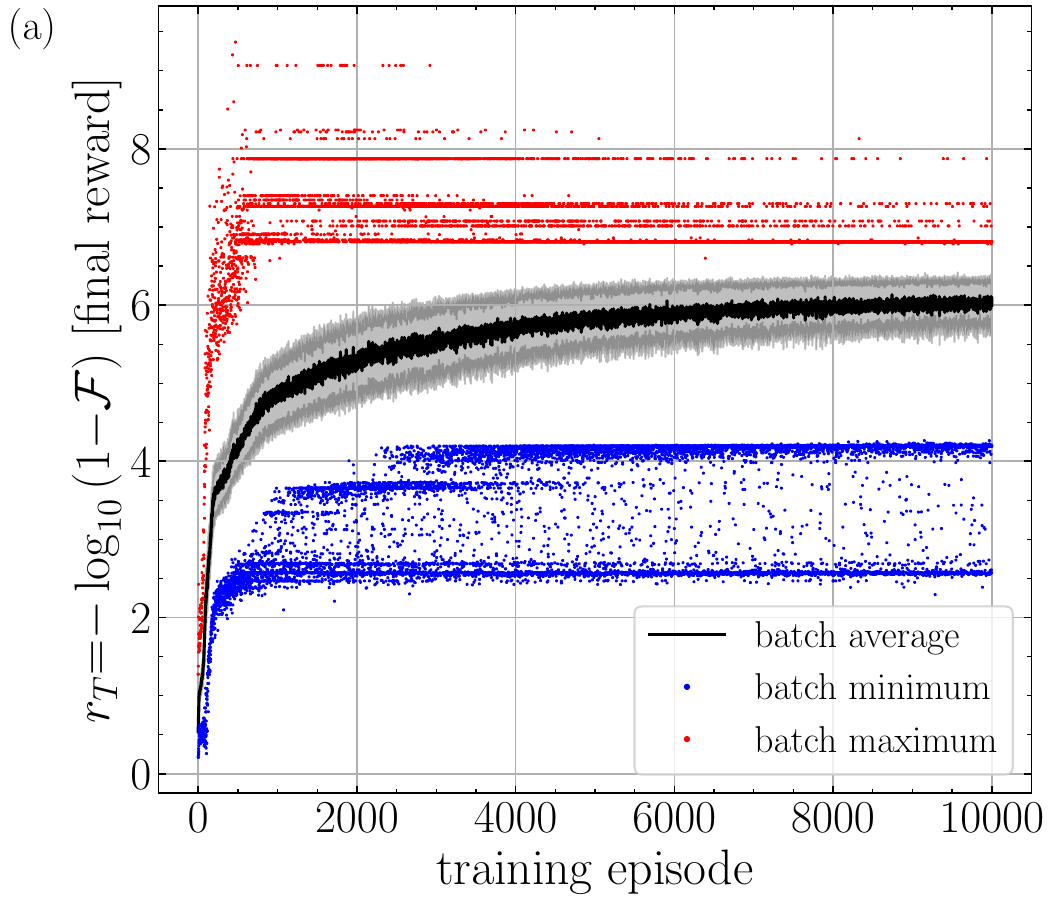}
\includegraphics[width=0.33\columnwidth]{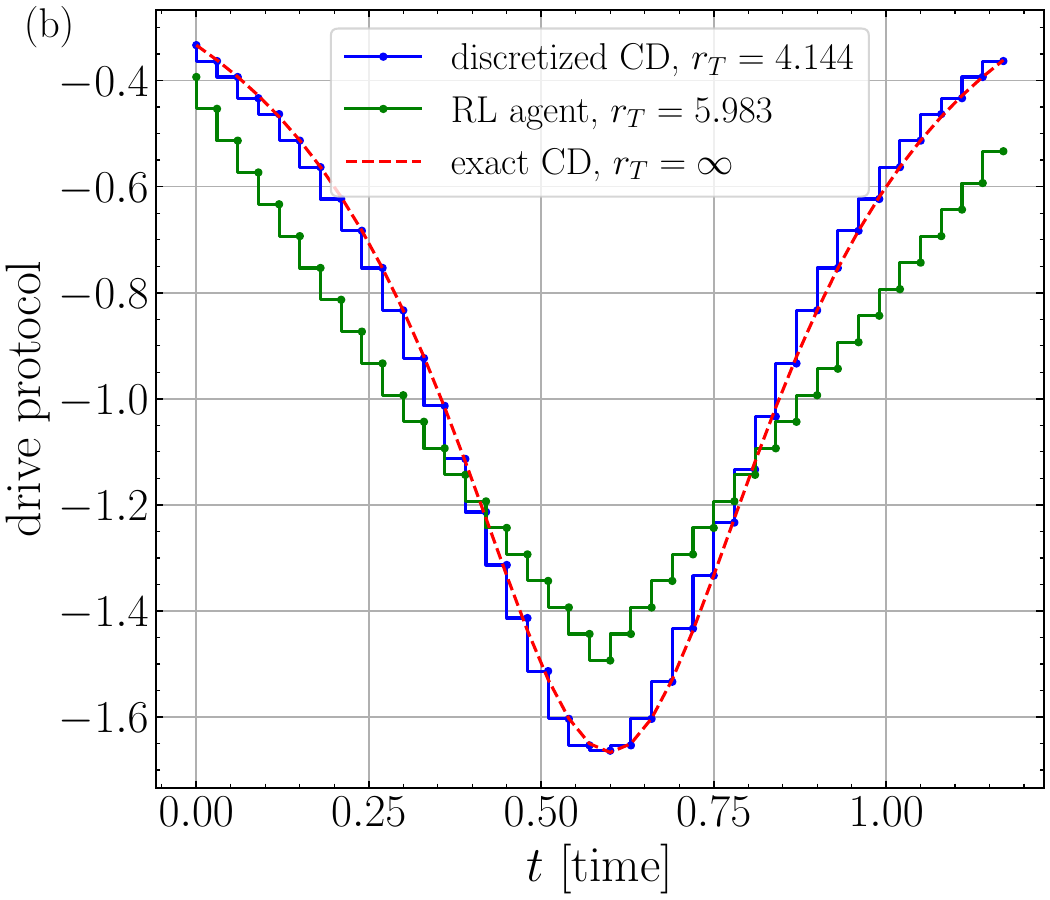}
\includegraphics[width=0.33\columnwidth]{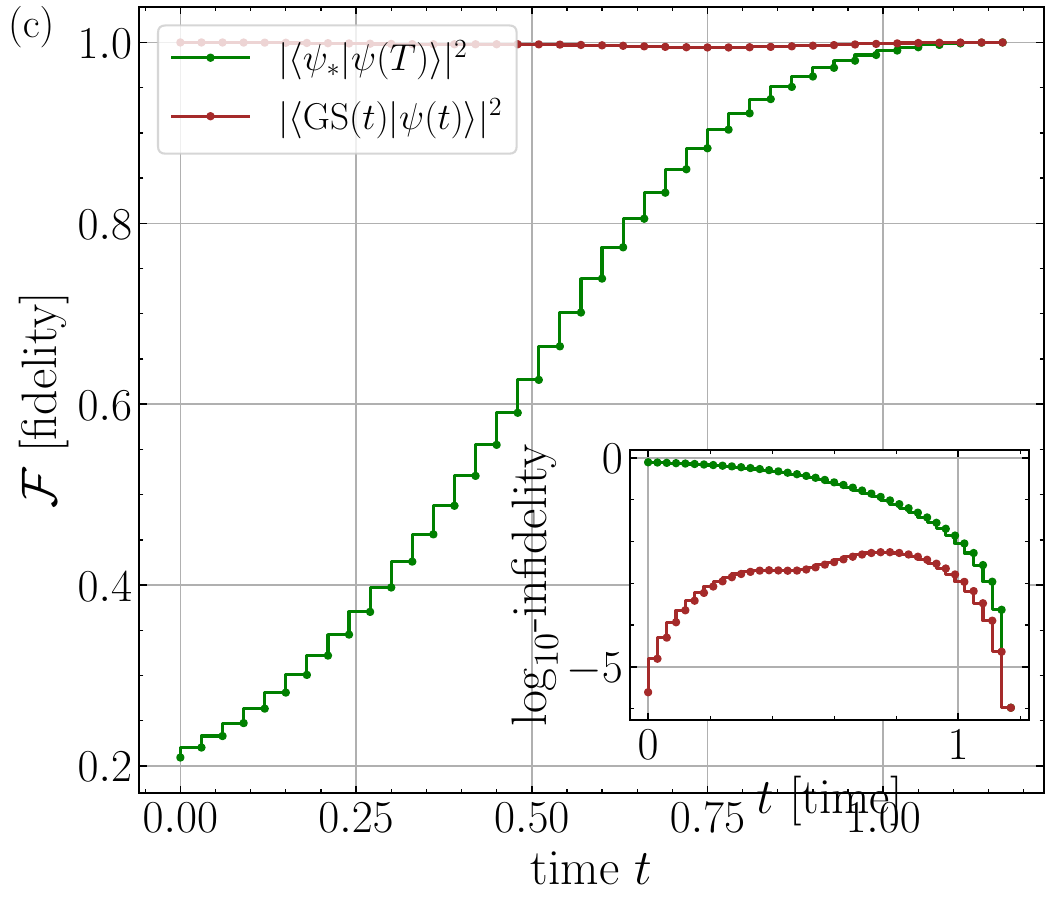}
\caption{
Reinforcement learning the CD control problem. 
\textbf{(a)} reward (logarithmic fidelity) at the end of the protocol against the number of training iterations (episodes): training is performed using a batch size of $256$. RL agent achieves an average fidelity $1-\mathcal{F}$ of about $10^{-6}$ (black line, with the shaded area showing the standard deviation); worst (best) protocols in the batch are shown in blue (red). 
\textbf{(b)} comparison of protocols and their final reward (corresponding to the number of decimals in $1-\mathcal{F}$). The exact continuous CD protocol (dashed red) performs worse than the RL agent (green) when time is discretized (blue).
\textbf{(c)} main plot shows the fidelity of the protocol learned by the RL agent (green) and the instantaneous fidelity (brown); inset shows the corresponding infidelities on a logarithmic scale.
}
\label{fig:RL-CD}
\end{figure}

We now apply the REINFORCE algorithm to this RL problem.
Figure~\ref{fig:RL-CD}(a) shows the training curves for the reward $r_T$ at the end of the episode (equal to the total return in this case). Like before, we consider the batch average (black), and the best (red) / worst (blue) protocols within the batch. Notice that achieving a 6-digit reward (i.e., a fidelity of $\mathcal F=0.999999$) requires about $8\times 10^3$ training episodes. We also see that there exist protocols with $r_T>8$ which the agent encounters during the exploration but fails to learn; this can be due to various reasons, e.g., the corresponding protocols may be unstable to small perturbations (recall that the agent takes actions probabilistically), or they may correspond to very deep and isolated minima of the control landscape (and hence, they are difficult to encounter often enough for the signal to accumulate and lead to stable learning).   

Figure~\ref{fig:RL-CD}(b) shows a comparison between the exact CD protocol (red dashed line), its discretization with the environment time step $\delta t$ (blue), and the protocol learned by the RL agent (green). Note that, while the continuous CD protocol is by construction perfect ($r_T=\infty)$, its discretized version has a reduced logarithmic fidelity $1-\mathcal{F}$ of $r_T=4.144$. For comparison, the RL agent achieves $r_T=5.983$. The RL agent achieves a better result, since it optimizes for preparing the target ground state only and pays no attention to the intermediate states of the system, see deviation from unity in the overlap with the instantaneous ground state (brown curve in Fig.~\ref{fig:RL-CD}(c)). By contrast, exact CD driving keeps the system in the instantaneous GS of $H(t)$ at \textit{any} time $t\in[0,T]$ with strictly unit fidelity. For example, at time $t/T=0.5$ in the middle of the protocol, the RL protocol achieves an instantaneous overlap with the GS of $r_T=2.678$ (brown curve at $t=0.6$, inset).
We invite the interested reader to check out the accompanying Jupyter notebook and explore the behavior of the agent further~\cite{github_code}.

\steve{The idea of using hybrid RL/OC algorithms extends immediately to multiqubit systems. The discrete nature of the control setup (time split into discrete timesteps and protocol values taken from a predefined finite discrete action set) suggests a natural extension in terms of quantum circuits built out of unitary gates.
It turns out, a suitable choice for actions are gate generators built out of the terms appearing in the expansion for the adiabatic gauge potential. For instance, in Ref.~\cite{yao2021reinforcement}, this idea was used to prepare the many-body ground state of Heisenberg and Ising chains, by using the (negative) energy of the final state as a reward function. 
}

\subsubsection{\label{subsec:RL_qubit_ancilla}Example: Continuous single-qubit feedback control using quantum data}

While the previous two examples demonstrated the basics of using RL, they employ RL frameworks that make it difficult to apply the corresponding agents in realistic experiments. The issue comes from the data used to train the agent, and is two-fold:
(i) at each timestep, the agent is assumed to observe the state of the qubit; this is problematic since quantum states are not observable. While it is not unimaginable to do a full-state tomography for a single qubit, besides the extra measurement overhead this would still result in some shot noise in the training data; moreover, tomography requires many repetitions of the experiment which can become an issue in the presence of noise because each trajectory is likely to end up in a different quantum state. We therefore need to revisit the definition of RL states if we want to run our agents on real quantum experiments. 
(ii) the RL agent uses the fidelity as a reward; similar to the issue above, the fidelity needs to be estimated from a number of binary measurements (following the projective measurement, the system is either found in the ground state, or in the excited state) taken sequentially and, at the very least, will also be noisy. 

To resolve these issues and design an RL framework that can interact with experiments, we have to replace the fully observable RL environment with a \textit{partially observable} environment. Hence, our agent will only have access to (partial) observations of the true RL state, which drastically reduces the amount of information it receives, and hence makes training more challenging; in fact, this is further exacerbated by single-shot qubit measurements that only contain binary information. For the implementation of our simulator, this means that we will have an internal variable that defines the exact RL state (here the full quantum wavefunction) and which undergoes the exact quantum dynamics; however, the agent-environment interface will only have access to observations of that state. Below, we will design an RL framework where the RL observations (now playing the role of the RL states) are all binary. 

\begin{figure}[t!]
\includegraphics[width=0.58\columnwidth]{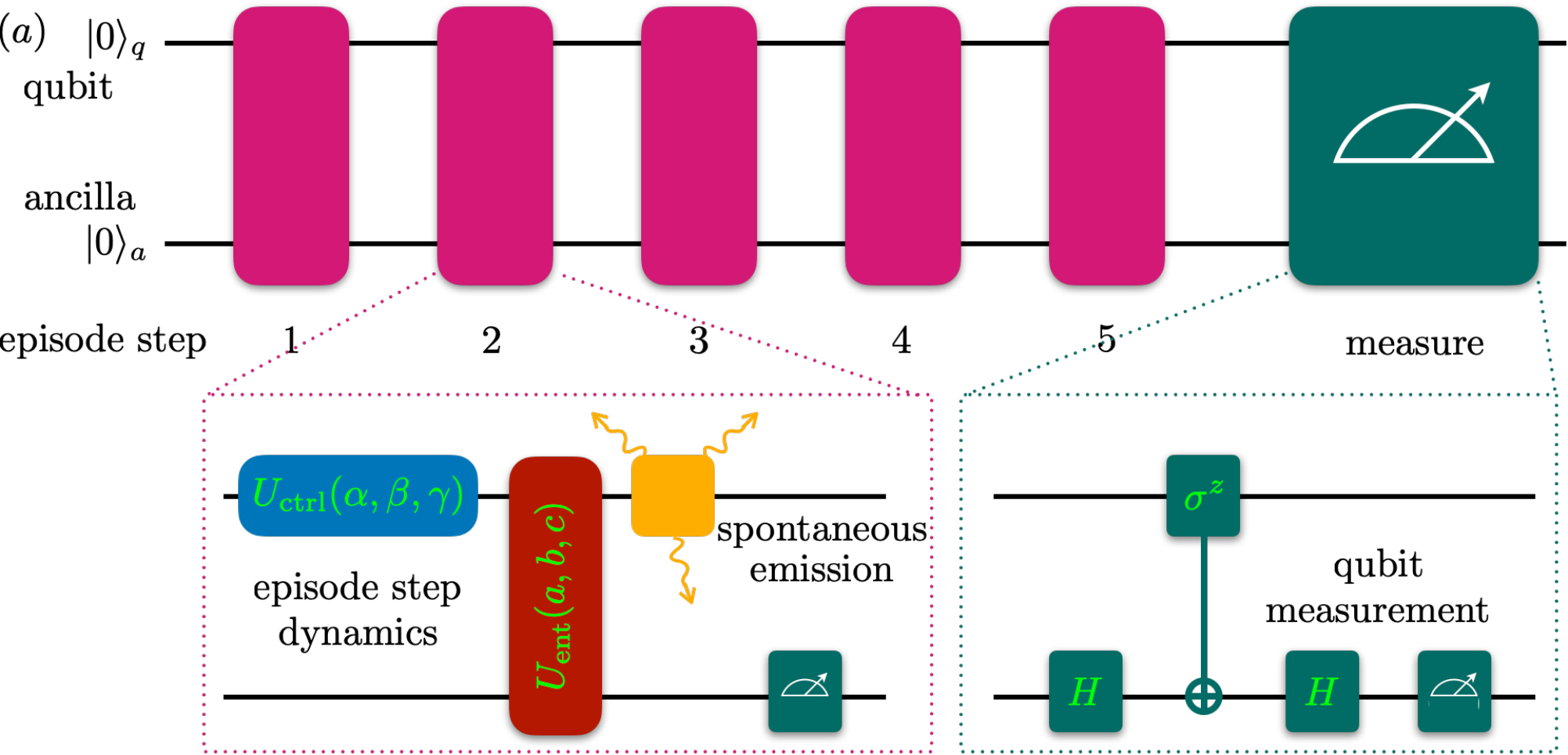}
\includegraphics[width=0.41\columnwidth]{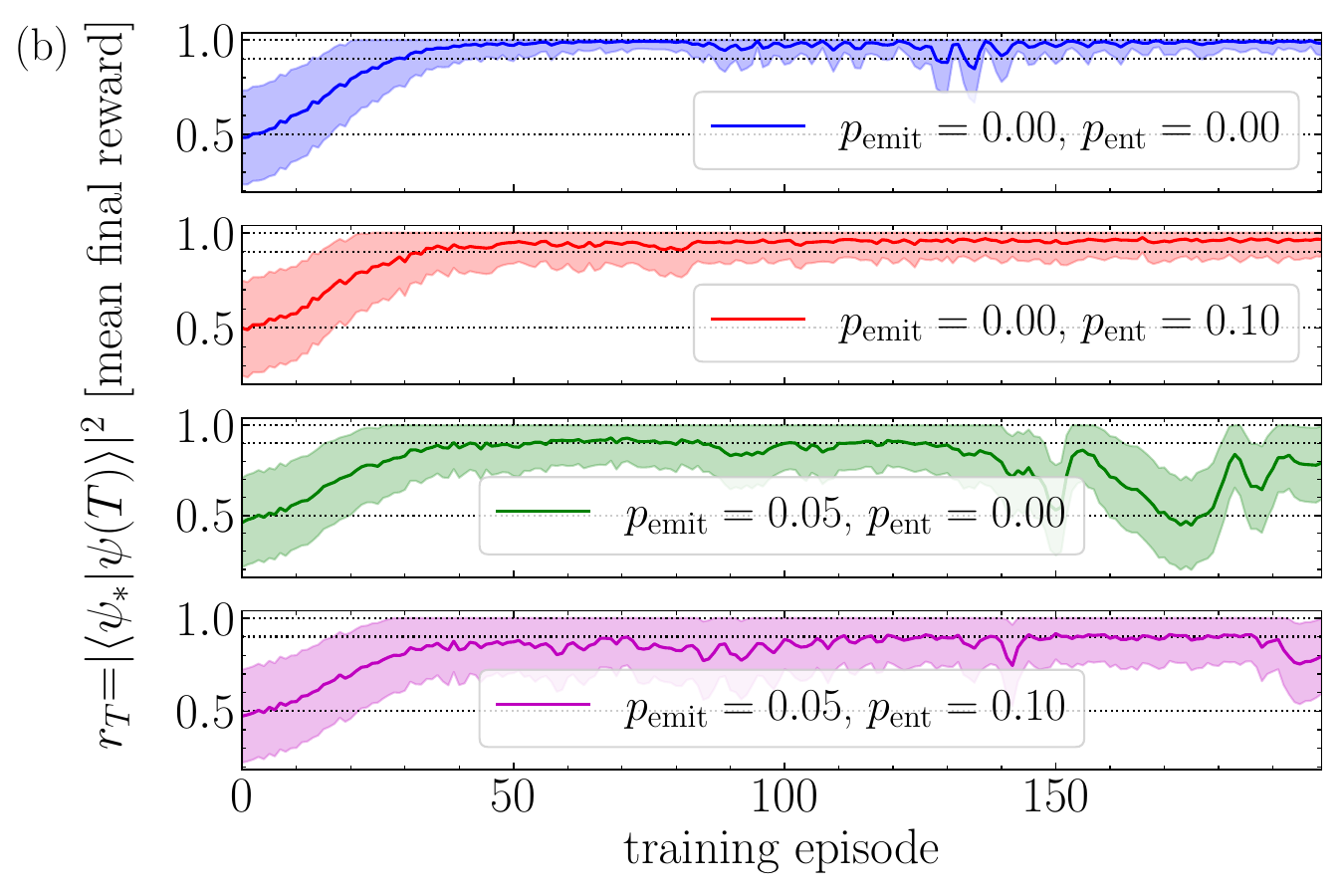}
\caption{
Reinforcement learning from quantum data. 
\textbf{(a)} 
We couple a qubit to an ancilla which we can measure projectively. 
At each episode step, the agent can apply a control gate $U_\text{ctrl}(\alpha,\beta,\gamma)$ (blue); however, the system is also subject to entangling noise $U_\text{ent}(a,b,c)$ (red) and spontaneous emission (orange). After each step we project the ancilla on the ground state. The time step, the measurement output, and the detected spontaneously emitted photon represent the RL observations; the reward is given at the end of the circuit by the binary output of the qubit collapse on the target state. At the end of the circuit, we use the ancilla to project the state of the qubit to the target state; this produces a binary output that defines the return.
\textbf{(b)} Training curves showing the final fidelity over a batch of $2048$ trajectories for:
(blue) the clean case (i.e., no entangling noise and no spontaneous emission);
(red) in the presence of entangling noise only; 
(green) in the presence of spontaneous emission only;
(magenta) in the presence of both noise and spontaneous emission. 
The shaded area marks the $1\sigma$ window over the batch. 
The horizontal dotted lines mark the $0.5, 0.8, 1.0$ fidelity thresholds, respectively.
We use a single-layer fully-connected neural network with $8$ hidden neurons. 
}
\label{fig:RL-qdata}
\end{figure}

More concerning is the quantum nature of the projective measurement itself, since it collapses the state of the qubit: this is similar to the state resetting procedure at the beginning of every episode, except now it occurs at each time step. The quantum Zeno effect then tells us that the state will barely evolve at all, and hence we may have a hard time bringing it to the desired target. To alleviate the measurement collapse problem while adhering to the laws of quantum mechanics, we introduce an ancilla qubit on which to perform measurements [see Fig.~\ref{fig:RL-qdata}(a)]. Whenever the controlled qubit and the ancilla are weakly entangled, measuring the ancilla will only weakly perturb the state of the qubit; that said, the weaker the entanglement, the smaller the amount of information we can extract about the state of the qubit from measuring the ancilla.  

Last, we turn to the issue of defining the reward. As already mentioned above, we can only afford to collapse the state of the qubit once per episode, and it is most natural to do this at the end (which sets the return to be equal to the final reward, a.k.a.~sparse reward function). However, instead of using the exact fidelity as a return, we will use the binary measurement output. 

At first sight, common sense may make it seem ludicrous that we can train any agent to learn from only binary information. Especially in realistic experiments where noise, decoherence, and spontaneous decay processes cause the qubit to end up in a different final state every time, even if we applied the same sequence of actions deterministically in an attempt to exactly repeat an episode. In such a scenario, estimating the qubit state and the fidelity requires using the theoretical framework of density matrices, whereas the RL agent is tasked to prepare a pure target state. 

To demonstrate that this is not as crazy as it may first sound, let us define an experimentally-friendly RL framework for a noisy qubit, and use the REINFORCE algorithm to train an RL agent to prepare a pure target state. As already advertized, we consider a qubit that can easily be reset to the ground state $\ket{1}_q$ at any time, and our goal is to transfer its population into a target state $\ket{\psi_\ast}_q$. To do this, this time we use continuous gates parametrized by the three angles $\alpha,\beta,\gamma\in[-\pi,\pi)$ (note that while one can also use an Euler angle parametrization, we prefer to work with a Tait-Bryan parametrization since it avoids solutions being degenerate in the case of $\beta=0$):
\begin{equation}
\label{eq:RL_ctrl_U}
    U_\text{ctrl}(\alpha,\beta,\gamma)= \mathrm{exp}(-i\gamma \sigma^z/2) \mathrm{exp}(-i\beta \sigma^y/2) \mathrm{exp}(-i\alpha \sigma^x/2)
\end{equation}
The goal of the RL agent is to learn to select the optimal angles to prepare the target starting from the fixed initial state. In addition, we will require that the agent can keep the state in the target over some fixed time.  

As we mentioned above, our qubit is paired with an ancilla initialized in the state $\ket{0}_a$. Crucially, any measurement on the system will be performed on the ancilla only, in order to avoid collapsing the state of the qubit; the binary output of these measurements will be used to define the observations and the return to train the RL agent. 

To make the problem more realistic, we now assume that the state of the qubit can undergo spontaneous emission and decay into its ground state. This spontaneous decay should in principle also be present for the ancilla, but we leave that for the interested reader to implement as an exercise. We also assume that some noise can lead to a small entanglement between the qubit and the ancilla, which we cannot control. In the following, we describe how to model these processes which define the RL environment.

\textbf{Spontaneous emission of qubit.---} Let us assume that the qubit state can undergo spontaneous emission with some probability $p_\text{emit}$; whenever it occurs, this process causes the qubit to decay to the ground state $\ket{1}_q$, no matter which state it was originally in.

We mention in passing that this is just one example of an error channel that is relevant to some (but not all) platforms; however, the procedure below can be generalized to other loss processes, depending on the particular system of interest.

Spontaneous emission is interesting for our RL setup since, by energy conservation, it is accompanied by a photon that can in principle be detected. 
Thus, although spontaneous emission is a stochastic process, we have a means to detect when it happened. We can then pass this information to our RL agent in the form of an observation to improve learning. 

We use a simplified model for spontaneous emission as follows:
\begin{equation}
    \ket{\psi}_q \longrightarrow 
\left.
  \begin{cases}
    \frac{P^z_q\ket{\psi}}{\sqrt{|\bra{\psi}P_z\ket{\psi}_q}}, & \text{with prob. } p_\text{emit}  \\
    \ket{\psi}_q, & \text{with prob. } 1-p_\text{emit},
  \end{cases}
  \right. 
\end{equation}
where $P^z_q = \frac{1}{2}(1 - \sigma^z)_q$ is the projector on the ground state $\ket{1}_q$ of the qubit.

\textbf{Entangling noise between qubit and ancilla.---} In realistic situations, the qubit and the ancilla are not perfectly decoupled from one another. Leftover terms in the implementation of the gates can lead to small entanglement between the qubit and the ancilla. 

We assume that this entangling noise is fixed, and can be parametrized by the parameters $a,b,c$, that we will refer to as angles, as 
\begin{equation}
    \ket{\psi}_q\ket{0}_a \longrightarrow  \exp(-i(a \sigma^x_q\sigma^x_a + b \sigma^y_q\sigma^y_a + c \sigma^z_q\sigma^z_a)) \ket{\psi}_q\ket{0}_a=U_\text{ent}(a,b,c)\ket{\psi}_q\ket{0}_a, 
\end{equation}
where $\sigma^\alpha_q\sigma^\alpha_a$ ($\alpha=x,y,z$) acts on the qubit-ancilla pair, respectively. One can check explicitly that these three operators commute and hence the exponential factors into three separate entangling gates that can be applied to the state consecutively (different Pauli matrices anti-commute for the same qubit but commute for different qubits). 

As a result of this noisy operation, the qubit and ancilla states may get entangled. Hence, measuring the ancilla to produce a binary partial observation, will perturb the state of the qubit; the strength of the perturbation is determined by the angles $a,b,c$ which are a property of the physical system (in our simulations, we use random numbers drawn uniformly from the interval $[-p_\text{ent}\times \pi, p_\text{ent}\times \pi]$ that are fixed across different episodes). It is this back-action that is difficult to model over multiple steps in a stochastic setting, and where RL can have an edge compared to more traditional control algorithms.    

\textbf{Measuring the state of the ancilla.---} Recall that we want our RL agent to receive as observations the binary output of ancilla measurements. We now discuss how to perform a measurement of the ancilla in the $z$-basis. Suppose the qubit-ancilla system is in the joint state $\ket{\psi}$. The ancilla measurement is then given by
\begin{equation}
    \ket{\psi} \longrightarrow 
\left.
  \begin{cases}
    \frac{P_a^z\ket{\psi}}{\sqrt{p}}, & \text{with prob. } p=|\bra{\psi}P_a^z\ket{\psi}|^2 \quad \text{\&\ measurement outcome } -1  \\
    \frac{(1-P_a^z)\ket{\psi}}{\sqrt{1-p}}, & \text{with prob. } 1-p \quad \text{\&\ measurement outcome } +1
  \end{cases}
  \right.
\end{equation}
Here the projector $P_a^z = 1_q\otimes \frac{1}{2}(1-\sigma^z)_a$ acts only on the ancilla subspace. 

Note that, in contrast to the spontaneous emission of the qubit, the measurement occurs with a probability that depends on the state of the system. Moreover, the state after the measurement always collapses. 

\textbf{Measuring the state of the qubit.---} Finally, we turn to the measurement that produces the reward data. To determine if the qubit at the end of the control circuit is in the target state, we can use the ancilla to perform a measurement of the operator 
\begin{equation}
    \sigma_\ast = \vec n_\ast \cdot \vec \sigma,
\end{equation}
where $\vec \sigma_q = (\sigma^x,\sigma^y,\sigma^z)_q$ is the vector of Pauli operators, and $\vec n_\ast  = \langle \psi_\ast|\vec \sigma|\psi_\ast\rangle_q$. The output of the measurement returns the $+1$-eigenvalue of $\sigma_\ast$ with probability $\mathcal F=|\langle\psi_\ast |\psi\rangle_q|^2$. After the measurement, the qubit is projected to the eigenstate corresponding to the measured eigenvalue.  

In practice, the measurement is performed using the ancilla, which is initialized in the $|0\rangle_a$ state. To this end, one first applies a Hadamard gate $H_a=(\sigma^x+\sigma^z)_a/\sqrt{2}$ on the ancilla, then a controlled-$\sigma_\ast$ gate which we denote by $C_{\sigma_\ast}$, followed by a second Hadamard gate on the ancilla [see Fig.~\ref{fig:RL-qdata}(a), green inset]. This circuit entangles the qubit state with the ancilla. Measuring the ancilla in the $z$-basis then returns the eigenvalue $+1$ with probability $\mathcal F=|\langle\psi_\ast |\psi\rangle_q|^2$, and collapses the state of the qubit on $|\psi_\ast\rangle_q$.

To see how this works, let us expand the state of the qubit in the eigenbasis of $\sigma_\ast$:
\begin{equation}
    |\psi\rangle_q = a|\psi_\ast\rangle_q + b|\psi_\ast^\perp\rangle_q,
\end{equation}
where $\sigma^\ast\ket{\psi_\ast}_q=+\ket{\psi_\ast}_q$ and $\sigma^\ast|\psi_\ast^\perp\rangle_q=-|\psi_\ast^\perp\rangle_q$.
The state of the qubit-ancilla system before the measurement circuit is thus 
\begin{equation}
    |\psi\rangle_q |0\rangle_a = a|\psi_\ast\rangle_q|0\rangle_a + b|\psi_\ast^\perp\rangle_q |0\rangle_a,
\end{equation}
where $q,a$ label the qubit/ancilla, respectively. After the first Hadamard on the ancilla, the state becomes (we use the convention $\ket{0}=(1,0)^t$ corresponding to the excited state):
\begin{equation}
    H_a|\psi\rangle_q |0\rangle_a = 
\frac{a}{\sqrt 2}|\psi_\ast\rangle_q(|0\rangle_a+|1\rangle_a) + \frac{b}{\sqrt 2}|\psi_\ast^\perp\rangle_q(|0\rangle_a+|1\rangle_a)
=\frac{1}{\sqrt 2}( a|\psi_\ast\rangle_q + b |\psi_\ast^\perp\rangle_q)\ket{0}_a + \frac{1}{\sqrt 2}( a|\psi_\ast\rangle_q + b |\psi_\ast^\perp\rangle_q )\ket{1}_a. 
\end{equation}
Next, we apply the controlled-$\sigma_\ast$ gate $C_{\sigma_\ast}$, i.e., we apply the $\sigma_\ast$ gate on the qubit if the state of the ancilla is $\ket{1}_a$, and we do not take any action if the ancilla is in $\ket{0}_a$; in doing so we recall that we decomposed the qubit state $\ket{\psi}_q$ in the eigenbasis of $\sigma_\ast$:
\begin{equation}
    C_{\sigma_\ast} H_a|\psi\rangle_q |0\rangle_a = \frac{1}{\sqrt 2}( a|\psi_\ast\rangle_q + b |\psi_\ast^\perp\rangle_q)\ket{0}_a + \frac{1}{\sqrt 2}( a|\psi_\ast\rangle_q - b |\psi_\ast^\perp\rangle_q )\ket{1}_a =\frac{a}{\sqrt 2}|\psi_\ast\rangle_q(|0\rangle_a+|1\rangle_a) + \frac{b}{\sqrt 2}|\psi_\ast^\perp\rangle_q(|0\rangle_a-|1\rangle_a).
\end{equation}
The second Hadamard gate on the ancilla then maps the state to
\begin{equation}
    H_aC_{\sigma_\ast} H_a|\psi\rangle_q |0\rangle_a = a|\psi_\ast\rangle_q|0\rangle_a + b|\psi_\ast^\perp\rangle_q |1\rangle_a.
\end{equation}
It now becomes clear that if after applying the circuit above we measure the ancilla in the $z$-basis, we obtain the measurement result $+1$ with probability $|a|^2=|\langle\psi_\ast |\psi\rangle_q|^2$, and $-1$ with probability $|b|^2$, as expected. Moreover, due to the entanglement between the qubit and the ancilla introduced by the $C_{\sigma_\ast}$ gate, the measurement of the ancilla automatically projects the state of the qubit in the corresponding eigenstate of $\sigma_\ast$. Note that this outcome is precisely what would come out if one could measure the operator $\sigma_\ast$ in the state $\ket{\psi}_q$.

\steve{We are now fully set to define} an episodic RL environment, which contains the laws of physics that govern the dynamics of the qubit-ancilla system (i.e., the application of the control gates and various sources of noise/decoherence to the qubit state). Our RL agent will interact with this environment to learn how to control the qubit to bring it from the initial ground state $\ket{1}_q$ to a prescribed target state $\ket{\psi_\ast}_q = \cos\frac{\theta}{2}\ket{0}+e^{i\phi}\sin\frac{\theta}{2}\ket{1}$ with $\theta=\pi/4$ and $\phi=\pi/3$. Each episode of the environment has $T=5$ steps; the agent applies a gate of the form in Eq.~\eqref{eq:RL_ctrl_U} at each step [see Fig.~\ref{fig:RL-qdata}(a)]. 

As discussed above, note that we cannot use the quantum state as an RL state since it is not observable. We therefore define RL observations using three ingredients: (i) information about the time step (the agent should know how many steps there are until the end of the episode), (ii) the binary measurement output of the ancilla measurement after each step, and (iii) the binary output of the photon detector which tells us whether a spontaneous emission event occurred or not. 

At each step within an episode, the agent will use its policy to generate the angles $\alpha,\beta,\gamma$ that define the control unitary from Eq.~\eqref{eq:RL_ctrl_U}. Because the angles can take on continuous values, we model the policy by an independent/uncorrelated Gaussian distribution for each angle (the agent will learn the mean and variance, respectively) $\pi_\theta\propto \mathcal{N}(\mu_\alpha,\sigma_\alpha^2)\mathcal{N}(\mu_\beta,\sigma_\beta^2)\mathcal{N}(\mu_\gamma,\sigma_\gamma^2)$; the angles to be applied will then be sampled from this policy. This is an example of a continuous control problem. In practice, we use a neural network (or any other variational ansatz) parametrized by the parameters $\theta$ to learn the optimal values of the means $\mu_k$ and the standard deviations $\sigma^2_k$: for a given observation $o_t$ which is an input to the neural network, the output is the tuples $(\mu_k, \sigma^2_k)$. 

Finally, the reward is set to zero at each time step (due to the character of projective measurements that destroy the state), except at the very end of the episode, when we use the ancilla to produce the binary reward corresponding to the quantum measurement. 

\begin{itemize}
    \item \textbf{observation space:} $\mathbf{S} = \mathbb{Z}_2^{3 \times T}$. The time step will be parsed to the agent as a one-hot encoding: the corresponding vector is initially set to zero. The ancilla measurements take on the value $+1$ if the ancilla is found in the state $\ket{0}_a$, and $-1$ otherwise, and are initialized to $+1$ for all episode steps. The ancilla is measured at the end of each step of the environment and is reset 
    to $\ket{0}_a$ afterwards. Finally, the photon detector gives $-1$ (i.e., qubit found in the ground state) if a photon has been detected at a given step and is initialized to $-1$ (no photon detected a priori). Hence, the observation consists of a $3 \times T$ binary vector. There are no well-defined terminal states in this task; instead, we consider a fixed number of $T=5$ time steps, after which the episode terminates deterministically. 

    \item \textbf{action space:} $\mathbf{A} = [-\pi,\pi]^3$. Note that the action space in this example is continuous; in particular, a single step is sufficient to reach any target state from any other state on the Bloch sphere. However, the presence of different types of noise makes this more challenging. Actions act on RL states as follows:
    \begin{enumerate}
        \item sample the angles $\alpha,\beta,\gamma$ from the Gaussian policy; 
        \item build the quantum gate $U(\alpha,\beta,\gamma)$, see Eq.~\eqref{eq:RL_ctrl_U}; 
        \item compute new state: $|\psi'\rangle = U(\alpha,\beta,\gamma)|\psi\rangle$. 
    \end{enumerate}
    Note that all actions are allowed from each state. 

    \item \textbf{reward space:} $\mathbf{R}=\{\pm 1\}$. We use sparse binary rewards obtained from the quantum measurement of the ancilla at the last episode step $T$ using the protocol described above.
\end{itemize}

Figure~\ref{fig:RL-qdata}(a) shows the control protocol applied to the qubit-ancilla system. The purple box represents one step, and is unpacked in the corresponding inset: first, the agent determines the optimal angles $\alpha,\beta,\gamma$ and applies the control unitary; the state then undergoes the entangling noise [blue] (whose strength is random, but fixed during training), followed by an eventual spontaneous emission of the qubit [orange]. Finally, we measure the ancilla in the $z$-basis, and use the binary measurement output as an observation. 
This protocol repeats for a total of $T=5$ steps, before we project the qubit to the target state [green box]. To do so, we use the Hadamard-$C_{\sigma_\ast}$-Hadamard sequence described above: measuring the ancilla in the $z$-basis then returns the correct output (as if we had measured the operator $\sigma_\ast$) and collapses the state of the qubit; it is this measurement data that we use as a reward to train the RL agent. 

We now parameterized the policy by a single-layer fully connected neural network with $8$ hidden neurons; the output of the network are the three mean and standard deviation values; we can sample actions (i.e., angles) from three independent corresponding normal distributions. To train the agent, we use a batch of $2048$ trajectories, over a total of $200$ episodes. Figure~\ref{fig:RL-qdata}(b) shows the corresponding training curves for four different types of noisy environments;
(i) when the entangling noise and spontaneous emission are off ($p_\text{emit}=0$, $p_\text{ent}=0$), the only source of noise is the shot noise when estimating the fidelity. We see that in this case, the agent learns to control the qubit after about $50$ episodes, reaching an almost perfect fidelity. Moreover, the $1\sigma$ uncertainty window (shaded area) shrinks as training progresses which indicates that the policy becomes more deterministic. 
(ii) we can then keep the spontaneous emission off and consider a weak entangling noise. This slightly reduces the achievable final fidelity to about $90\%$, but does not degrade the overall performance of the agent. 
(iii) in the opposite case, with spontaneous emission replacing the weakly entangling gate as a dominant noise source, we find that the agent reaches about $80\%$ fidelity. Following a stable learning stage, we notice large fluctuations in the training curve: these are presumably due to the small neural network used. We expect that training the agent longer will eventually stabilize the curve close to its maximum value.
(iv) finally, we turn all sources of noise on. The overall fidelity achieved by the agent in this case lies at around $80\%$. 
We invite the interested reader to check out the accompanying Jupyter notebook and explore the behavior of the agent further~\cite{github_code}.

\steve{Generalizing this setup beyond single-qubit control is a topic of forefront research. For instance, there was a recent demonstration of how RL agents can discover error-correcting codes in few-qubit systems by using an ancilla to infer partial information about the state of the system~\cite{foesel2018reinforcement}. 
Similarly, a recent experimental realization of a GKP code built from the states of a cavity resonator and manipulated using a transmon qubit as an ancilla~\cite{sivak2023real}, provides another real-world application of the framework introduced above. For details, we refer the interested readers to the growing body of literature on using RL for quantum error correction. 
}

\subsection{\label{subsec:RL_exp}Implementations of reinforcement learning in experiment}

We close this section by mentioning three applications of RL in experiments with quantum systems. The field is rapidly evolving; hence, the selection below is to be taken as exemplary rather than exhaustive. 


In Ref.~\cite{reuer2023realizing} the authors used an RL agent to initialize the state of a single transmon qubit. This state preparation task requires real-time control on time scales much shorter than the coherence time of a qubit. Thus, for the purpose of efficiency, the RL agent has to be integrated with the experimental device requiring low-latency feedback. The authors achieved this by using a sub-microsecond-latency neural network on a field-programmable gate array (FPGA) which interacts directly with the transmon qubit. The agent was then trained using quantum measurements. 
This work laid the foundations for resolving the challenge of developing and training a reinforcement learning agent able to operate using real-time feedback.


A more difficult, yet closely related, problem is the preparation of quantum gates. Quantum gates are implemented using control pulses, much like those discussed in Sec.~\ref{sec:QOC}. Gate synthesis can be viewed as a state preparation problem, but for a collection of a complete basis set for the Hilbert space; the challenge is to prepare the basis states with the correct relative phases among them. The agent typically has access to the control fields of a Hamiltonian, the time evolution of which generates the quantum gate.
RL has proven useful in tasks involving the design of low-level pulse controls that can compensate for hardware errors. In Ref.~\cite{baum2021experimental}, the authors demonstrate an RL agent that can design a universal set of error-robust quantum logic gates. In particular, they also train their agent in runtime on a superconducting quantum computer to implement a cross-resonance two-qubit gate. The agent does not require knowledge of a Hamiltonian model for the system, its controls, or any underlying error processes. The pulse sequences produced by the RL agent for the cross-resonance gate are up to three times faster than state-of-the-art pulses. Moreover, RL gates were found to be robust against calibration drifts.

And finally, a generalization of the RL framework used in Sec.~\ref{subsec:RL_qubit_ancilla} was recently used in an experiment to demonstrate quantum error correction for a cavity-QED qubit. The cavity mode defines a harmonic oscillator from which a logical qubit is built using a Gottesman-Kitaev-Preskill (GKP) code. The oscillator is coupled to a transmon which allows for readout and control of the logical qubit. The RL agent is trained to preserve the logical subspace using measurements of the code stabilizers. As a result, the trained RL agent increases the lifetime of the GKP qubit by more than a factor of 2, using an active quantum error correction protocol~\cite{sivak2023real}.

\clearpage



\section{Advantages, Limitations, and hybrid algorithms}
\label{sec:comparison}

As should be evident from the preceding sections, each technique has its own strengths and weaknesses: e.g., the simplicity in the formulation of counterdiabatic driving comes at the price of the often highly complex control terms it necessitates, optimal quantum control produces high-fidelity protocols so long as it does not get stuck in local minima, whereas the ability of RL agents to generalize typically requires large amounts of data. In Table~\ref{table:comparisons} we provide our opinion on some of the key advantages and limitations for each of the approaches discussed in Secs.~\ref{sec:STA}-\ref{sec:RL_theory}.

\begin{table}[h!]
\centering
\begin{tabular}{|p{2.5cm}|p{5.5cm}|p{5.5cm}|p{1.5cm}|}
\hline
Techniques & Advantages \& focus & Limitations \& Scope of applicability & References \\
\hline
\hline
\multicolumn{4}{|c|}{\textbf{Shortcuts-to-adiabaticity} -- Sec.~\ref{sec:STA}} \\
\hline
\hline
Counterdiabatic driving & Analytically derived which guarantees effectiveness and also allows \steve{us} to provide physical insight into fundamental controllability of the system & Only strictly applicable to exactly solvable models and often results in highly nonlocal control terms limiting direct experimental implementability & Refs.~\cite{STAreview,Berry2009} \\
\hline
Variational counterdiabatic driving & Can be approached semi-analytically or fully numerically and lends itself to a hybrid approach combined with quantum optimal control. Allows \steve{us} insight into the adiabatic gauge potential in regimes where its exact calculation is difficult. & The local approximation of counterdiabatic terms tends to be poor when approaching critical and/or transition points in the spectrum, i.e., where energy gaps close. & Ref.~\cite{sels2017minimizing,kolodrubetz2017geometry} \\
\hline
\hline
\multicolumn{4}{|c|}{\textbf{Quantum optimal control} -- Sec.~\ref{sec:QOC}} \\
\hline
\hline
Open-loop optimal control & Systematic approach that can be applied to any setting and allows \steve{us} to impose physically motivated constraints, e.g. allowable operations, time scale etc. Benign landscape structure gives certain guarantees to find good, if not necessarily the absolutely optimal, solution & Does not generalize as system parameters change - each configuration requires solving a new optimization problem. Numerically obtained control solutions are hard to interpret physically. & Refs.~\cite{Glaser2015} \\
\hline
\hline
\multicolumn{4}{|c|}{\textbf{Machine Learning based quantum control} -- Sec.~\ref{sec:RL_theory}} \\
\hline
\hline
Reinforcement Learning & Versatile in scope of application and like OC can accommodate any constraints from the outset. Can generalize to unseen setups (e.g., starting from different initial states). (Typically) does not require gradients of the function to be optimized. Does not require information about the controlled quantum system. & May require a lot of training data. (Typically) does not come with formal convergence guarantees. May get stuck in a suboptimal local minimum. Does not require information about the controlled quantum system. & Refs~\cite{GiannelliPLA} \\
\hline
Other ML~approaches & Train once, apply in diverse setups. Whenever optimization is time-consuming, learn the correct optimized output to avoid running it every time. & Generalization (i.e., extrapolation and interpolation) has limits and depends on the data used in training. Requires static training data distribution (i.e., characteristic system parameters should not change over time). & Refs.~\cite{mehta2019high,carleo2019machine} \\
\hline
\end{tabular}
\caption{Comparative table highlighting some key advantages and limitations of the various control techniques presented throughout the tutorial and other closely related approaches.}
\label{table:comparisons}
\end{table}

While we have presented and discussed each control technique in isolation, with an aim to provide a clear pedagogical introduction to the basic mathematical formulation of quantum control protocols, it is worth highlighting that significant benefits can be achieved by {\it combining} these techniques together. The effectiveness of such an approach can already be seen in fairly rudimentary settings e.g. Sec.~\ref{ex:LMGmodel} where counterdiabatic driving informed the choice of operators to employ for the control, leaving the pulse shape to be optimised. More sophisticated approaches, although similar in spirit, have been used to control many-body systems by combining shortcuts-to-adiabaticity with optimal control to design more experimentally favourable control protocols. The most intuitive approach starts by constraining the available set of controllable operators. By optimising the operators' time dependence, this technique can allow for remarkably good performance when traversing a critical point~\cite{Saberi2014, CampbellPRL} so long as we stay away from the thermodynamic limit when the spectral gap closes and the norm of the AGP diverges. As we will discuss more below, variational counterdiabatic driving allows \steve{us} to formalise and significantly expand this approach as shown in Ref.~\cite{COLD_PRXQ}. Beyond the clear practical advantage that hybrid approaches can provide vis-\'a-vis designing easier to implement control protocols, it can also allow to explore the fundamental limits of controllability. Hybrid control approaches allow to determine the true minimal control time under a given set of constraints and, therefore, explore the attainability of fundamental bounds such as the quantum speed limit and Lieb-Robinson bounds~\cite{Kiely2021NJP, Kiely2022PRR}. Significant efficiencies in the optimisation procedure itself are also attainable by combining gradient-free with gradient-based optimisation procedures~\cite{KochEPJQT2015}, the crucial insight here being that once the need for numerical optimisation enters the problem, it can be highly advantageous to ensure that a `good' initial seed is employed. This becomes particularly relevant when exploring rugged control landscapes~\cite{Chakrabarti2007, Beato2024a, Beato2024b, Fentaw2025}.

For the other considered control techniques, as we have seen, both optimal control and reinforcement learning offer numerical tools to devise quantum control protocols, and a natural question arises as to when one gains an edge over the other. A major difference between RL and optimal control is that optimal control focuses on the optimal solution and any acquired information about the physical system used during optimization is lost. RL, on the other hand, extracts essential information for maximizing the reward, and stores it in the policy -- a left-over product. Reinforcement learning algorithms are capable of re-using this information at a later stage, e.g.~starting from a different initial state, self-correcting for exploration mistakes made during training or caused by noise/uncertainty, or even trying to control a related but different system. However, this comes at a cost: RL is not computationally as efficient as optimal control and takes more iterations to find an optimal solution, especially if it starts without prior knowledge of the controlled system (cf.~model-free vs.~model-based RL~\cite{sutton_barto_rl}). Another advantage of RL over established quantum control methods, which rely on built-in optimization algorithms, is the balance between exploitation of already obtained knowledge, and exploration in uncharted parts of the control landscape (a.k.a.~the Exploration-Exploitation dilemma~\cite{sutton_barto_rl}, which was mentioned also in Sec. \ref{subsec:QOC_intro}). Below the quantum speed limit, decision-based exploration offers an alternative to multi-starting local gradient optimizers. Unlike such methods, the RL agent progressively becomes familiar with the control landscape, and finds protocols robust to sampling noise. Due to exploration noise, RL is not suitable for finding global minima of the cost function; in return, the minima it finds are more stable/robust to changes in the protocol. In difficult optimization problems, however, optimal control is never guaranteed to find the globally optimal protocol either, and we usually have no way of verifying if it did. Anticipating possible future applications of RL in physics, we note that RL brings in three key ideas:
(i) \emph{model-free} RL requires no pre-knowledge of the controlled physical system;
(ii) \emph{adaptive} RL transfers acquired knowledge to new setups and can point to hidden relations between nonequilibrium phenomena;
(iii) \emph{autonomous} RL provides novel insights into automating complex manipulation protocols. \\

Hybridized procedures allow to circumvent pitfalls associated with a particular approach, ultimately pushing the limits of controllability. As previously mentioned, a drawback of counterdiabatic driving (and associated analytical techniques) is the requirement to exactly solve the model, thus significantly limiting its range of applicability. However, it is evident that many physically relevant settings can be approximated by models with known analytical solutions. From the above examples and discussion, it follows that using the information accessible by determining the control protocol for the analytically tractable approximation, can serve as a highly insightful seed which can then be optimised using, e.g. optimal control theory and/or machine learning. Such an approach has already been demonstrated to be highly effective in a diverse range of settings, we refer the reader to Refs.~\cite{Hegade2022portfolio,Chandarana2023digitized,wurtz2022counterdiabaticity,yao2021reinforcement, Ref1}, for some further examples. 
To highlight the benefit that hybridized control strategies offer we mention three recent approaches that are showing particular promise: 
\begin{itemize}
\item {\it Enhanced shortcuts-to-adiabaticity (eSTA)}. The analytical nature of STA methods means that they have typically been applied to low-dimensional, idealised models. Thus, while the models serve as approximations to real physical systems, it is evident that directly applying the corresponding control terms to the actual system would, necessarily, not achieve perfect finite-time adiabatic dynamics. However, given that the formulation of STA methods guarantees that a counterdiabatic control term exists, it follows that if the Hamiltonian description of the real physical system can be considered a perturbation away from the idealised model, the correct control term is also likely to be closely related to the one derived for the idealised model. This insight served as the starting motivation for eSTA techniques~\cite{eSTA1, eSTA2}, which combines the knowledge learned from the STA methods together with gradient based optimisation to achieve high fidelity control in several physically relevant settings, in particular transport~\cite{eSTA1, eSTA2, eSTA6} and expansion~\cite{eSTA3} in anharmonic trapping potentials, manipulation and transport in double well~\cite{eSTA7}, driven qubits beyond the rotating wave approximation~\cite{eSTA1}, and bosonic Josephson junctions~\cite{eSTA5}. The computed control terms have also been shown to be robust to systematic errors and noise sources~\cite{eSTA2, eSTA5}. 

\item {\it Counterdiabatic local optimised driving}. The approach of variational counterdiabatic driving introduced in Sec.~\ref{subsec:varl_AGP} enabled the construction of approximate CD terms through minimisation without the need to diagonalise the Hamiltonian \cite{sels2017minimizing,kolodrubetz2017geometry}. This minimisation procedure can be tackled analytically for a limited set of scenarios, including the Ising model discussed in Sec.~\ref{subsec:varl_AGP}, but often requires numerical minimisation protocols to obtain the coefficients of the approximate CD for a particular protocol, akin to those discussed in the context of quantum optimal control in Sec.~\ref{sec:QOC}. The most natural way to enhance the performance of variational CD protocols is to introduce higher-order CD terms with operators that involve more bodies, i.e., by adding up to $N$-body interactions, but these are inherently nonlocal and often difficult to engineer in experiments or on quantum hardware. Note, these nonlocal and many-body terms are also what we are precisely wanting to avoid calculating when motivating the introduction of variational counterdiabatic driving. Counterdiabatic local optimised driving~\cite{COLD_PRXQ} takes a different approach by fixing the variational CD at low, local, orders of approximation while adding control fields that extend the family of dynamical Hamiltonians that can be explored. This allows \steve{us to use} counterdiabatic optimised local driving to find the optimal dynamical path for the local approximated CD. Recent work has extended this to determine the additional control fields and associated optimal set of operators necessary for a given problem~\cite{morawetz2024efficient}. Furthermore, this approach allows one to optimise the dynamical protocol without direct simulation, by calculating the ignored nonlocal CD terms and minimising their contribution, and this method could be further enhanced by combining it with the numerical approaches that have since been developed to calculate the CD terms, which we will discuss in Sec.~\ref{subsec:numagp}.

\item As a framework providing optimization algorithms capable of learning, RL presents an ideal playground for developing concepts for hybrid quantum control. Whereas RL does not magically produce high-fidelity protocols, this can be greatly facilitated by combining it with various STA methods, including CD driving, explored in Refs.~\cite{yao2021reinforcement,XiChenRL1}. The key idea to apply RL to real-world NISQ devices is to exploit the Quantum Approximate Optimization Algorithm (QAOA) which defines a variational quantum circuit with unknown parameters. The RL agent has to learn either the gate angles (continuous) or the circuit architecture (discrete) and, ideally, both. Since solving both the continuous and discrete optimization involves mixed continuous-discrete action spaces which can be challenging to deal with, often one resorts to RL for one of these two problems, leaving the other to optimal control methods (Sec.~\ref{sec:QOC}), such as gradient descent, or tree searches~\cite{yao2022monte}. To design the circuit architecture ansatz, one can use the terms appearing in the nested commutator ansatz~\cite{Claeys2019Floquet} for the variational AGP, see Sec.~\ref{subsec:varl_AGP}. While this does not implement CD driving itself, these terms generate unitary dynamics that unlocks shortcuts in the Hilbert space, leading to efficient state preparation away from the adiabatic regime. 
Using this approach Ref.~\cite{yao2021reinforcement} developed RL agents to prepare ground states of many-body spin chains on NISQ devices, while the RL agent of Ref.~\cite{XiChenRL1} is trained to design optimal pulses for single-qubit gates applicable, e.g., on a superconducting qubit device. 
We have explored in detail such hybrid CD-RL control shortly, see Sec.~\ref{sec:RL_2q}.
We mention in passing that, besides CD driving, RL can be combined in a hybrid approach to optimize dynamical invariants -- a different STA technique~\cite{XiChenRL3}.
\end{itemize}

\clearpage

\section{Future challenges and prospects for quantum control}
\label{sec:Outlook}

\begin{figure}[h]
\label{fig:outlook}
\includegraphics[width=0.85\columnwidth]{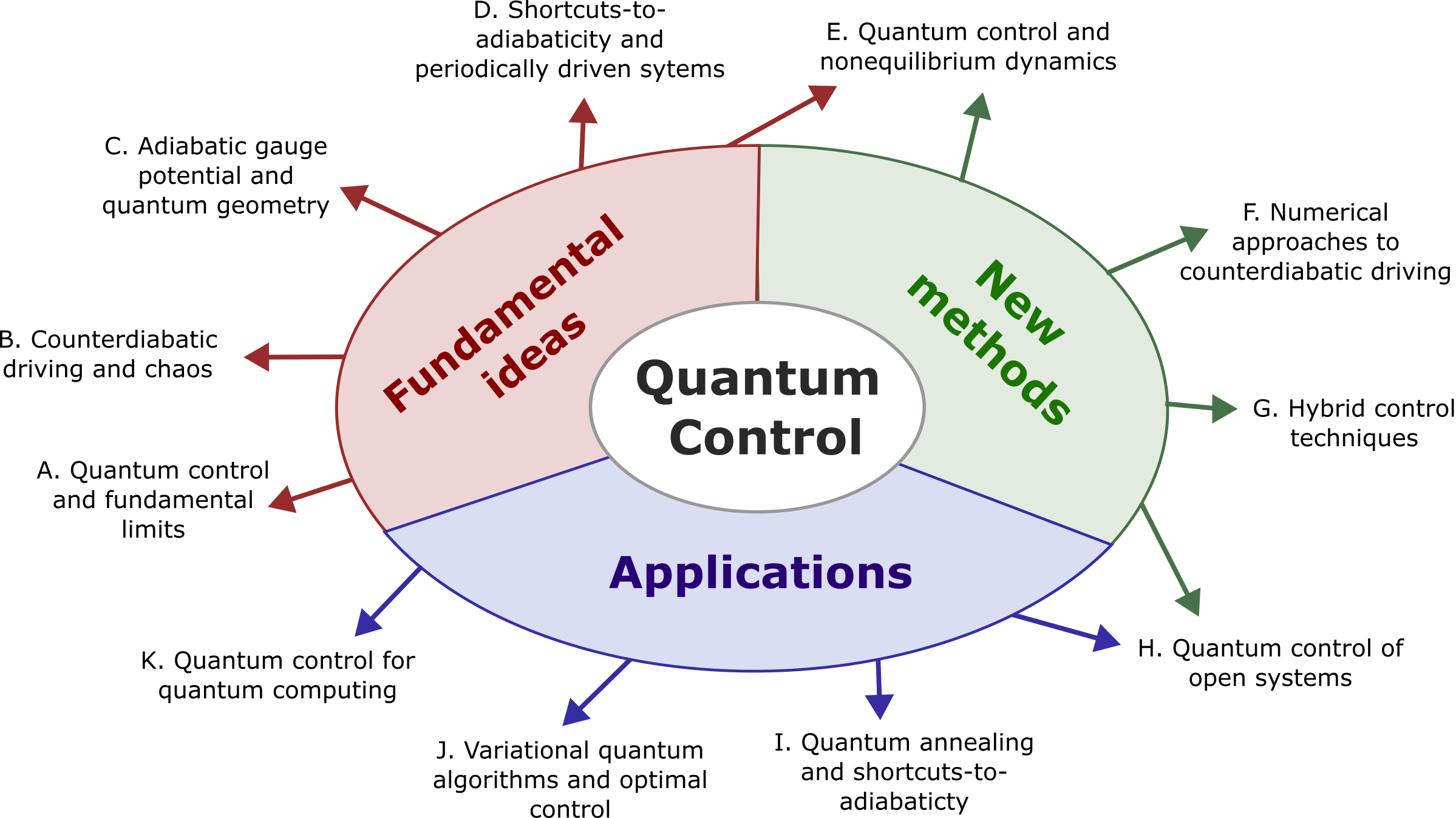}
\caption{Overview of the topics discussed in Sec.~\ref{sec:Outlook} and their relation to fundamental ideas, new methods, and applications of quantum control tools.}
\end{figure}

The ubiquity of control means that the techniques and tools presented above can find fertile application in a wide variety of places. In this section, we conclude this tutorial by briefly touching on several areas where quantum control is prevalent and where there are potentially significant benefits to be gained from the continued development of more advanced techniques for coherent control of complex quantum systems, see Fig.~\ref{fig:outlook}.\\

\subsection{Quantum control to push fundamental limits in quantum dynamics}
Given a model for a quantum-dynamical system and a target state or unitary transformation, the tools of optimal control allow us to search for a protocol that achieves such a process. If the optimization systematically fails to find a solution, this could indicate that the process was not achievable, with the resources available, in the first place, thus revealing a fundamental limitation of the dynamics. In the examples of Sec. \ref{sec:QOC}, this was related to restrictions placed by the quantum speed limit, and the fact that there is a minimum time required to perform a given quantum transformation, which in turn can be traced back to fundamental uncertainty relations. For quantum control problems of the form described in Sec. \ref{sec:STA}, where the initial and target states can be adiabatically connected, it has been shown that quantum speed limit bounds can be significantly refined to yield more informative bounds on the systems dynamics \cite{mandelstam_45,fleming_73,bhattacharyya1983, anandan_90,margolus_98, bukov2019geometric,FunoPRL2017}, which in turn are linked with the geometric picture described in Sec.~\ref{outlook_sec_geometry} and can help in developing alternative strategies to achieve control even for complex systems~\cite{Balducci2}.

The example above suggests that optimal control could be a useful tool to reveal and probe other fundamental bounds related to quantum dynamics. One such example is the Cram\'er-Rao bound \cite{pezze2018}, which limits the precision with which a quantum state can act as a probe of a weakly encoded phase. Efforts towards using QOC for quantum sensing have focused on optimizing the quantum Fisher information \cite{pang2017,liu2017,lin2021,Gietka2021}, and related protocols based on variational quantum circuits (see also Sec.~\ref{outlook_sec_qc}) have looked at optimising other types of metrics \cite{kaubruegger2021}. Instead of maximizing the sensitivity of the quantum state to an external perturbation, optimal control methods could also be applied to minimize such sensitivity, i.e., yielding \textit{robust} quantum processes \cite{poggi2023_urc}. Naturally, robust quantum control is highly desirable in several applications, for instance, in quantum computing where extreme precision at the quantum level is typically required, \steve{or in quantum thermodynamics where control techniques can be leveraged to designing protocols for the efficient charging of quantum batteries~\cite{ref7, ref8, ref9, ref10, ref11, ref12}}.

In the realm of many-body systems, different fundamental bounds limit the rate at which quantum information spreads through the degrees of freedom of a system. These include Lieb-Robinson bounds \cite{chen2023}, which determine the existence of light-cone-like dynamics in a system with local interactions, as well as fast-scrambling bounds \cite{lashkari2013}, related to the maximum speed of operator growth. It is an open question how driving and optimal control could be used to create protocols that scramble quantum information optimally. Even more fundamentally, optimal control tools for observable evolution (i.e., in the Heisenberg picture) have not been studied in depth up to now.  Protocols that manipulate quantum information optimally could be of use for certain quantum algorithms, as well as for quantum simulation of complex models related to black hole information dynamics.

\subsection{Connections between counterdiabatic driving, the adiabatic gauge potential, and chaos}

Adiabatic gauge potentials underlie quantum geometry, cf.~Sec.~\ref{outlook_sec_geometry}. The latter encodes how weakly the wavefunction changes, as measured by the fidelity susceptibility. It was pointed out that this connection can be used, e.g., to detect phase transitions without the need to define any order parameters~\cite{venuti2007quantum,gu2010fidelity}. 
Recently, it was proposed to use the sensitivity of eigenstates to external perturbations to define quantum (and classical) chaos~\cite{pandey2020adiabatic}. Indeed, eigenstate deformations are quantified by the metric tensor $g_{\lambda\mu}= \text{Re}\; \langle n|\mathcal{A}_\lambda \mathcal{A}_\mu|n\rangle_c$; in chaotic systems, one typically performs an extra average over the eigenstates which corresponds to an infinite-temperature ensemble. Hence, eigenstate deformations can be used as a probe for quantum chaos.
To see this connection from a different perspective, recall that the correction to the eigenstates in the presence of perturbations is determined by the AGP $\mathcal{A}$. Moreover, as it satisfies the operator-valued equation $[G(\mathcal A),\mathcal H]=0$ [cf.~Eq.~\eqref{eq:glambda}], one can associate adiabatic transformations (i.e., the transformations generated by the AGP) with emergent conservation laws whose generators also commute with the Hamiltonian $\mathcal{H}$. This leads to an intricate relationship between the AGP and integrability. Since chaos is a manifestation of the lack of integrability, it turns out it can be quantified by analyzing the stability of time-averaged trajectories to small deformations in the Hamiltonian~\cite{lim2024defining}. We note in passing that chaos and ergodicity are, in general, different notions~\cite{kim2024integrability}; the AGP can help identify the presence of chaotic behavior. 
Finally, recall that the fidelity susceptibility can be measured in experiments since it is related to the response function of the system~\cite{kolodrubetz2017geometry}; in this sense, the more slowly a system relaxes and the stronger its sensitivity is to low-frequency response, the more chaotic it is. 

\subsection{Adiabatic gauge potential and quantum geometry} 
\label{outlook_sec_geometry}

One of the central objects of this tutorial -- the adiabatic gauge potential -- is inextricably related to quantum geometry~\cite{kolodrubetz2017geometry}. Indeed, its connected correlation function in the eigenstate $\ket{n}$ of the Hamiltonian, $\chi_{\mu\nu}=\langle n|\mathcal{A}_\lambda \mathcal{A}_\mu|n\rangle_c$, defines the quantum geometric tensor $\chi_{\mu\nu}$. 
The real part, $g_{\mu\nu} = \text{Re}\; \chi_{\mu\nu}$, is the metric tensor which determines the distance between states on the manifold parametrized by $\lambda$ (the Fubini-Study metric). It can be shown that adiabatic evolution follows the geodesics (i.e., the minimal-distance curves) w.r.t.~$g_{\mu\nu}$. The metric tensor is also known as the quantum Fisher information or fidelity susceptibility, and finds ample applications in quantum parameter estimation and quantum sensing. 
On the other hand, the imaginary part of the quantum metric tensor is proportional to the Berry curvature $F_{\mu\nu} = -2\text{Im}\; \chi_{\mu\nu}$. It is often used to determine the Chern number -- a topological invariant that lies at the heart of understanding topological insulators, and is invoked in the explanation of the integer quantum Hall effect~\cite{bradlyn2022lecture}. Moreover, the eigenstate expectation value of the gauge potential, $\langle n|\mathcal{A}_\lambda|n\rangle$ is the Berry connection which defines parallel transport on the parameter manifold; it is frequently used to compute the geometric (Berry) phase accumulated by the wavefunction upon parallel transport along a closed path in the parameter manifold. 
Unlike the Berry connection, both the quantum metric and the Berry curvature are invariant under changing the phase of the wavefunction, and are therefore measurable quantities. In particular, the metric tensor is related to the spectral function via linear response theory, while the Berry curvature can be measured from the leading order nonadiabatic response of observables~\cite{kolodrubetz2017geometry}. 

\subsection{Shortcuts-to-adiabaticity and periodically driven (Floquet) systems}
As we discussed in Sec.~\ref{subsec:varl_AGP}, real-valued Hamiltonians have purely imaginary-valued adiabatic gauge potentials. For non-interacting systems, these take the form of $y$-magnetic fields or current terms; interactions modify these currents and can even make them less local (think of interaction-dependent hopping terms). Thus, implementing even approximate CD driving requires experimental control over such generalized current operators. Moreover, in systems with broken time-reversal symmetry, the Hamiltonian is complex-valued and even more complicated terms appear in the AGP. It is therefore important to find ways to implement the AGP in experiments, at least approximately. 
One way to do this is to use periodic (Floquet) drives~\cite{Goldman2014_FloquetGaugeFields,Bukov_2015_general_HFE,Eckardt2017_FloquetGases_Review,Aidelsburger2018_Review,Oka2019_FloquetMaterials,Weitenberg2021_FloquetGases}. Recently, it was shown that the AGP can be generated by taking nested commutators of the non-driven Hamiltonian $\mathcal{H}(\lambda=0)$ and the generalized force $\partial_\lambda \mathcal{H}(\lambda)$~\cite{Claeys2019Floquet}. Incidentally, the same nested commutators appear in the Floquet-Magnus expansion (or its step-drive version, the Baker-Campbell-Hausdorff expansion), used to find approximations to the time-independent Floquet Hamiltonian $\mathcal{H}_F$ which governs the stroboscopic dynamics of periodically driven systems. This opens up the intriguing possibility of implementing an effective stroboscopic dynamics generated by $\mathcal{H}_F = \mathcal{H}+\dot\lambda \mathcal{A}_\lambda$~\cite{Claeys2019Floquet} (a related idea was used to enhance energy minimization algorithms on digital quantum devices using CD driving, giving rise to the CD quantum approximate optimization algorithm (CD-QAOA)~\cite{wurtz2022counterdiabaticity}). A proof of principle realization of counter-diabatic driving using periodic drives has recently been reported in an experiment using NV centers~\cite{boyers2019floquet}. 

Floquet drives, on their own, are typically used to alter the properties of static systems via so-called Floquet engineering. The key idea is to use a strong time-periodic drive to dress the states of the system; the resulting stroboscopic dynamics generated by the Floquet Hamiltonian can be used to explore dynamically states and/or parameter regimes accessible only in so-called synthetic quantum matter (e.g., ultracold or Rydberg atoms, trapped ions, superconducting qubits, etc.)~\cite{Goldman2014_FloquetGaugeFields,Bukov_2015_general_HFE,Eckardt2017_FloquetGases_Review,Aidelsburger2018_Review,Oka2019_FloquetMaterials,Weitenberg2021_FloquetGases}. However, using periodic drives as an engineering toolbox brings up the question of how to control the population of stationary states in the presence of the periodic drive that created them in the first place; commonly used protocols in Floquet engineering include ramps of the drive amplitude, modulation of the drive frequency (so-called chirps), or protocols changing the phase of the periodic drive. 
Floquet control is a daunting challenge since any external control will violate the periodicity of the Hamiltonian and render Floquet's theorem inapplicable; moreover, due to drive-induced resonances, the adiabatic limit for the effective Floquet Hamiltonian does not exist (though there can be adiabatic regimes at small but finite ramp speeds)~\cite{eckardt2008avoided,weinberg2017adiabatic}. Nevertheless, the concept of local CD driving can be extended to Floquet states: besides the contribution from the Floquet eigenstates, the AGP associated with the Floquet Hamiltonian contains an additional term due to the micromotion dynamics (i.e., the dynamics in between the stroboscopic times). Curiously, there also exists a generalization of the variational principle we derived in Sec.~\ref{sec:varcd}~\cite{schindler2024counterdiabatic}. In that sense, finding the Floquet Hamiltonian (i.e., applying Floquet engineering) can be cast as an inverse counter-diabatic problem~\cite{schindler2024geometric}.
Floquet systems demonstrate that the ideas of CD driving are more general and apply to systems taken far out of equilibrium; moreover, CD driving proves instrumental in telling inherently nonequilibrium effects apart from equilibrium-like phenomena. Indeed, CD driving offers an entirely new, geometric perspective on Floquet theory~\cite{schindler2024geometric}. The question of how to efficiently control generic nonequilibrium steady states is one of the open frontiers of present-day quantum dynamics.

\subsection{Quantum control to characterise non-equilibrium dynamics} \label{outlook_sec_agp_insight}
While the principal goal of quantum control is to achieve a particular aim, it is useful to acknowledge that this paradigm can be inverted and the control of a system can be used as a remarkably versatile tool for understanding the underlying physics. In the case of control via counterdiabatic techniques the intuition is simple: the control allows \steve{us} to circumvent or suppress the non-equilibrium dynamics; therefore, the control itself contains all the relevant information about the dynamical response of the system to an, in principle, arbitrary perturbation. This can readily be seen for example in assessing characteristics of the controlled dynamics which allows \steve{us} to study the Kibble-Zurek mechanism~\cite{Puebla2020, CampbellEPL2023}. 
Furthermore, while counterdiabatic driving predicts that highly complex and non-local terms are generally necessary for many-body systems, the techniques outlined previously provide a versatile toolbox in order to build up the control while respecting given experimental constraints. This not only allows for these techniques to be scaled up, and thus enhance their practical utility, it also \steve{provides us with} a framework to consider the relevance of particular interactions~\cite{COLD_PRXQ, lawrence2024numerical} or to determine the minimal operator set necessary for control~\cite{delCampoKrylovPRX}, ultimately providing valuable insight into the fundamental controllability of complex systems.

\subsection{Numerical approaches to the calculation of counterdiabatic driving and the adiabatic gauge potential} \label{subsec:numagp}
In Sec.~\ref{sec:STA}, we exposed the reader to a number of approaches for obtaining the adiabatic gauge potential (AGP), or counterdiabatic driving term, which were analytic in nature. The final variational approach \cite{sels2017minimizing} outlined in Sec.~\ref{subsec:varl_AGP} was until relatively recently the state-of-the-art, with the advancement of the known form of the ansatz given in Ref.~\cite{Claeys2019Floquet}. However, with the advancement of adiabatic quantum computing, and the later connection of the AGP to the digital variational quantum approximate optimisation algorithm (QAOA) \cite{wurtz2022counterdiabaticity}, there was growing interest in developing numerical procedures to calculate or approximate the AGP in many-body quantum systems beyond that which can be tackled with the techniques outlined in Sec.~\ref{sec:STA} and which are complex systems from the point of view of optimal control. To-date, this has led to three approaches which tackle this issue from different angles. The orthogonal commutator expansion method \cite{lawrence2024numerical} takes a similar approach to that of the variational adiabatic gauge potential of Sec.~\ref{subsec:varl_AGP} but recasts solving for the AGP into that of solving a block-tridiagonal system of equations, which can be solved efficiently numerically. This method is particularly targeted at obtaining the exact AGP numerically in complex scenarios. The second method is a Krylov approach which uses a standard approach to solving large quantum many-body problems numerically by introducing a Krylov basis which spans the minimal operator subspace to describe the AGP \cite{Claeys2019Floquet, delCampoKrylovPRX}. This method is particularly useful for finding approximations of the AGP at moderate orders of the commutator expansion of Ref.~\cite{Claeys2019Floquet} for complicated many-body Hamiltonians. The third method uses matrix product operators -- a tensor network technique~\cite{kim2024variational,mckeever2024towards}, and allows one to compute the AGP in large quantum systems. It can outperform both the orthogonal commutator and the Krylov ansatz methods, although it is currently unclear if this advantage is generic. The three methods are equivalent in the limit in which they all solve for the exact AGP, if they can reach that limit, in a given problem. However, as all three methods have only recently been introduced, it is unclear how the approximate forms of each compare and this is a current area of active research.

\subsection{Hybrid control techniques}
As discussed in Sec.~\ref{sec:comparison}, it is clear that hybridised protocols offer the most versatile approach to achieve coherent control. The performance of such techniques also indicates that determining the ultimate limits of controllability is a multifaceted problem and one which will require input from the various sub-communities. This is particularly relevant as quantum devices continue to be developed and there is a particular need for effective and robust control. Indeed, the utility and scalability of novel quantum technologies is likely to be strongly impacted by our ability to coherently manipulate the components of the devices, particularly in the presence of unwanted, or potentially unknown, sources of noise, as will be discussed in the proceeding Sec.~\ref{outlook_sec_opensystems}.

\subsection{Quantum control of open systems} \label{outlook_sec_opensystems}
A simple fact is that no system is ever truly shielded from the surrounding environment and indeed virtually all experimental setups go to great lengths to offset, or at least mitigate, the unavoidable interaction between a system and, e.g., stray electric and magnetic fields or unwanted seismic activity (such as a subway passing by), etc. If the timescales of the noise are sufficiently slow compared to the system's evolution or if they host so-called decoherence-free subspaces or can be dynamically decoupled from the environment~\cite{DFSreview}, then often the control protocol for the closed unitary dynamics are put to work, i.e. precisely the techniques that have been the focus of this tutorial. Indeed, speeding up the unitary dynamics such that it achieves a given evolution faster than the noise can act is one of the core drivers in the development of quantum control techniques. However, increasingly quantum control is becoming an integral part of emerging quantum technologies themselves and this has resulted in a steady growth in examining control of genuinely open quantum systems~\cite{Koch2016OpenSys}. The simplest such studies involve employing the same Hamiltonian control techniques already discussed while explicitly in the presence of unwanted and/or unavoidable environmental noise commonly modelled using phenomenological approaches, particularly Markovian master equations~\cite{Koch2016OpenSys}. Control strategies, particularly those following the QOC theory and ML type prescriptions, are determined while explicitly including the open system effects as part of the optimisation, aiming to find the optimal path in state space that minimises the impact of the environment.

The most general picture for control is when both the system's Hamiltonian and interaction with its environment can be manipulated, so-called reservoir engineering approaches, see e.g.~\cite{Kapit2017ResEng}. Such a setting encompasses and extends the above to the non-Markovian regime which, owing to the inherent complexity of characterising and understanding such open system dynamics, means comparatively little is known regarding the typically achievable levels of control. For control techniques based on shortcuts-to-adiabaticity the issue becomes perhaps even more fundamental since the very definition of adiabaticity must be carefully reconsidered when the system exchanges energy with the environment~\cite{AdiabaticityOpenSys1, AdiabaticityOpenSys2}. Nevertheless some progress in extending shortcuts-to-adiabaticity to open systems has been made~\cite{VacantiNJP, JunJing1}. In addition to these general considerations, it is worth noting that noisy intermediate-scale quantum (NISQ) devices have recently opened up a new kind of non-unitary way to control quantum systems: these so-called feed-forward protocols rely on mid-circuit projective measurements the results of which are fed back to the system and determine the subsequent structure of the applied unitary circuit, giving rise to a mixed quantum-classical control. If accessible, feed-forward protocols can provide a way to cut down the preparation time for various many-body states since they can evade some constraints on information spreading imposed by locality, such as Lieb-Robinson bounds or quantum speed limits~\cite{verresen2021efficiently,piroliquantum2021,smith2023deterministic,zhu2023nishimori,smith2024constant,iqbal2024topological}. 

The complexity associated with developing general control protocols that are either robust to, or leverage, environmental noise indicates that hybrid approaches are likely to be the most fruitful path forward and therefore a greater integration of the various control sub-communities is vital moving forward.

\subsection{Quantum annealing and shortcuts-to-adiabaticity}
Quantum annealers are analog quantum devices designed to access low-energy states of \steve{arbitrary} Hamiltonians on specific interaction graphs which can encode solutions to practical problems, for example, in combinatorial optimization~\cite{AdiabaticComp}. These devices largely rely on being able to adiabatically transform simple Hamiltonians to more complex ones, in the way described in Sec.~\ref{sec:CD}. The success of annealing protocols is predominantly limited by the coherence time of the device, and the scaling of the minimum energy gap for a given problem. Counterdiabatic driving can be a useful tool to lower the total evolution time used by a control protocol, to make it fit within the coherence time of the device. These tools have recently begun to be implemented in quantum annealing prototypes, as well as their gate-based extensions (like QAOA, see Sec.~\ref{outlook_sec_qc}).

Problems, where the minimum energy gap is too small or scales unfavourably with system size, represent a major challenge for quantum annealers. From a physics perspective, it is known that minimum gaps scaling polynomially with system size are typically associated with second-order (continuous) phase transitions, while exponentially vanishing gaps are linked to first-order (discontinuous) phase transitions \cite{jorg2010}. To mitigate the gap scaling, it has been proposed to include so-called catalysts in the annealing protocol \cite{ghosh2024} or employing augmented counterdiabatic techniques~\cite{Balducci1}. These are additional terms in the Hamiltonian that provide intermediate stops in the annealing schedule. It has been shown that such strategies can substantially improve gap scaling. Unfortunately, it is hard to know a priori what kind of catalyst Hamiltonian will be needed for a given problem. Connecting CD driving to the Catalyst formalism could help provide a more formal ground to these tools, and improve their usefulness \cite{prielinger2021}. More generally, the properties of CD protocols could be used to probe interesting connections between computational complexity, gap scaling, and quantum phase transitions (see also the discussion in Sec.~\ref{outlook_sec_agp_insight}).

\subsection{Variational quantum algorithms and quantum optimal control}

More generally, an important connection has been recognised between the formalism of quantum optimal control and variational quantum algorithms (VQAs) \cite{cerezo2021}, which have led the charge in the search of useful applications of near-term NISQ computers. In a generic VQA, a circuit is formed of a series of parameter-dependent gates $U_i(\{\theta_k\})$, which act upon a fiduciary state, typically $\ket{0}^{\otimes N}$. Measuring the qubits at the end of the circuit allows \steve{us} to evaluate some cost function, for instance, associated with the energy of a target model. This cost function is then fed to a classical optimizer, which searches for a new set of parameters $\{\theta_k\}$ to try in the next cycle of the hybrid quantum-classical procedure.

The VQA framework is the foundation for potential applications such as the quantum approximate optimization algorithm (QAOA), the variational quantum eigensolver (VQE), and many quantum machine learning (QML) approaches. The connection between the QOC formalism laid out in Sec.~\ref{sec:QOC} and VQAs is evident; the main difference being that the quantum evolution necessary to evaluate the cost function is implemented in an actual quantum processor in the latter, instead of numerically simulated classically in the former. VQAs have been intensely explored in recent years and have revealed important properties of what we could expect in quantum optimal control when scaling up system size and in \steve{the} presence of noise \cite{magann2021,larocca2022}. An example of these is the phenomenon of barren plateaus, which predict that certain highly expressible (i.e. controllable \cite{dalessandro_book}) quantum circuits lead to cost function gradients that decrease exponentially with system size, meaning that the resulting optimization being performed by the classical computing resource becomes prohibitive at large scales. The presence of noise has been shown to aggravate this problem. Workarounds to this issue imply the use of less expressible VQA ansatze, which in quantum control language set the stage for studying complex quantum systems that are not fully, but partially controllable \cite{ragone2024}. 

\subsection{Quantum control for quantum computing} \label{outlook_sec_qc}

As we have mentioned throughout this tutorial, quantum optimal control already plays a major role in the development of quantum computers, as it has allowed to improve the fidelities of entangling gates in many physical platforms. 

One prominent example is the use of the so-called cross resonance (CR) gate \cite{rigetti2010fully,groszkowski2011tunable}. This has been a driving protocol behind many  advances in quantum computing based on superconducting technology \cite{sheldon2016procedure}. The CR gate provides a protocol that requires only microwave control. This reduces the need for magnetic flux lines in superconducting circuit designs that could otherwise become sources of additional noise. By employing various types of optimization, very high fidelities for the CR gates have been achieved~\cite{kandala2021demonstration,wei2022hamiltonian}, as well as extension of the CR gate based on the same including the so-called cross-cross resonance gate~\cite{heya2021cross}. This work has used optimal control in a closed loop that takes into account the hardware specifics \cite{werninghaus2021}. At this point, the high CR gate fidelities are typically measured in systems with a limited number of qubits~\cite{malekakhlagh2022mitigating}, and it is not given that the control protocols used here will also apply as one scales the system to a large number of qubits. This has led to proposals for new types of pulse shaping in terms of DRAG protocols designed to work in the presence of a large number of qubits \cite{li2024experimental}. 

Other major commercial team building superconducting quantum computers have also made strides towards high-fidelity gates by using quantum control. New ways to perform DRAG pulse shaping have been implemented, resulting in large reductions in single qubit gate errors for transmon systems~\cite{hyyppa2024reducing}. Furthermore, new approaches to reduce errors associated with readout and reset using quantum optimal control can be found in \cite{gautier2025optimal} with key examples from transmon systems. 

A drawback of transmon systems is their low anhoarmonicity in the spectrum that leads to physical limitations on the speed of quantum gates. This has led to the pursuit of other qubit modalities, and a prominent recent example is flux qubits that may allow for very fast gates. The regime of interest is influenced by driving pulses in several important ways, including inducing AC Stark shifts of the energy levels and unintended leakage to levels outside the computational (two-level) space. Countering these effects is a main venue for pulse shaping techniques. Experimental work \cite{rower2024suppressing} has shown how to mitigate these effects and achieve very high single-qubit gate fidelities. However, this comes at the cost of additional hardware requirements. In a very recent paper \cite{zwanenburg2025single} it was shown how one may avoid hardware complexity by rethinking the pulse shaping schemes. This requires a reworking of the rotating wave approximation that is applied widely in the superconducting circuit community, but that is not expected to hold when targeting very fast gates. 

For trapped ion quantum computing, a major ingredient in many experimental implementations is the entangling gate proposed by S{\"o}rensen and M{\"o}lmer \cite{sorensen2000entanglement}. This has been hugely successful as a protocol for achieving high-fidelity gates in ion trap quantum computers \cite{schafer2018fast,gerster2022experimental,moses2023race}. In a recent work by Kirchhoff and co-workers \cite{kirchhoff2025correction}, a new theoretical analysis of the scheme in \cite{sorensen2000entanglement} shows that there may be errors that have not previously been accounted for and that when included these properly, one finds a lower limit on the error that matches well with recent experimental errors in the different ion trap quantum computing implementations. The work of \cite{kirchhoff2025correction} then goes on to suggest how to proceed using control techniques to further reduce errors using, e.g., pulse shaping. 

A key point in which control has yielded decisive benefits for quantum computing has been in the optimization of performance for near-term noisy quantum devices. In the context of variational quantum circuits (see the previous section), this has led to exploration of how to use parametrized quantum gates to shorten circuit depth and reduce noise for increased performance. One such effort has been directed at quantum computing to simulate the behavior of fermionic systems in Ref.~\cite{barends2015digital}, that took advantage of fast adiabatic control~\cite{martinis2014fast}. Further work has since demonstrated how these features can be used for quantum chemistry and quantum materials calculations on quantum computers~\cite{ganzhorn2019gate, foxen2020demonstrating}. Parametrized two-qubit gates are expected to be the better option for these quantum simulation tasks on quantum computers \cite{lacroix2020improving,rasmussen2022parameterized} and it is expected to bring further challenges and opportunities for quantum control \cite{shi2020resource}.

Beyond the two-qubit gates, there are many instances in which gates acting on three or more qubits can be a major benefit in the implementation of quantum algorithms on quantum computers \cite{shi2020resource}. The controlled operations of basic algorithms such as Grover search and other oracle-based designs, as well as key steps in Shor's factoring algorithm, will imply an overhead in decomposition into single- and two-qubit gates. Even in the presence of error correction, this overhead will influence the execution time as the problem size grows. Here, the availability of $N$-qubit operations can be used to reduce the overhead, and in many cases by significant amounts. Furthermore, three-qubit gates are crucial ingredients of quantum error correction protocols \cite{nielsen2010quantum}, and as such are highly desirable operations to have in any quantum computing architecture with low overhead and high quality. Hence, major efforts have been put into realizing multi-qubit gates with high fidelity.

Optimized multi-qubit quantum gates have been considered in several platforms. As an example, the three-qubit Toffoli gate has been considered using a single-shot protocol within superconducting transmon qubits \cite{zahedinejad2015high}. This can be generalized to multi-qubit gates beyond three qubits such as the $n$-bit Toffoli or related controlled gates that may be realized using Rydberg atoms \cite{khazali2020fast}, trapped ions \cite{rasmussen2020single} and superconducting circuits \cite{khazali2020fast,rasmussen2020single,warren2023extensive}. Other controlled three-qubit gates realized with quantum optimal control have been proposed that target time-optimal implementations \cite{jandura2022time} in neutral atoms that may be used for quantum computing \cite{henriet2020quantum} and quantum simulation tasks \cite{morgado2021quantum} .
For trapped ion quantum computing, the Grover search algorithm has been realized on a small set of qubits \cite{figgatt2017complete} using three- and four-qubit gates \cite{katz2023demonstration}.

The topic of multiqubit gates is intimately connected to the exploration of systems beyond qubits, i.e., systems in which the local Hilbert space has dimension $d>2$, so-called qudits. One may also view this from the point of view of breaking the abstraction that the assumption of having qubits in the system binds us to \cite{shi2020resource}. 
Here one can implement quantum algorithms in a more versatile fashion \cite{wang2020qudits,rasmussen2020reducing}. The gates acting on qudits are more involved and can present new challenges for control techniques as multiple levels are now active and it may be advantageous to activate several transitions at once to realize qudit gates \cite{baekkegaard2019realization,shi2020resource}. This may in turn introduce new issues as there are now multiple ways in which leakage can occur and it is can be more challenging to design quantum control protocols to mitigate these effects.
Using different types of protocols it has by now been possible to realize qudit-based quantum processors in several platforms, including cold atoms \cite{davis2019photon}, superconducting circuits \cite{blok2021quantum}, trapped ions \cite{ringbauer2022universal}, and photonics \cite{reimer2019high,chi2022programmable}.

There are other important connections between quantum control and quantum computing development which have been more recently proposed, and are currently being exploited. A natural extension is to use quantum control to perform quantum tasks that are specific to different tasks in the quantum computing stack. For instance, developing fault-tolerant quantum algorithms can benefit from using optimal control to generate magic states~\cite{omanakuttan2024} or parity check unitaries \cite{lewis2024} for quantum error correction, developing robust implementations of quantum gates~\cite{CarolanPRA2023}, and for the manipulation of surface code defects~\cite{raii2024}. It has also been demonstrated that quantum optimal control can be used to synthetize gates for digital quantum simulation circuits~\cite{lysne2020,kairys2021}. The lessons learned from STA, the impact of ML to train for general scenarios, and, importantly, the framework of quantum optimal control are crucial for the development and optimisation of quantum error suppression, mitigation, and correction which are a critical part of the strategies and roadmaps being pursued throughout the quantum computing ecosystem, in both experiments and theory, across government, academia, and the growing quantum computing industry.

\section*{Jupyter Notebooks}

Jupyter notebooks are available on GitHub~\cite{github_code} for interested readers to explore different code parameters in the Examples discussed in the main text. These notebooks can also be used as a starting point to apply the control techniques introduced in this tutorial to open problems at the forefront of research. 

\acknowledgments
We thank all the wonderful people we've worked with on quantum control over the years. All authors thank Andrew J. Daley, Christiane Koch, and Anatoli Polkovnikov for helpful discussions and advice throughout the writing of this tutorial.
SC is particularly grateful to Eoin Carolan, Anthony Kiely, and George Mihailescu for valuable discussions and feedback. MB thanks Paul M.~Schindler and F.~Balducci for numerous related discussions. CWD thanks Ewen D. C. Lawrence, Ieva \v Cepait\.e, and Stewart Morawetz for numerous related discussions. PMP is grateful to Anupam Mitra, Nathan Lysne, Sivaprasad Omanakuttan, Poul Jessen, and Ivan Deutsch for many helpful optimal control discussions over the years. NTZ would like to thank Stig Elkj{\ae}r Rasmussen, Kasper Sangild Christensen, Kasper Poulsen, and in particular Alan Costa dos Santos for collaboration and discussions on projects involving quantum control.
SC acknowledges support from the Science Foundation Ireland Starting Investigator Research Grant ``SpeedDemon" No. 18/SIRG/5508 and the Alexander von Humboldt Foundation. PMP acknowledges support by U.S. National Science
Foundation (grant number PHY-2210013). CWD was supported by the Engineering and Physical Sciences Research Council through Grant No. EP/Y005058/2. MB was funded by the European Union (ERC, QuSimCtrl, 101113633). Views and opinions expressed are however those of the authors only and do not necessarily reflect those of the European Union or the European Research Council Executive Agency. Neither the European Union nor the granting authority can be held responsible for them. NTZ is supported in part by the Novo Nordisk Foundation, the Innovation Fund Denmark, as well as the Independent Research Fund Denmark. \\

\begin{center}{\bf DATA AVAILABILITY} 
\end{center}
The data and the notebooks associated with this manuscript version are openly available on Zenodo~\cite{zenodo_record}.

\newpage
\appendix*
\section*{List of abbreviations}
\begin{table}[!h]
\centering
\begin{tabular}{|c|p{7.5cm}|}
\hline
Abbreviation & Definition \\
\hline\hline
2LS & two-level system \\
\hline
AC & actor-critic \\
\hline
AGP & adiabatic gauge potential \\
\hline
BEC & Bose-Einstein condensate \\
\hline
BFGS & Broyden-Fletcher-Goldfarb-Shannon \\
\hline
CD & counterdiabatic \\
\hline
CRAB & chopped randomised basis \\
\hline
DDGP & deep deterministic policy gradient \\
\hline
DQN & deep Q-netowrk \\
\hline
DRAG & derivative removal by adiabatic gate \\
\hline
FPGA & field-programmable gate array \\
\hline
GKP & Gottesman-Kitaev-Preskill \\
\hline
GRAPE & gradient-ascent pulse engineering \\
\hline
L-BFGS-B & limited memory variant of BFGS \\
\hline
LMG & Lipkin-Meshkov-Glick \\
\hline
LZ & Landau-Zener \\
\hline
ML & machine learning \\
\hline
NISQ & noisy intermediate-scale quantum \\
\hline
NMR & nuclear magnetic resonance \\
\hline
NV & Nitrogen-vacancy \\
\hline
PG & policy gradient \\
\hline
PPO & proximal policy optimisation \\
\hline
QAOA & quantum approximate optimisation algorithm \\
\hline
QML & quantum machine learning \\
\hline
QOC & quantum optimal control \\
\hline
QSL & quantum speed limit \\
\hline
RL & reinforcement learning \\
\hline
STA & shortcuts to adiabaticity \\
\hline
STIRAP & stimulated rapid adiabatic passage \\
\hline
TRPO & trust-region policy optimisation \\
\hline
VQA & variational quantum algorithm \\
\hline
VQE & variational quantum eigensolver \\
\hline

\hline
\end{tabular}
\caption{Summary of abbreviations used throughout.}
\label{table:Abbreviations}
\end{table}

\clearpage


\bibliography{references}

\end{document}